\documentclass{rsproca}

\usepackage{color}

\begin{document}

\title{Turbulent Reconnection and Its Implications}

\author{A. Lazarian$^{1}$, G. Eyink$^{2}$, E. Vishniac$^{3}$ and G. Kowal$^4$}

\address{\tiny
$^{1}$ Department of Astronomy, University of Wisconsin, 475 North Charter Street, Madison, Wisconsin 53706, USA\\
$^{2}$ Department of Applied Mathematics and Statistics, The Johns Hopkins University, Baltimore, Maryland 21218, USA\\
$^{3}$ Department of Physics and Astronomy, McMaster University, 1280 Main Street West, Hamilton, Ontario L8S 4M1, Canada\\
$^{4}$ Escola de Artes, Ci\^{e}ncias e Humanidades, Universidade de S\~{a}o Paulo, Av. Arlindo B\'{e}ttio, 1000 -- Ermelino Matarazzo, CEP
03828-000, S\~{a}o Paulo, SP, Brazil
}

\subject{xxxxx, xxxxx, xxxx}


\corres{A. Lazarian\\
\email{lazarian@astro.wisc.edu}}

\begin{abstract}
\small
Magnetic reconnection is a process of magnetic field topology change, which is
one of the most fundamental processes happening in magnetized plasmas.  In most
astrophysical environments the Reynolds numbers corresponding to plasma flows
are large and therefore the transition to turbulence is inevitable.  This
turbulence, which can be pre-existing or driven by magnetic reconnection itself,
must be taken into account for any theory of magnetic reconnection that attempts
to describe the process in the aforementioned environments.  This necessity is
obvious as 3D high resolution numerical simulations show the transition to the
turbulence state of initially laminar reconnecting magnetic fields.  We discuss
ideas of how turbulence can modify reconnection with the focus on the Lazarian
\& Vishniac (1999) reconnection model.  We present numerical evidence supporting
the model and demonstrate that it is closely connected to the experimentally
proven concept of Richardson dispersion/diffusion as well as to more recent
advances in understanding of the Lagrangian dynamics of magnetized fluids.  We
point out that the Generalized Ohm's Law that accounts for turbulent motion
predicts the subdominance of the microphysical plasma effects for reconnection
for a realistically turbulent media.  We show that on of the most dramatic
consequences of turbulence is the violation of the generally accepted notion of
magnetic flux freezing.  This notion is a corner stone of most theories dealing
with magnetized plasmas and therefore its change induces fundamental shifts in
accepted paradigms, for instance, turbulent reconnection entails reconnection
diffusion process that is essential for understanding star formation.  We argue,
that at sufficiently high Reynolds numbers the process of tearing reconnection
should transfer to turbulent reconnection.  We discuss flares that are predicted
by turbulent reconnection and relate this process to solar flares and gamma ray
bursts.  With reference to experiments, we analyze solar observations, in-situ
as measurements in the solar wind or heliospheric current sheet and show the
correspondence of data with turbulent reconnection predictions.  Finally, we
discuss First Order Fermi acceleration of particles that is a natural
consequence of the turbulent reconnection.
\end{abstract}



\begin{fmtext}
\end{fmtext}


\maketitle

%
\section{Problem of Magnetic Reconnection in Realistically Turbulent Plasmas}

Magnetic fields are known to critically modify the dynamics and properties of
magnetized plasmas.  It is generally accepted that magnetic fields embedded in a
highly conductive fluid retain their topology for all time due to the magnetic
fields being frozen-in \cite{Alfven:1943, Parker:1979}. This concept of
frozen-in magnetic fields is a basis of many theories, e.g. of the theory of
star formation in magnetized interstellar medium.

In spite of this, there is ample evidence that magnetic fields in highly
conducting ionized astrophysical objects, like stars and galactic disks, show
evidence of changes in topology, i.e. ``magnetic reconnection'', on dynamical
time scales \cite{Parker:1970, Lovelace:1976, PriestForbes:2002}.  Historically,
magnetic reconnection research was motivated by observations of the solar corona
\cite{Innes_etal:1997, YokoyamaShibata:1995, Masuda_etal:1994} and this
influenced attempts to find peculiar conditions conducive for flux conservation
violation, e.g. special magnetic field configurations or special plasma
conditions.  For instance \cite{PriestForbes:2002} showed examples of magnetic
configurations that produce fast reconnection and much work has been done
showing how reconnection can be accelerated in plasmas with very small collision
rates \cite{Shay_etal:1998, Drake:2001, Drake_etal:2006, Daughton_etal:2006,
UzdenskyKulsrud:2006} (see also reviews \cite{Bhattacharjee_etal:2003,
ZweibelYamada:2009, Yamada_etal:2010} and references therein).  However, it is
clear that reconnection is a ubiquitous process taking place in various
astrophysical environments.  For instance, magnetic reconnection can be inferred
from the existence of large-scale dynamo activity inside stellar interiors
\cite{Parker:1993, Ossendrijver:2003}. We would argue that it is also required
to enable the eddy-type motions in magnetohydrodynamic (MHD) turbulence, e.g. in
the Goldreich \& Sridhar turbulence \cite{GoldreichSridhar:1995}.  In fact, it
is easy to show that without fast magnetic reconnection magnetized fluids would
behave like Jello or felt, rather than as a fluid.

It is clear that solar flares \cite{Sturrock:1966} are just one vivid example of
reconnection activity.  Other dramatic reconnection events attributed to
reconnection include $\gamma$-ray bursts (see \cite{LyutikovLazarian:2013} for a
review), while reconnection routinely takes place in essentially everywhere both
in collisional and collisionless magnetized plasmas.  Incidentally, magnetic
reconnection occurs rapidly in computer simulations due to the high values of
resistivity (or numerical resistivity) that are employed at the resolutions
currently achievable.  Therefore, if there are situations where magnetic fields
reconnect slowly, numerical simulations do not adequately reproduce the
realities of astrophysical plasmas. This means that if collisionless
reconnection is the only way to make reconnection rapid, then numerical
simulations of many astrophysical processes including those of the interstellar
medium (ISM), which is collisional, are in error.  Fortunately, observations of
collisional solar photosphere indicate that the reconnection is fast in these
environments (see \cite{ShibataMagara:2011}), which contradicts to the idea that
being collisionless is the prerequisite for plasma to reconnect fast.

What makes reconnection challenging to explain is that it is not possible to
claim that reconnection must always be rapid empirically, as solar flares
require periods of flux accumulation time, which corresponds to slow
reconnection.  Thus magnetic reconnection should have some sort of trigger,
which should not depend on the parameters of the local plasma.  In this review
we argue that the trigger is turbulence. This opens a wide avenue for the
application of turbulent reconnection theory to explain the astrophysical
explosions, e.g. solar and stellar flares and superflares, as well as gamma ray
bursts.

A lot of support to models of reconnection based on plasma physics comes from
the {\em in situ} measurements of magnetospheric reconnection. While important
for some practical purposes, e.g. for some aspects of Space Weather program,
this reconnection happens on scales comparable to the ion inertial length and
therefore is atypical for large scale reconnection that happens in most
astrophysical systems. We argue that the large scale magnetic reconnection is
based on MHD turbulence physics making small scale plasma reconnection processes
irrelevant for the reconnection rates that are attained.

With the advent of numerical simulations it becames clear that the regular
schemes of reconnection, like classical Sweet-Parker or Petschek reconnections,
do not work. Instead, the reconnecting systems transfer to a more chaotic state
which is characterized by the hierarchy of magnetic islands in 2D
\cite{Loureiro_etal:2007, Lapenta:2008, Daughton_etal:2009a,
Daughton_etal:2009b, Bhattacharjee_etal:2009, Cassak_etal:2009} or a more
complex chaotic state in 3D \cite{KarimabadiLazarian:2013}.  We argue that as
the scale of reconnection layers increases the turbulent reconnection will
inevitably take over, modifying and suppressing the plasmoid instability that
gives rise to currently observed picture.

Turbulence generation has long been associated with magnetic reconnection
processes (see \cite{LaRosaMoore:1993}). This review, however, is mostly dealing
with how turbulence changes the rates of magnetic reconnection, although we also
consider turbulence generation by reconnection.

Magnetic reconnection is a ubiquitous process in turbulent media, but it is not
easy to observe as reconnection transfers most of the energy into kinetic motion
related to smaller scale eddies, thus supporting the energy cascade. Apart from
Solar flares, which dynamics can be compared with the predictions of turbulent
reconnection, {\em in-situ} measurement of reconnection available for the solar
wind provide ways of testing theoretical predictions. We show that both sets of
data are consistent with turbulent MHD based magnetic reconnection.

It is worth noting, that our discussion addresses 3D magnetic reconnection. The
change of dimensionality of physical problems changes frequently the physics
involved. For the theory based on MHD turbulence, this is very important to note
that the properties of MHD turbulence are very different in 2D and 3D.

The theory of turbulent reconnection that we describe is based on the Lazarian
\& Vishniac work (\cite{LazarianVishniac:1999}, henceforth LV99) and the
extensions of the original model in subsequent publications, for instance in
Eyink, Lazarian \& Vishniac (\cite{Eyink_etal:2011}, henceforth ELV11).  The
original LV99 model was supported by numerical simulations, some results of
which have been published (see \cite{Kowal_etal:2009, Kowal_etal:2012,
Eyink_etal:2011}), as well as different pieces of observational evidences that
we describe in the review.  Additional theoretical support for the model comes
from very recent work by \cite{Eyink:2014}.  While our review reflects our
optimism based on the successes of the LV99 model in explaining different
astrophysical phenomena, e.g. gamma ray bursts (\cite{ZhangYan:2011} and ref.
therein), removal of magnetic fields from molecular clouds (e.g.
\cite{Lazarian:2012} and ref. therein), we feel that the challenges presented by
the variety of astrophysical conditions are very stimulating for further studies
of magnetic reconnection.  We also accept the limitations of our model that is
intended for describing the astrophysical phenomena at large scales and
therefore adopting MHD approximation. Therefore, magnetic reconnection happening
at the scale of ion Larmor radius, as is the case of the Earth magnetosphere
cannot be described by the model. Important cases of magnetic reconnection in
the presence of plasma effects as well as plasmoid instabilities are described
in an extensive review by \cite{Yamada_etal:2010}.  Other cases when magnetic
reconnection can be fast in MHD regime without turbulence are discussed at
length e.g. in an excellent book by \cite{PriestForbes:2000}.  Thus this review
should be viewed as a personalized outlook on the reconnection problem by the
authors who are exploring the connection of the two ubiquitous processes,
namely, magnetic reconnection and astrophysical turbulence, while many issues of
the problem are far from being finally settled and different ideas and
alternative models are being tested and explored by different research groups.
We accept that magnetic reconnection, similar to magnetic turbulence, is a very
deep subject where the synergy of different approaches and techniques may prove
to be beneficial eventually. We also note that the claim that turbulence can
accelerate magnetic reconnection predates the LV99 model (see
\cite{Speiser:1970, JacobsonMoses:1984, BhattacharjeeHameiri:1986,
MatthaeusLamkin:1986, Strauss:1986}). Some new approaches to turbulent
reconnection were formulated more recently (see \cite{Guo_etal:2012}). In the
review we provide a comparison of LV99 with these approaches.

In what follows, we argue that turbulent reconnection is the generic process
taking place in astrophysical environments which are turbulent due to the huge
Reynolds numbers of the flows involved. The turbulence can be pre-existing and
also self-generated by the reconnection process. We provide the MHD description
of astrophysical turbulence in section 2, describe LV99 model of turbulent
reconnection in section 3, provide its elaboration and extension in section 4,
demonstrate examples of the numerical testing of the model in section 5, discuss
the observational testing of the model with solar data and solar wind data in
section 6, outline the implications of the model in section 7, provide a
comparison of the model with other ideas of fast stochastic reconnection in
section 8. We conclude by discussing the general tendency of models of
reconnection to get more stochastic in section 9.

%
\section{Astrophysical Turbulence and Its MHD Description}
\label{sec:turbulence}

Observations of the interstellar medium reveal a Kolmogorov spectrum of electron
density fluctuations (see \cite{Armstrong_etal:1995, ChepurnovLazarian:2010}) as
well as steeper spectral slopes of supersonic velocity fluctuations (see
\cite{Lazarian:2009} for a review).  Measurements of the solar wind fluctuations
also reveal turbulence power spectrum \cite{Leamon_etal:1998}).  Ubiquitous
non-thermal broadening of spectral lines as well as measures obtained by other
techniques (see \cite{Burkhart_etal:2010}) confirm that turbulence is present
everywhere in astrophysical environments where we test for its existence.  This
is not surprising as magnetized astrophysical plasmas generally have very large
Reynolds numbers due to the large length scales involved and the fact that the
motions of charged particles in the direction perpendicular to magnetic fields
are constrained.  Laminar plasma flows at these high Reynolds numbers are prey
to numerous linear and finite-amplitude instabilities, from which turbulent
motions readily develop\footnote{In addition, the mean free path of particles
can also be constrained by the instabilities developed on the collisionless
scales of plasma (see \cite{Schekochihin_etal:2009, LazarianBeresnyak:2006,
BrunettiLazarian:2011}).  In this situation not only Alfv\'enic but also
compressible turbulent modes can survive.}.

\begin{figure}[!t]
\centering
\includegraphics[width=1.0 \columnwidth]{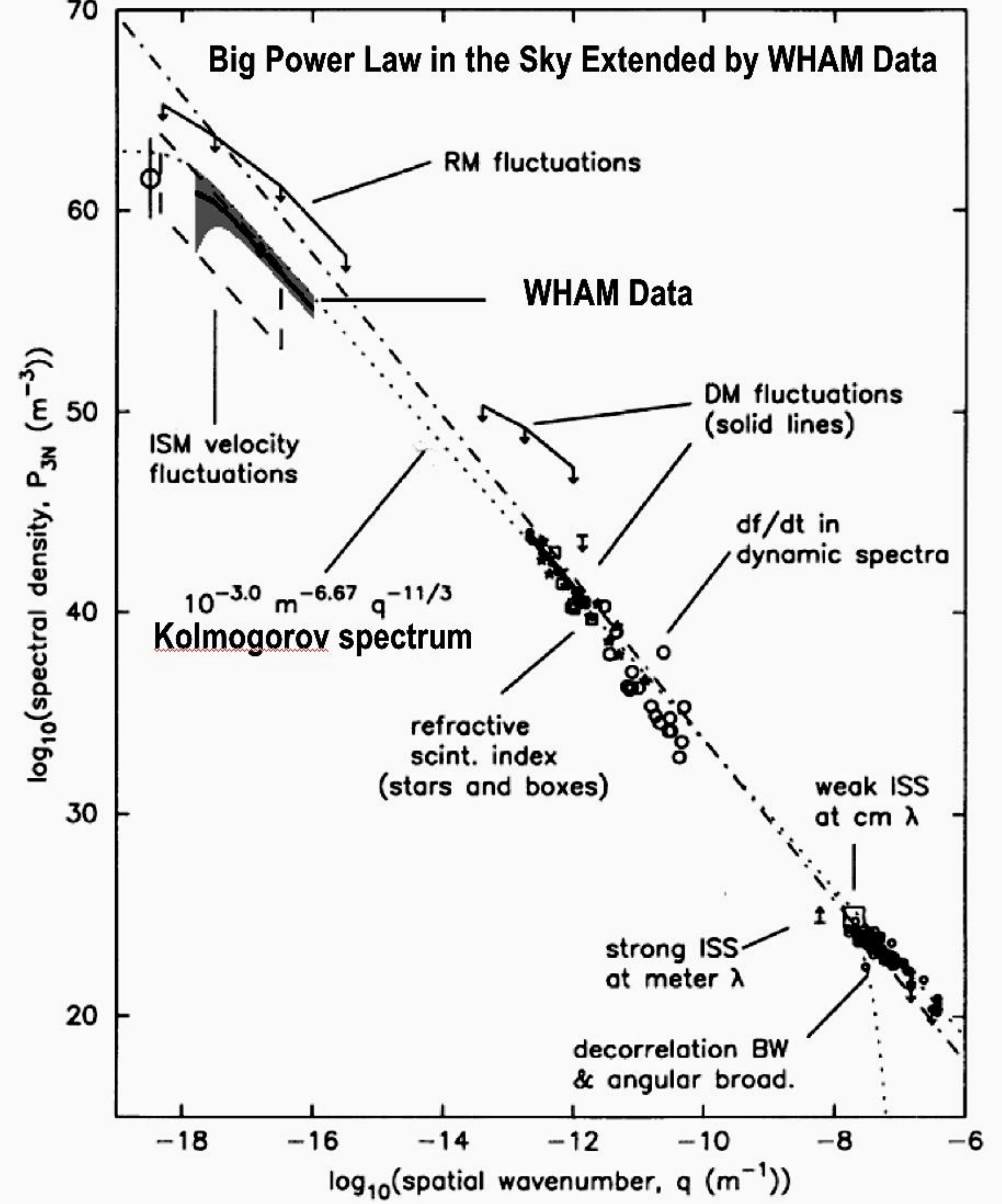}
\caption{Big power law in the sky from \cite{Armstrong_etal:1995} extended to
scale of parsecs using WHAM data. From \cite{ChepurnovLazarian:2010}.
\label{f1}}
\end{figure}

Indeed, observations show that turbulence is ubiquitous in all astrophysical
plasmas.  The spectrum of electron density fluctuations in Milky Way is
presented in Figure~\ref{f1}, but similar examples are discussed in
\cite{Leamon_etal:1998, Bale_etal:2005} for solar wind, \cite{Padoan_etal:2006}
for molecular clouds and \cite{VogtEnsslin:2005} for the intracluster medium.
The plasma turbulence is sometimes driven by an external energy source, such as
supernova in the ISM \cite{NormanFerrara:1996, Ferriere:2001}, merger events and
active galactic nuclei outflows in the intercluster medium (ICM)
\cite{Subramanian_etal:2006, EnsslinVogt:2006, Chandran:2005}, and baroclinic
forcing behind shock waves in interstellar clouds.  In other cases, the
turbulence is spontaneous, with available energy released by a rich array of
instabilities, such as magneto-rotational instability (MRI) in accretion disks
\cite{BalbusHawley:1998}, kink instability of twisted flux tubes in the solar
corona \cite{GalsgaardNordlund:1997, GerrardHood:2003}, etc.  In all these
cases, turbulence is not driven by reconnection.  Nevertheless, we would like to
mention that an additional driving of turbulence through the energy release in
the reconnection zone can sometimes be important, especially in magnetically
dominated low beta plasmas.  We discuss the case of turbulence driven by
reconnection in section 4c.  All in all, whatever its origin, the signatures of
plasma turbulence are seen throughout astrophysical media.

As turbulence is known to change dramatically many processes, in particular,
diffusion and transport processes, it is natural to pose the question to what
extent the theory of astrophysical reconnection must take into account the
pre-existing turbulent environment.  We note that even if the plasma flow is
initially laminar, kinetic energy release by reconnection due to some plasma
process, e.g. tearing and related plasmoid generation, is expected to generate
vigorous turbulent motion in high Reynolds number fluids.

Turbulence in plasma happens at many scales, from the largest to those below the
proton Larmor radius.  To understand at what scales the MHD description is
adequate one needs to reiterate a few known facts \cite{Kulsrud:1983,
Eyink_etal:2011}.  Indeed, to describe magnetized plasma dynamics one should
deal with three characteristic length-scales: the ion gyroradius $\rho_i$, the
ion mean-free-path length $\ell_{mfp,i}$ arising from Coulomb collisions, and
the scale $L$ of large-scale variation of magnetic and velocity fields.

The MHD approximation is definitely applicable to ``strongly collisional''
plasma with $\ell_{mfp,i}\ll \rho_i$.  This is the case, e.g. of star interiors
and most accretion disk systems.  For such ``strongly collisional'' plasmas a
standard Chapman-Enskog expansion provides a fluid description of the plasma
\cite{Braginskii:1965},  with a two-fluid model for scales between
$\ell_{mfp,i}$ and the ion skin-depth $\delta_i= \rho_i/\sqrt{\beta_i}$ and an
MHD description at scales much larger than $\delta_i$.

Hot and rarefied astrophysical plasmas are often ``weakly collisional'' with
$\ell_{mfp,i}\gg \rho_i$.  Indeed, the relation that follows from the standard
formula for the Coulomb collision frequency (e.g. see \cite{Fitzpatrick:2011})
is
\begin{equation}
\frac{\ell_{mfp,i}}{\rho_i}\propto \frac{\Lambda}{\ln\Lambda}\frac{V_A}{c}, \label{lmfp-rho}
\end{equation}
where  $\Lambda=4\pi n\lambda_D^3$ is the plasma parameter, or the number of
particles within the Debye screening sphere.  For some media that $\Lambda$ can
be large.

For the ``weakly collisional'' but well magnetized plasmas one can invoke the
expansion over the small ion gyroradius.  This results in the ``kinetic MHD
equations'' for lengths much larger than $\rho_i$.  The difference between these
equations and the MHD ones is that the pressure tensor in the momentum equation
is anisotropic, with the two components $p_\|$ and $p_\perp$ of the pressure
parallel and perpendicular to the local magnetic field direction
\cite{Kulsrud:1983}.  In ``weakly collisional'', i.e.  $L\gg \ell_{mfp,i}.$, and
collisionless, i.e.  $\ell_{mfp,i}\gg L$ systems turbulence is bound to induce
instabilities that limit the effective mean free path $[\ell_{mfp,i}]_{eff}$ by
magnetically mediated scattering of particles \cite{SchekochihinCowley:2006,
LazarianBeresnyak:2006}.  This effective mean free path is a game changer and it
is not surprising that numerical simulations in \cite{SantosLima_etal:2013} that
accounted for this effect demonstrated that turbulence in ``collisionless
plasmas'' of galaxy clusters is very similar to MHD turbulence on the scales
larger than $[\ell_{mfp,i}]_{eff}$.

We can also note that additional simplifications that justify the MHD approach
occur if the turbulent fluctuations are small compared to the mean magnetic
field, and having length-scales parallel to the mean field much larger than
perpendicular length-scales.  Treating wave frequencies that are low compared to
the ion cyclotron frequency we enter the domain of ``gyrokinetic approximation''
which is commonly used in fusion plasmas e.g. \cite{Schekochihin_etal:2007,
Schekochihin_etal:2009}, for which at length-scales larger than the ion
gyroradius $\rho_i$ the incompressible shear-Alfv\'{e}n wave modes get decoupled
from the compressive modes and can be described by the simple ``reduced MHD''
(RMHD) equations.  These Alfv\'{e}n modes are most important for fast magnetic
reconnection, what we discuss later in the review.

In short, our considerations above confirm the generally accepted notion that
the MHD approximation is adequate for most astrophysical turbulent plasmas at
sufficiently large scales.  In particular, the Goldreich-Srindhar
\cite{GoldreichSridhar:1995} (henceforth GS95) theory of Alfv\'enic turbulence
should be true for describing Alfv\'enic part of the MHD turbulent
cascade\footnote{We will concentrate on Alfv\'{e}nic modes, while disregarding
the slow and fast magnetosonic modes of MHD turbulence \cite{ChoLazarian:2002,
ChoLazarian:2003, KowalLazarian:2010}, which is possible as the backreaction of
fast and slow modes on Alfv\'enic cascade is insignificant
\cite{ChoLazarian:2002, GoldreichSridhar:1995, LithwickGoldreich:2001}.}.  For
Alfv\'enic turbulence the eddies are elongated along magnetic field with the
relation between the parallel and perpendicular dimensions due to the critical
balance condition, namely,
\begin{equation}
\ell_{\|}^{-1}V_A\sim \ell_{\bot}^{-1}\delta u_\ell,
\label{crit}
\end{equation}
where $\delta u_\ell$ is the eddy velocity, while $\ell_{\|}$ and $\ell_{\bot}$
are eddy scales parallel and perpendicular to the {\it local} direction of
magnetic field, respectively.  The notion of local magnetic field is the
essential part of the modern understanding of Alfv\'enic turbulence and it was
added to the GS95 picture by the later studies (LV99, \cite{ChoVishniac:2000,
MaronGoldreich:2001}).  The use of local magnetic field is expected as at small
scale eddies can be influenced only by the magnetic field around them and not by
the global mean field.

A description of MHD turbulence that incorporates both weak and strong regimes
was presented in LV99.  In the range of length-scales where turbulence is
strong, this theory implies that
\begin{equation}
\ell_{\|}\approx L_i \left(\frac{\ell_{\bot}}{L_i}\right)^{2/3} M_A^{-4/3}
\label{Lambda}
\end{equation}
\begin{equation}
\delta u_{\ell}\approx u_{L} \left(\frac{\ell_{\bot}}{L_i}\right)^{1/3} M_A^{1/3},
\label{vl}
\end{equation}
when the turbulence is driven isotropically on a scale $L_i$ with an amplitude
$u_L$.  As we see further, driving of turbulence by reconnection may be
different from the isotropic driving assumed for the derivation of the
expressions above.

We do not discuss theories of Alfv\'enic turbulence that were develop to obtain
the spectral index of $-3/2$ which was suggested by limited-resolution numerical
simulations, e.g. in \cite{MaronGoldreich:2001}\footnote{Low resolution
numerical simulations are notorious in being ambiguous in terms of spectral
slope.  For instance, the initial compressible simulations suggested the
spectral index of high Mach number hydrodynamic turbulence to be $-5/3$, which
prompted theoretical attempts to explain this, e.g. \cite{Boldyrev:2002}.
However, further high resolution research \cite{Kritsuk_etal:2007} revealed that
the flattering of the spectrum observed was the result of a bottleneck effect,
which is more extended in compressible than in incompressible fluids.  In the
MHD simulations that are indicative of $-3/2$ spectrum, similar to the
aforementioned low resolution hydrodynamic simulations showing $-5/3$ no
bottleneck effect is seen.  As the bottleneck is a physical effect, the fact
that it is not seen in simulations to our mind means that it is just extended
and higher resolution simulations are necessary.  Therefore, choosing between
theories on the basis of just spectral slope of low resolution simulations may
be tricky.}.

The additional physics that was considered included, e.g. dynamical alignment
\cite{Boldyrev:2006}, polarization intermittency \cite{BeresnyakLazarian:2006},
turbulence non-locality \cite{Gogoberidze:2007}.  In particular
\cite{Boldyrev:2006} study predicts the the Kraichnan index of $-3/2$
\cite{Iroshnikov:1964, Kraichnan:1965} rather than Kolmogorov index $-5/3$ that
follows from GS95.  We feel that more recent high resolution numerical
simulations (see \cite{Beresnyak:2013,Beresnyak:2014}) provide results in
agreement with the GS95 expectations, while the more shallow spectra are likely
to be due to the bottleneck effect arising from MHD turbulence being less local
compared to hydrodynamic one \cite{BeresnyakLazarian:2010}.

The measurement in the solar wind show evidence for the $-3/2$ spectrum at the
1AU from the Sun and $-5/3$ spectrum at distances larger than 1AU
\cite{Roberts:2010}.  We believe that the more relevant to MHD turbulence is the
spectrum measured at larger distances where the there is less influence from the
imbalance as well as the transient processes of spectrum evolution.  While the
discussion of the exact scaling of MHD turbulence is ongoing (see papers and
comments by Perez et al. \cite{Perez_etal:2012, Perez_etal:2014} and answers to
them in \cite{Beresnyak:2013, Beresnyak:2014}), we would like to stress that our
results on reconnection marginally depend on the exact spectral index of
turbulence.  In LV99, which was developed when GS95 theory was far from being
accepted, in the Appendix the reconnection rates were provided for arbitrary
spectral indexes of turbulence and scale dependent anisotropies.

More discussions of astrophysical turbulence can be found in recent reviews,
e.g. \cite{BrandenburgLazarian:2013, Lazarian:2013, BeresnyakLazarian:2015}.  In
particular, there many additional effects are discussed, e.g. compressibility,
effect of partial ionization as well as the effect of imbalance of turbulence.
The latter may be a consequence of having sources and sinks of turbulent energy
that are not coincident in space.  All these effects are not of principal
importance for our discussion of turbulent reconnection and therefore we do not
discuss them here.

Finally, we point out that we concentrate our attention on subAlfv\'enic
turbulence as the reconnection of weakly perturbed magnetic fields is the
natural generalization of the classical formulation of the reconnection problem.
The opposite extreme is the turbulence in the dynamically unimportant magnetic
field, where the magnetic field are reversing at the resistive dissipative
scale.  This is a degenerate example employed in the model of kinetic dynamo and
it is of no interest for the reconnection research.  If turbulence is
superAlfv\'enic, magnetic field becomes dynamically important and stiff at the
scale of $L_i M_A^{-3}$ \cite{Lazarian:2006} and the reconnection ideas below
can be applied to such fields.

The most important points of this section are
\begin{itemize}
\item astrophysical fluids are generically turbulent,
\item MHD description is of Alfv\'enic turbulence is valid at sufficiently large
scales,
\item we have an adequate theory of Alfvenic turbulence.
\end{itemize}
In what follows we refer to these points dealing with the problem of turbulent
reconnection.

\section{Turbulent Reconnection Model}

The model of turbulent reconnection in LV99 generalizes the classical
Sweet-Parker model \cite{Parker:1957, Sweet:1958}\footnote{The basic idea of the
model was first discussed by Sweet and the corresponding paper by Parker refers
to the model as ``Sweet model''.}.  In the latter model two regions with uniform
{\it laminar} magnetic fields are separated by thin current sheet.  The speed of
reconnection is given roughly by the resistivity divided by the sheet thickness,
i.e.
\begin{equation}
V_{rec1}\approx \eta/\Delta.
\label{eq.1}
\end{equation}
For {\it steady state reconnection} the plasma in the current sheet must be
ejected from the edge of the current sheet at the Alfv\'{e}n speed, $V_A$.  Thus
the reconnection speed is
\begin{equation}
V_{rec2}\approx V_A \Delta/L_x,
\label{eq.2}
\end{equation}
where $L_x$ is the length of the current sheet, which requires $\Delta$ to be
large for a large reconnection speed.  As a result, the overall reconnection
speed is reduced from the Alfv\'{e}n speed by the square root of the Lundquist
number, $S\equiv L_xV_A/\eta$, i.e.
\begin{equation}
V_{rec, SP}=V_A S^{-1/2}.
\label{SP}
\end{equation}
The corresponding Sweet-Parker reconnection speed is negligible in astrophysical
conditions as $S$ may be $10^{16}$ or larger.

\begin{figure}
\centering
\includegraphics[width=0.65\textwidth]{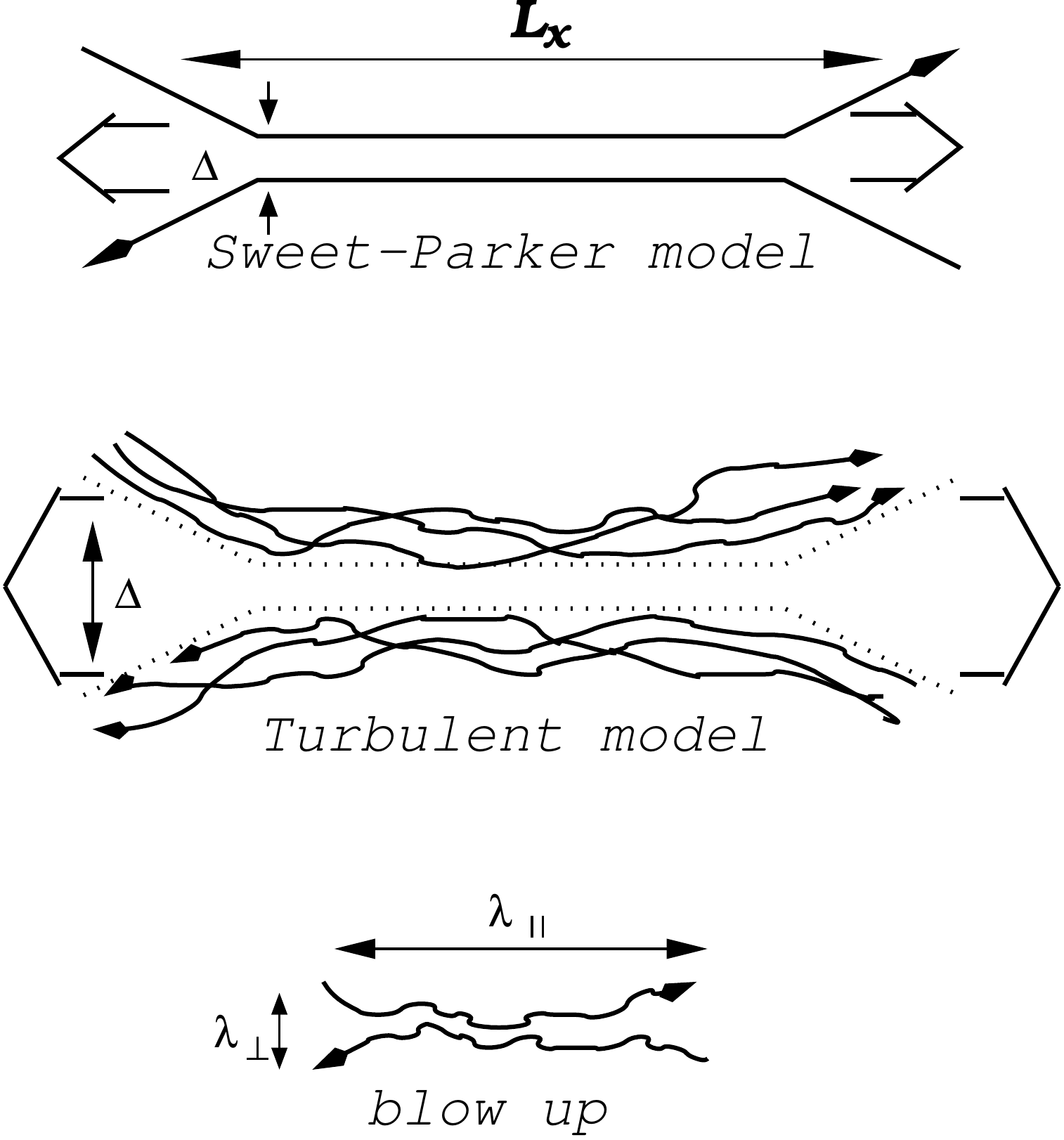}
\caption{{\it Upper plot}: Sweet-Parker model of reconnection.  The outflow is
limited to a thin width $\delta$, which is determined by Ohmic diffusivity.  The
other scale is an astrophysical scale $L_x \gg \delta$.  Magnetic field lines
are laminar.
{\it Middle plot}: Turbulent reconnection model that accounts for the
stochasticity of magnetic field lines.  The stochasticity introduced by
turbulence is weak and the direction of the mean field is clearly defined.  The
outflow is limited by macroscopic field line wandering.
{\it Low plot}: An individual small scale reconnection region.
\cite{Lazarian_etal:2004}.}
\label{recon}
\end{figure}

The corresponding model of magnetic reconnection is illustrated by
Figure~\ref{recon}.

Similar to the Sweet-Parker model, the LV99 model deals with a generic
configuration, which should arise naturally as magnetic flux tubes try to make
their way one through another.  However, in the LV99 model the large-scale
magnetic field wandering determines the thickness of outflow.  Thus LV99 model
does not depend on resistivity and can provide both fast and slow reconnection
rates depending on the level of turbulence.

To obtain the reconnection rate in LV99 model one should use the scaling
relations for Alfv\'{e}nic turbulence from \S~\ref{sec:turbulence}.  A bundle of
field lines confined within a region of width $y$ at some particular point
spreads out perpendicular to the mean magnetic field direction as one moves in
either direction following the local magnetic field lines.  The rate of field
line diffusion is given by
\begin{equation}
{d\langle y^2\rangle\over dx}\sim {\langle y^2\rangle\over \lambda_{\|}},
\end{equation}
where $\lambda_{\|}^{-1}\approx \ell_{\|}^{-1}$, $\ell_{\|}$ is the parallel
scale and the corresponding transversal scale, $\ell_{\perp}$, is $\sim \langle
y^2\rangle^{1/2}$, and $x$ is the distance along an axis parallel to the
magnetic field.  Therefore, using equation (\ref{Lambda}) one gets
\begin{equation}
{d\langle y^2\rangle\over dx}\sim L_i\left({\langle y^2\rangle\over L_i^2}\right)^{2/3}
\left({u_L\over V_A}\right)^{4/3}
\label{eq:diffuse}
\end{equation}
where we have substituted $\langle y^2\rangle ^{1/2}$ for $\ell_{\perp}$.  This
expression for the diffusion coefficient will only apply when $y$ is small
enough for us to use the strong turbulence scaling relations, or in other words
when $\langle y^2\rangle < L_i^2(u_L/V_A)^4$.

When the turbulence injection scale is less than the extent of the reconnection
layer, i.e. $Lx\gg L_i$ magnetic field wandering obeys the usual random walk
scaling with $L_x/L_i$ steps and the mean squared displacement per step equal to
$L_i^2(u_L/V_A)^4$.  Therefore
\begin{equation}
\langle y^2\rangle^{1/2}\approx (L_i x)^{1/2} (u_L/V_A)^2~~~x>L_i
\label{eq:diffuse3}
\end{equation}

Combining Eqs. (\ref{eq:diffuse}) and (\ref{eq:diffuse3}) one can derive the
thickness of the outflow $\Delta$ and obtain (LV99):
\begin{equation}
V_{rec}\approx V_A\min\left[\left({L_x\over L_i}\right)^{1/2},
\left({L_i\over L_x}\right)^{1/2}\right]
M_A^2,
\label{eq:lim2a}
\end{equation}
where $V_AM_A^2$ is proportional to the turbulent eddy speed.  This reconnection
rate represents a large fraction of the Alfv\'{e}n speed when $L_i$ and $L_x$
are not too different and $M_A$ is not too small.

Due to the importance of the turbulent reconnection, it is advantageous to
consider re-deriving the reconnection rates in another way.  This was is based
on the Lagrangian properties of magnetized plasma, in particular on the
Richardson dispersion (see \cite{EyinkBenveniste:2013} and references therein).

Richardson diffusion/dispersion can be illustrated with a simple hydrodynamic
model.  Consider the growth of the separation between two particles
$dl(t)/dt\sim v(l),$ which for Kolmogorov turbulence is $\sim \alpha_t l^{1/3}$,
where $\alpha_t$ is proportional to the energy cascading rate, i.e.
$\alpha_t\approx V_L^3/L$ for turbulence injected with superAlv\'{e}nic velocity
$V_L$ at the scale $L$.  The solution of this equation is
\begin{equation}
l(t)=[l_0^{2/3}+\alpha_t (t-t_0)]^{3/2},
\label{sol}
\end{equation}
which at late times leads to Richardson diffusion/dispersion or $l^2\sim t^3$
compared with $l^2\sim t$ for ordinary diffusion.  Both terms ``diffusion and
dispersion'' can be used interchangeably, but keeping in mind that the
Richardson process results in superdiffusion (see \cite{LazarianYan:2014} and
references therein) we feel that it is advantageous to use the term
``dispersion''. Although the Richardson dispersion process was introduced for
hydrodynamic turbulence a similar process is valid for magnetized fluids.  We
will not distinguish the magnetized and not magnetized case by name and instead
of magnetic Richardson dispersion will use just Richardson dispersion.  In
magnetized turbulence Richardson dispersion is important in terms of spreading
magnetic fields which provides a way to re-derive the LV99 relations.

The fact that time dependence of the magnetic field diffusion induces magnetic
reconnection, can be illustrated with the Sweet-Parker reconnection.  There
magnetic field lines are subject to Ohmic diffusion.  The latter induces the
mean-square distance across the reconnection layer that a magnetic field-line
can diffuse by resistivity in a time $t$ given by
\begin{equation}
\langle y^2(t)\rangle \sim \lambda t.
\label{diff-dist}
\end{equation}
where $\lambda=c^2/4\pi\sigma$ is the magnetic diffusivity.  The field lines are
advected out of the sides of the reconnection layer of length $L_x$ at a
velocity of order $V_A$.  Therefore, the time that the lines can spend in the
resistive layer is the Alfv\'{e}n crossing time $t_A=L_x/V_A$. Thus, field lines
that can reconnect are separated by a distance
\begin{equation}
\Delta = \sqrt{\langle y^2(t_A)\rangle} \sim \sqrt{\lambda t_A} = L_x/\sqrt{S},
\label{Delta}
\end{equation}
where $S$ is Lundquist number.  Combining Eqs. (\ref{eq.2}) and (\ref{Delta})
one gets again the well-known Sweet-Parker result, $v_{rec}=V_A/\sqrt{S}$.

The difference with the turbulent case is that instead of Ohmic diffusion one
should use the Richardson one \cite{Eyink_etal:2011}.  In this case the mean
squared separation of particles $\langle |x_1(t)-x_2(t)|^2 \rangle\approx
\epsilon t^3$, where $t$ is time, $\epsilon$ is the energy cascading rate and
$\langle...\rangle$ denote an ensemble averaging (see \cite{Kupiainen:2003}).
For subAlfv\'{e}nic turbulence $\epsilon\approx u_L^4/(V_A L_i)$ (see LV99) and
therefore analogously to Eq. (\ref{Delta}) one can write
\begin{equation}
\Delta\approx \sqrt{\epsilon t_A^3}\approx L(L/L_i)^{1/2}M_A^2
\label{D2}
\end{equation}
where it is assumed that $L<L_i$.  Combining Eqs. (\ref{eq.2}) and (\ref{D2})
one obtains
\begin{equation}
v_{rec, LV99}\approx V_A (L/L_i)^{1/2}M_A^2.
\label{LV99}
\end{equation}
in the limit of $L<L_i$.  Similar considerations allow to recover the LV99
expression for $L>L_i$, which differs from Eq.~(\ref{LV99}) by the change of the
power $1/2$ to $-1/2$ and recover Eq.~(\ref{eq:lim2a}).

\section{Extending LV99 Reconnection Theory}
\label{sec:lv99_extension}

\subsection{Recent theoretical developments: rigorous mathematical approach}

Recently the LV99 notion of magnetic line wandering has played a central role in
the extension of ``general magnetic reconnection'' (GMR) theory to turbulent
plasmas.  Recall that GMR theory \cite{Schindler_etal:1988, HesseSchindler:1988}
attempts to quantify the changes of magnetic connections between plasma
elements.  It is assumed in the standard approach to GMR that such changes occur
only in narrow, sparsely distributed current layers or ``diffusion regions'' of
small total volume.  This assumption is invalid for turbulent plasmas.  By
tracking along field-lines wandering in space, \cite{Eyink:2014} has developed
an extended version of GMR theory valid for both laminar and turbulent plasmas.

The study in \cite{Eyink:2014} provides a rigorous mathematical treatment of the
motion of magnetic field lines in turbulent plasmas.  The slip source vector
which is defined as the ratio of the curl of the non ideal electric field in the
Generalized Ohm's Law and the magnetic field strength was introduced and it was
demonstrated that this vector gives the rate of development of slip velocity per
unit arc length of field line.  It diverges at magnetic nulls, unifying GMR with
magnetic null-point reconnection.  In a turbulent inertial range the curl
becomes extremely large while the parallel component is tiny, so that line
slippage occurs even while ideal MHD is accurate.  This means that ideal MHD is
valid for a turbulent inertial-range only in a weak sense which does not imply
magnetic line freezing (see also section 7).  By rigorous estimates of the terms
in the Generalized Ohm's Law for an electron-ion plasma the paper shows that all
of the non-ideal terms (from collisional resistivity, Hall field, electron
pressure anisotropy, and electron inertia) are irrelevant compared with the
effects of turbulence and large-scale reconnection is thus governed solely by
ideal dynamics.  It is encouraging that in terms of magnetic reconnection the
results of this study correspond to LV99 model and thus provide more rigorous
theoretical foundations for turbulent reconnection.  The results for the
slippage velocity in \cite{Eyink:2014} are identical to the expression of the
reconnection velocities in LV99. Together with the earlier discussed results on
Richardson dispersion in magnetic turbulence, these provide new outlook on the
nature of magnetic reconnection in turbulent fluids.

\subsection{Effect of energy dissipation in the reconnection layer}

In LV99 expressions were derived assuming that only a small fraction of the
energy stored in the magnetic field is lost during large-scale reconnection and
the magnetic energy is instead converted nearly losslessly to kinetic energy of
the outflow.  This can only be true, however, when the Alfv\'enic Mach number
${\cal M}_A=u_L/V_A$ is small enough.  If ${\cal M}_A$ becomes large, then
energy dissipation in the reconnection layer becomes non-negligible and there is
a reduction of the outflow velocity (see ELV11).  Note that even if ${\cal M}_A$
is initially small, reconnection may drive stronger turbulence (see section 4c)
and increase the fluctuation velocities $u_L$ in the reconnection layer.  This
scenario may be relevant to post-CME reconnection, for example, where there is
empirical evidence that the energy required to heat the plasma in the
reconnection layer (``current sheet'') to the observed high temperatures is from
energy cascade due to turbulence generated by the reconnection itself
\cite{Susino_etal:2013}. In addition, $V_A$ within the reconnection layer will
be smaller than the upstream values, because of annihilation of the
anti-parallel components, which will further increase the Alfv\'enic Mach
number.

The effect of turbulent dissipation can be estimated from steady-state energy
balance in the reconnection layer:
\begin{equation}
\frac{1}{2}v_{out}^3 \Delta = \frac{1}{2}V_A^2 v_{ren} L_x - \varepsilon L_x \Delta,  \label{Ebal}
\end{equation}
where kinetic energy carried away in the outflow is balanced against magnetic
energy transported into the layer minus the energy dissipated by turbulence.
Here we estimate the turbulent dissipation using the formula $\varepsilon =
u_L^4/V_A L_i$ for sub-Alfv\'enic turbulence \cite{Kraichnan:1965}. Dividing
(\ref{Ebal}) by $\Delta=L_x v_{rec}/v_{out}$, we get
\begin{equation}
     v_{out}^3 = V_A^2 v_{out} - 2 \frac{u_L^4}{V_A} \frac{L_x}{L_i},
\end{equation}
which is a cubic polynomial for $v_{out}$. The solutions are easiest to obtain
by introducing the ratios $f=v_{out}/V_A$ and $r= 2 {\cal M}_A^4 (L_x/L_i)$ which
measure, respectively, the outflow speed as a fraction of $V_A$ and the energy
dissipated by turbulence in units of the available magnetic energy, giving
\begin{equation}
    r = f - f^3.
\label{cubic}
\end{equation}
When $r=0$, the only solution of (\ref{cubic}) with $f>0$ is $f=1,$ recovering
the LV99 estimate $v_{out}=V_A$ for ${\cal M}_A\ll 1$.  For somewhat larger
values of $r,$ $f\simeq 1-(r/2)$, in agreement with the formula $f=(1-r)^{1/2}$
that follows from Eq.~(65) in ELV11, implying a slight decrease in $v_{out}$
compared with $V_A$.  Note that formula (\ref{cubic}) cannot be used to
determine $f$ for too large $r$, because it has then no positive, real
solutions! This is easiest to see by considering the graph of $r$ vs. $f$.  The
largest value of $r$ for which a positive, real $f$ exists is $r_{max} =
2/\sqrt{27}\approx 0.385$ and then $f$ takes on its minimum value $f_{min} =
1/\sqrt{3}\approx 0.577$. This implies that the LV99 approach is limited to
${\cal M}_A$ sufficiently small, because of the energy dissipation inside the
reconnection layer and the consequent reduction of the outflow velocity.  This
is not a very stringent limitation, however, because $r$ is proportional to
${\cal M}_A^4$.  If one assumes $L_x \simeq L_i$, one may consider values of
${\cal M}_A$ up to $0.662$.  Given the neglect of constants of order unity in
the above estimate, we may say only that the LV99 approach is limited to ${\cal
M}_A\lesssim 1$. At the extreme limit of applicability of LV99, $v_{out}$ is
still a sizable fraction of $V_A$, i.e. 0.577, not a drastically smaller value.

The effect of the reduced outflow velocity may be, somewhat paradoxically, to
{\it increase} the reconnection rate.  The reason is that field-lines now spend
a time $L_x/v_{out}$ exiting from the reconnection layer, greater than assumed
in LV99 by a factor of $1/f$.  This implies a thicker reconnection layer
$\Delta$ due to the longer time-interval of Richardson dispersion in the layer,
greater than LV99 by a factor of $(1/f)^{3/2}$.  The net reconnection speed
$v_{rec}=v_{out}\Delta/L_x$ is thus larger by a factor of $(1/f)^{1/2}$.  The
increased width $\Delta$ more than offsets the reduced outflow velocity
$v_{out}$.  However, this effect can give only a very slight increase, at most
by a factor of $3^{1/4}\simeq 1.31$ for $f_{min}=1/\sqrt{3}$.  We see that for
the entire regime ${\cal M}_A\lesssim 1$ where LV99 theory is applicable, energy
dissipation in the reconnection layer implies only very modest corrections.  It
is worth emphasizing that ``large-scale reconnection'' in super-Alfv\'enic
turbulence with ${\cal M}_A>1$ is a very different phenomenon, because magnetic
fields are then so weak that they are easily bent and twisted by the turbulence.
 Any large-scale flux tubes initially present will be diffused by the turbulence
through a process much different than that considered by LV99. For a discussion
of this regime, see \cite{KimDiamond:2001}.

\subsection{Reconnection in partially ionized gas}
\label{ssec:ionized}

On sufficiently small scales Alfv\'enic turbulence in the partially ionized gas
is differs from our description provided in section 2.  Due to viscosity caused
by neutral atoms, the fluid viscosity becomes substantially larger than the
fluid resistivity, which means that the Prandtl number of the fluid is high.
Turbulence in high Prandtl number fluids has been studied numerically in
\cite{Cho_etal:2002, Cho_etal:2003, SchekochihinCowley:2006} and theoretically
in \cite{Lazarian_etal:2004}.  However, for our present discussion it is
important that for scales larger than the viscous damping scale the turbulence
follows the usual GS95 scaling and the considerations about Richardson
dispersion and magnetic reconnection that accompany are valid at these scales.

In high Prandtl number media the GS95-type turbulent motions decay at the scale
$l_{\bot, crit}$, which is much larger than the scale of Ohmic dissipation.
Thus over a range of scales less than $l_{\bot, crit}$ to some much smaller
scale magnetic field lines preserve their identity. To establish the range of
scales at which magnetic fields perform Richardson diffusion one can observe
that the transition to the Richardson dispersion is expected to happen when the
field line separation reaches the perpendicular scale of the critically damped
eddies $l_{\bot, crit}$.  The separation in the perpendicular direction starts
with the scale $r_{init}$ following the Lyapunov exponential growth with the
distance $l$ measured along the magnetic field lines, i.e. $r_{init}
\exp(l/l_{\|, crit})$, where $l_{\|, crit}$ corresponds to critically damped
eddies with $l_{\perp, crit}$.  It seems natural to associate $r_{init}$ with
the separation of the field lines arising from the action of Ohmic resistivity
on the scale of the critically damped eddies
\begin{equation}
r_{init}^2=\eta l_{\|, crit}/V_A,
\label{int}
\end{equation}
where $\eta$ is the Ohmic resistivity coefficient.

At scales smaller than $l_{\perp, crit}$ the magnetic line separation obeys the
laws established by Rechester \& Rosenbluth \cite{RechesterRosenbluth:1978}.
The distance along the local magnetic field field over which anisotropic
turbulence separates the magnetic field lines by $l_{\perp, crit}$ is the
Rechester-Rosenbluth length (see \cite{Lazarian:2006}):
\begin{equation}
L_{RR}\approx l_{\|, crit} \ln (l_{\bot, crit}/r_{init})
\label{RR}
\end{equation}
Taking into account Eq. (\ref{int}) and that
\begin{equation}
l_{\bot, crit}^2=\nu l_{\|, crit}/V_A,
\end{equation}
where $\nu$ is the viscosity coefficient, one can rewrite Eq. (\ref{RR}) as:
\begin{equation}
L_{RR}\approx l_{\|, crit}\ln Pt
\label{RR2}
\end{equation}
where $Pt=\nu/\eta$ is the Prandtl number. This means that when the current
sheets are much longer than $L_{RR}$, then magnetic field lines undergo
Richardson dispersion and according to \cite{Eyink_etal:2011} the reconnection
follows the laws established in LV99.  At the same time on scales less than
$L_{RR}$ magnetic reconnection may be slow\footnote{Incidentally, this can
explain the formation of density fluctuations on scales of thousands of
AU, that are observed in the ISM.}.

\subsection{Self-sustained Turbulent Magnetic Reconnection}

Reconnection releases energy and induces outflows. Even if the initial magnetic
field configuration is laminar, magnetic reconnection ought to induce turbulence
due to the outflow (LV99, \cite{LazarianVishniac:2009}).  This effect was confirmed
by observing the development of turbulence both in recent 3D Particle in Cell
(PIC) simulations \cite{Karimabadi_etal:2013} and 3D MHD simulations
\cite{Beresnyak:2013b, Kowal_etal:2015}.

In terms of MHD simulations, Beresnyak \cite{Beresnyak:2013b} was the first to
study turbulent reconnection with turbulence arising from the reconnection
itself. However, the periodic boundary conditions adopted in
\cite{Beresnyak:2013b} limited the time span over which magnetic reconnection can
be studied and therefore the simulations focus on the process of establishing
reconnection.

Analytical description of the results in the framework of LV99 model was adopted
by Beresnyak (\cite{Beresnyak:2013b}, private communication).  Below we provide
our theoretical account of the results in \cite{Beresnyak:2013b} using our
understanding of LV99 turbulent reconnection.  We obtain expressions which are
different from those by \cite{Beresnyak:2013b}.

The logic of the derivation below is straightforward.  As the opposite magnetic
fluxes enters in contact, the width of the reconnection layer $\Delta$ grows.
The rate at which this happens is limited by the mixing rate induced by the
eddies at the scale $\Delta$, i.e.
\begin{equation}
\frac{1}{\Delta} \frac{d \Delta}{dt}\approx g \frac{V_{\Delta}}{\Delta}
\label{Delta_start}
\end{equation}
with a factor $g$ which takes into account possible inefficiency in the
diffusion process. As $V_{\Delta}$ is a part of the turbulent cascade, i.e. the
mean value of $V_{\Delta}^2\approx \int \Phi(k_1) dk_1$, where
\begin{equation}
\Phi=C_k \epsilon^{2/3}k^{-5/3}_1,
\label{Phi1}
\end{equation}
and $C_k$ is a Kolmogorov constant, which for ordinary MHD turbulence is
calculated in \cite{Beresnyak:2012}, but in our special case may be different.  If
the energy dissipation rate $\varepsilon$ were time-independent, then the layer
width would be implied by Eqs.~(\ref{Delta_start}) and (\ref{Phi1}) to grow according to
Richardson's law $\Delta^2 \sim \varepsilon t^3.$ However, in the transient
regime considered, energy dissipation rate is evolving. If the y-component of
the magnetic field is reconnecting and the cascade is strong, then the mean
value of the dissipation rate $\epsilon$ is
\begin{equation}
\epsilon\approx \beta V_{Ay}^2/(\Delta/V_{\Delta}),
\label{eps}
\end{equation}
where $\beta$ is another coefficient measuring the efficiency of conversion of
mean magnetic energy into turbulent fluctuations. This coefficient can be obtained from
numerical simulations.

The ability of the cascade to be strong from the very beginning follows from the large
perturbations of the magnetic fields by magnetic reconnection, while magnetic energy
can still dominate the kinetic energy. The latter factor that can be experimentally measured
is given by a parameter $r_A$.
With this factor and making use of Eqs.(\ref{Phi1})
and (\ref{eps}),  the expression for
$V_{\Delta}$ can be rewritten in the following way:
\begin{equation}
V_\Delta\approx C_k r_A (V_{Ay}^2 V_\Delta \beta)^{2/3}
\label{V_Delta}
\end{equation}
where the dependences on $k_1\sim 1/\Delta$ cancel out.

This provides the expression for the turbulent velocity at the injection scale $V_\Delta$
\begin{equation}
V_{\Delta}\approx (C_K r_A)^{3/4} V_{Ay} \beta^{1/2}
\label{V_Delta_final}
\end{equation}
as a function of the experimentally measurable parameters of the system. Thus
the growth of the turbulent reconnection zone is according to
Eq.(\ref{Delta_start})
\begin{equation}
\frac{d\Delta}{dt}\approx g \beta^{1/2} (C_K r_A)^{3/4} V_{Ay}
\label{growth}
\end{equation}
which predicts the nearly constant growth of the outflow region as seen in Fig.3
in \cite{Beresnyak:2013b}.

For the steady state regime, one expects the outflow to play an important role.
The equations for the reconnection rate were obtained in LV99 for the isotropic
injection of energy. For the case of anisotropic energy injection of turbulence
we should apply the following approach. Using Eq. (\ref{delta_obs}) and
identifying $V_\Delta$ with the total velocity dispersion, which is similar to
the use of $U_{obs, turb}$ in Eq. (\ref{obs}) one can get
\begin{equation}
V_{rec}\approx V_\Delta (\Delta/L_x)^{1/2}
\label{prel}
\end{equation}
where the mass conservation condition provides the relation $V_{rec} L_x\approx V_{Ay} \Delta$.
Using the latter condition one gets
\begin{equation}
V_{rec} \approx V_{Ay} (C_K r_A)^{3/2} \beta
\label{rec_self}
\end{equation}
which somewhat slower than the rate at which the reconnection layer was growing
initially.

\subsection{Flares of Turbulent Reconnection}

On the basis of LV99 theory a simple quantitative model of flares
was presented in
\cite{LazarianVishniac:2009}.  There it is assumed that since stochastic
reconnection is expected to proceed unevenly, with large variations in the
thickness of the current sheet, one can expect that some unknown fraction of
this energy will be deposited inhomogeneously, generating waves and adding
energy to the local turbulent cascade.

For the sake of simplicity, the plasma density is assumed to be uniform so that the
Alfv\'{e}n speed and the magnetic field strength are interchangeable.  The
nonlinear dissipation rate for waves is
\begin{equation}
\tau_{nonlinear}^{-1}\sim\max\left[ {k_\perp^2 v_{wave}^2\over k_\|V_A},k_\perp^2 VL\right],
\end{equation}
where the first rate is the self-interaction rate for the waves and the second
is the dissipation rate induced by the ambient turbulence (see
\cite{BeresnyakLazarian:2008}).  The important point here is that
$k_\perp$ for the waves falls somewhere in the inertial range of the strong
turbulence.  Eddies at that wavenumber will disrupt the waves in one eddy
turnover time, which is necessarily less than $L/V_A$.  Therefore, the bulk of
the wave energy will go into the turbulent cascade before escaping from the
reconnection zone.

An additional simplification is achieved by assuming that some fraction
$\epsilon$ of the energy liberated by stochastic reconnection is fed into the
local turbulent cascade.  The evolution of the  turbulent energy density per
area is
\begin{equation}
{d\over dt}\left(\Delta V^2\right)=\epsilon V_A^2 V_{rec}-V^2\Delta {V_A\over L_x},
\end{equation}
where the loss term covers both the local dissipation of turbulent energy, and
its advection out of the reconnection zone.  Since $V_{rec}\sim v_{turb}$  and
$\Delta\sim L_x(V/V_A)$,  it is possible to rewrite this by defining
$\tau\equiv tV_A/L_x$ so that
\begin{equation}
{d\over d\tau}M_A^3\approx \epsilon M_A-M^3_A.
\end{equation}
If $\epsilon$ is a constant then
\begin{equation}
V\approx V_A\epsilon^{1/2}(1-e^{-2\tau/3})^{1/2}.
\end{equation}
This implies that the time during which reconnection rate rises to
$\epsilon^{1/2}V_A$ is comparable to the ejection time from the reconnection
region ($\sim L_x/V_A$).

Within this toy model $\epsilon$ is not defined.  Its value can be constrained
through observations.  Given that reconnection events in the solar corona seem
to be episodic, with longer periods of quiescence, this is suggestive that
$\epsilon$ is very small, for example, depends strongly on the ratio of
the  thickness of the current sheet to $L_x$.  In particular, if it scales as
$M_A$ to some power greater than two then initial conditions dominate the early
time evolution.

Another route by which magnetic reconnection might be self-sustaining via
turbulence injection would be in the context of a series of topological knots in
the magnetic field, each of which is undergoing reconnection.  For simplicity,
one can assume that as each knot undergoes reconnection it releases a
characteristic energy into a volume which has the same linear dimension as the
distance to the next knot.  The density of the energy input into this volume is
roughly $\epsilon V_A^2 V/L_x$, where here $\epsilon$ is defined as the
efficiency with which the magnetic energy is transformed into turbulent energy.
Thus one gets
\begin{equation}
\epsilon {V_A^2V\over L_x}\sim {v'^3\over L_k},
\end{equation}
where $L_k$ is the distance between knots and $v'$ is the turbulent velocity
created by the reconnection of the first knot.  This process will proceed
explosively if $v'>V$ or
\begin{equation}
V_A^2 L_k\epsilon> V^2 L_x.
\end{equation}
The condition above is easy to fulfill.  The bulk motions created by
reconnection can  generate turbulence as they interact with their surrounding,
so $\epsilon$ should be of order unity.  Moreover the length of any current
sheet should be at most comparable to the distance to the nearest distinct
magnetic knot.  The implication is that each magnetic reconnection event will
set off its neighbors, boosting their reconnection rates from $V_L$, set by the
environment, to $\epsilon^{1/2}V_A(L_k/L_x)^{1/2}$ (as long as this is less than
$V_A$).  The process will take a time comparable to the crossing time $L_x/V_L$
to begin, but once initiated will propagate through the medium with a speed
comparable to speed of reconnection in the individual knots.  The net effect can
be a kind of modified sandpile model for magnetic reconnection in the solar
corona and chromosphere.  As the density of knots increases, and the energy
available through magnetic reconnection increases, the chance of a successfully
propagating reconnection front will increase.

\subsection{Relativistic reconnection}

Magnetic turbulence in a number of astrophysical highly magnetized objects,
accretion disks near black holes, pulsars, gamma ray bursts may be in the
relativistic regime when the Alfv\'{e}n velocity approaches that of light.  The
equations that govern magnetized fluid in this case look very different from the
ordinary MHD equations.  However, studies by \cite{Cho:2005} and
\cite{ChoLazarian:2014} show that for both balanced and imbalanced turbulence,
the turbulence spectrum and turbulence anisotropies are quite similar in this
regime and the non-relativistic one.  This suggests that the Richardson
dispersion and related processes of LV99-type magnetic reconnection should cary
on to the relativistic case (see \cite{LazarianYan:2012}). This prediction was
confirmed by the recent numerical simulations Makoto Takomoto (2014, private
communication) who with his relativistic code adopted the approach in
\cite{Kowal_etal:2009} and showed that the rate of 3D relativistic magnetic
reconnection gets independent of resistivity.

The suggestion that LV99 is applicable to relativistic reconnection motivated
the use of the model for explaining gamma ray bursts in
\cite{Lazarian_etal:2003} and \cite{ZhangYan:2011} studies and in accretion
disks around black holes and pulsars studies
\cite{deGouveiaDalPinoLazarian:2005, Giannios:2013}. Now, as the extension of
the model to relativistic case has be confirmed these and other cases where the
relativistic analog of LV99 process was discussed to be applicable (see
\cite{LyutikovLazarian:2013}) are given numerical support.

Naturally, more detailed studies of both relativistic MHD turbulence and
relativistic magnetic reconnection are required. It is evident that in
magnetically-dominated, low-viscous plasmas turbulence is a generic ingredient
and thus it must be taken into account for relativistic magnetic reconnection.
As we discuss elsewhere in the review the driving of turbulence may by external
forcing or it can be driven by reconnection itself.

\section{Numerical Testing of Turbulent Reconnection Theory}

\begin{figure*}
\centering
\raisebox{-0.5\height}{\includegraphics[trim = 20mm -10mm 20mm 0mm, clip, width=0.48\textwidth]{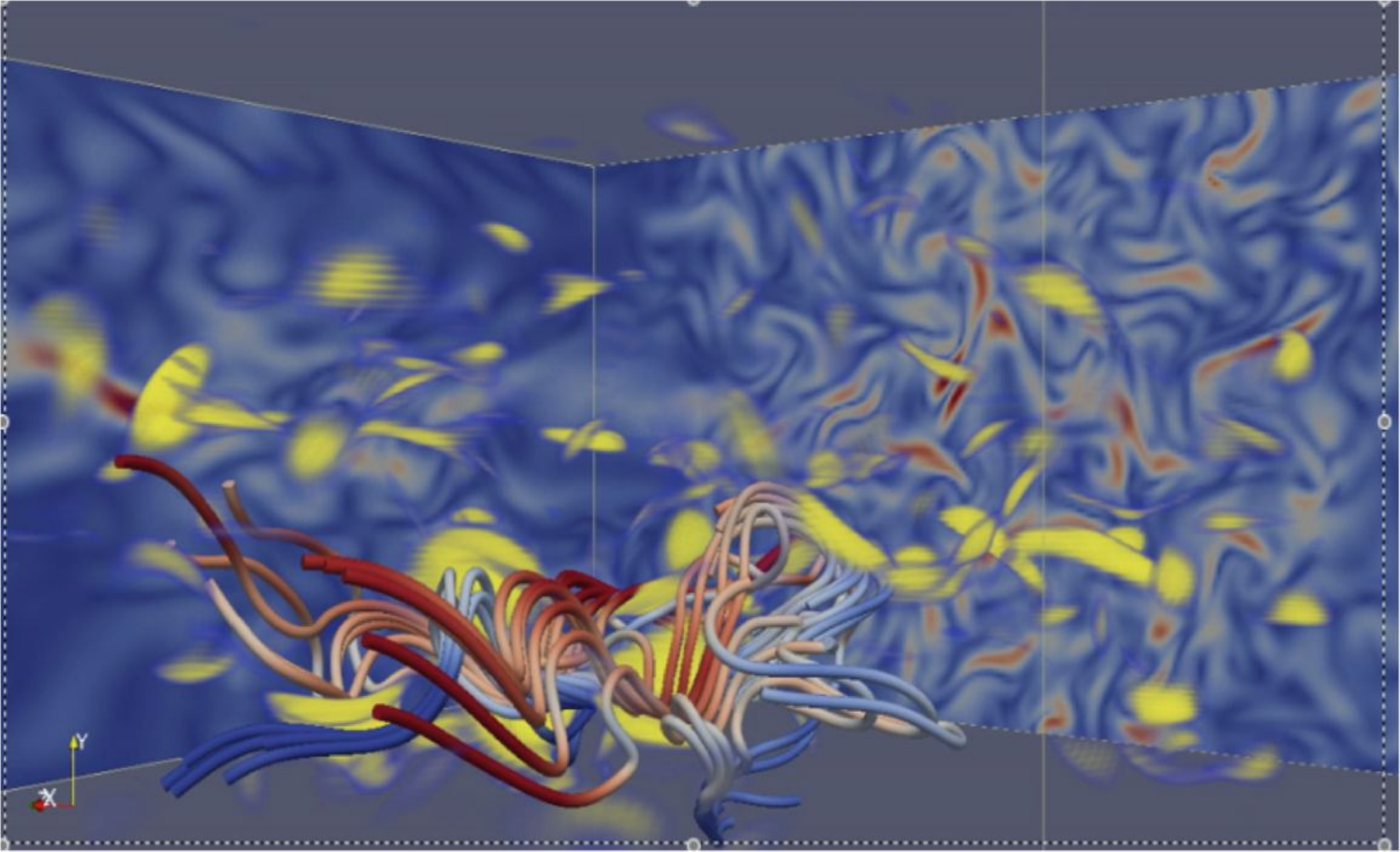}}
\raisebox{-0.5\height}{\includegraphics[width=0.25\textwidth]{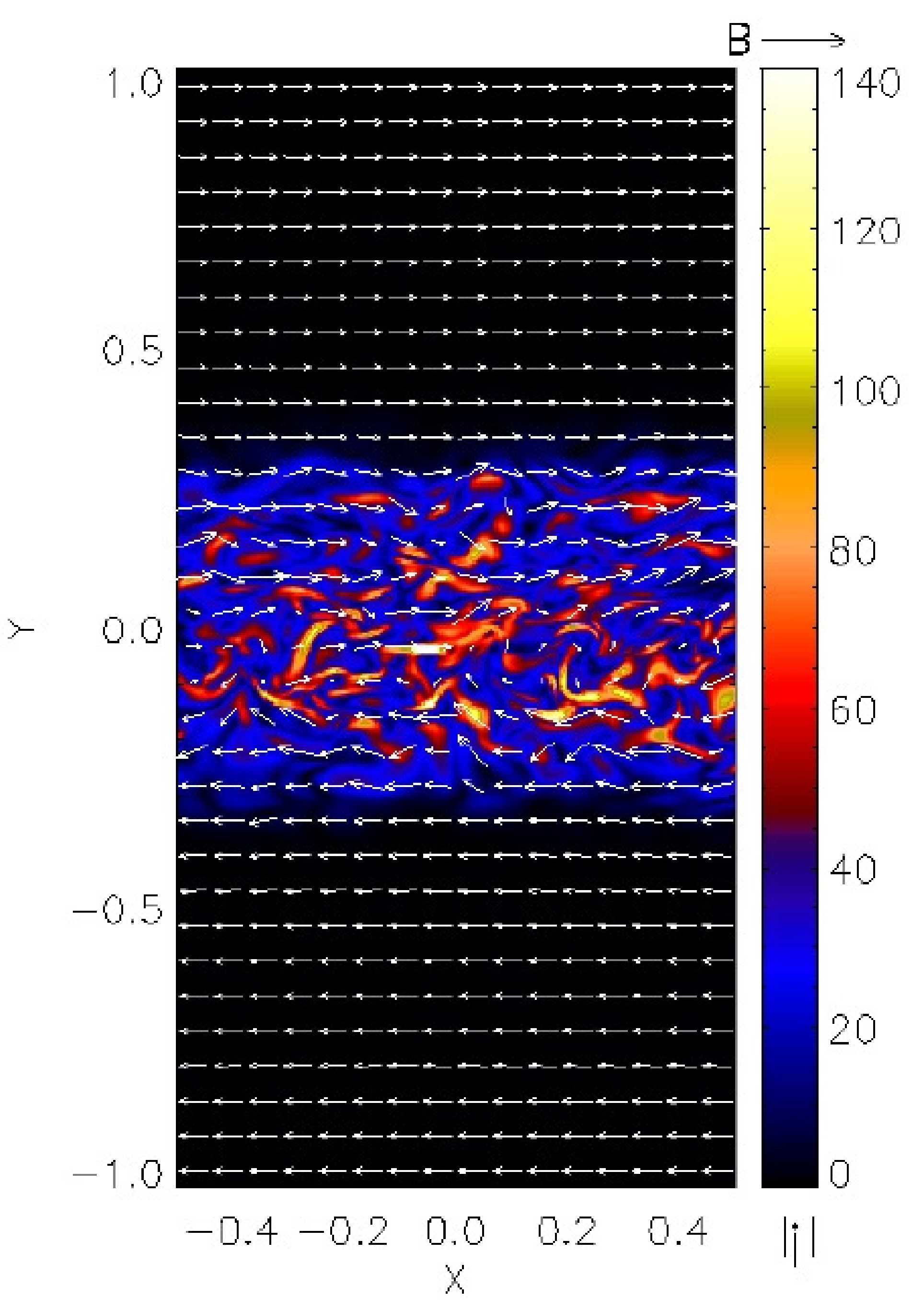}}
\raisebox{-0.5\height}{\includegraphics[width=0.25\textwidth]{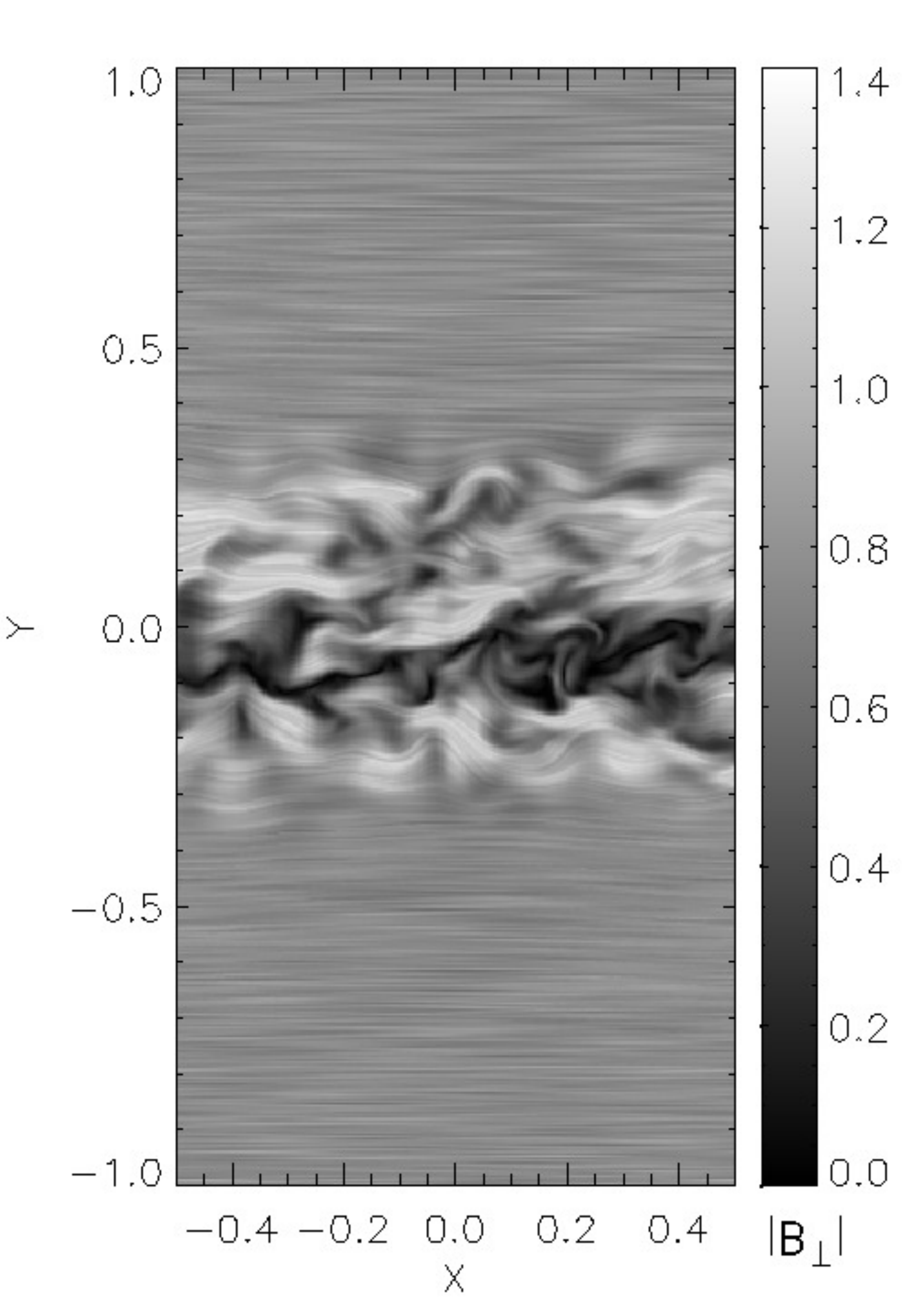}}
\caption{ Visualization of reconnection simulations in \cite{Kowal_etal:2009, Kowal_etal:2012}.
{\it Left panel}: Magnetic field in the reconnection region.
{\it Central panel}: Current intensity and magnetic field configuration during
stochastic reconnection.  The guide field is perpendicular to the page. The
intensity and direction of the magnetic field is represented by the length and
direction of the arrows.  The color bar gives the intensity of the current.
{\it Right panel}: Representation of the magnetic field in the reconnection zone
with textures.
\label{visual}}
\end{figure*}

Figure~\ref{visual} illustrates results of numerical simulations of turbulent
reconnection with turbulence driven using wavelets in \cite{Kowal_etal:2009} and
in real space in \cite{Kowal_etal:2012}.

\begin{figure}
\centering
\raisebox{-0.5\height}{\includegraphics[width=0.48\textwidth]{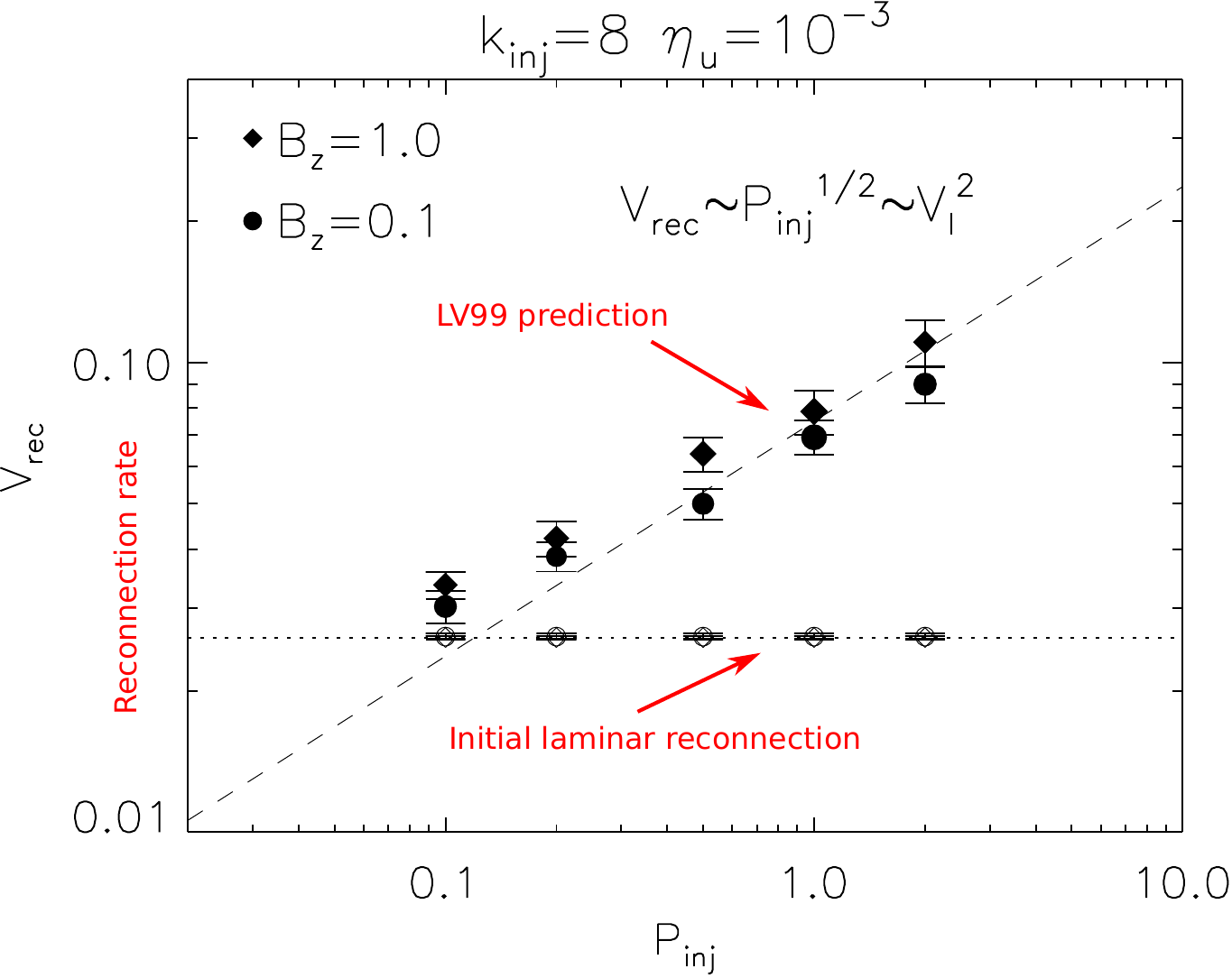}}
\raisebox{-0.5\height}{\includegraphics[width=0.48\textwidth]{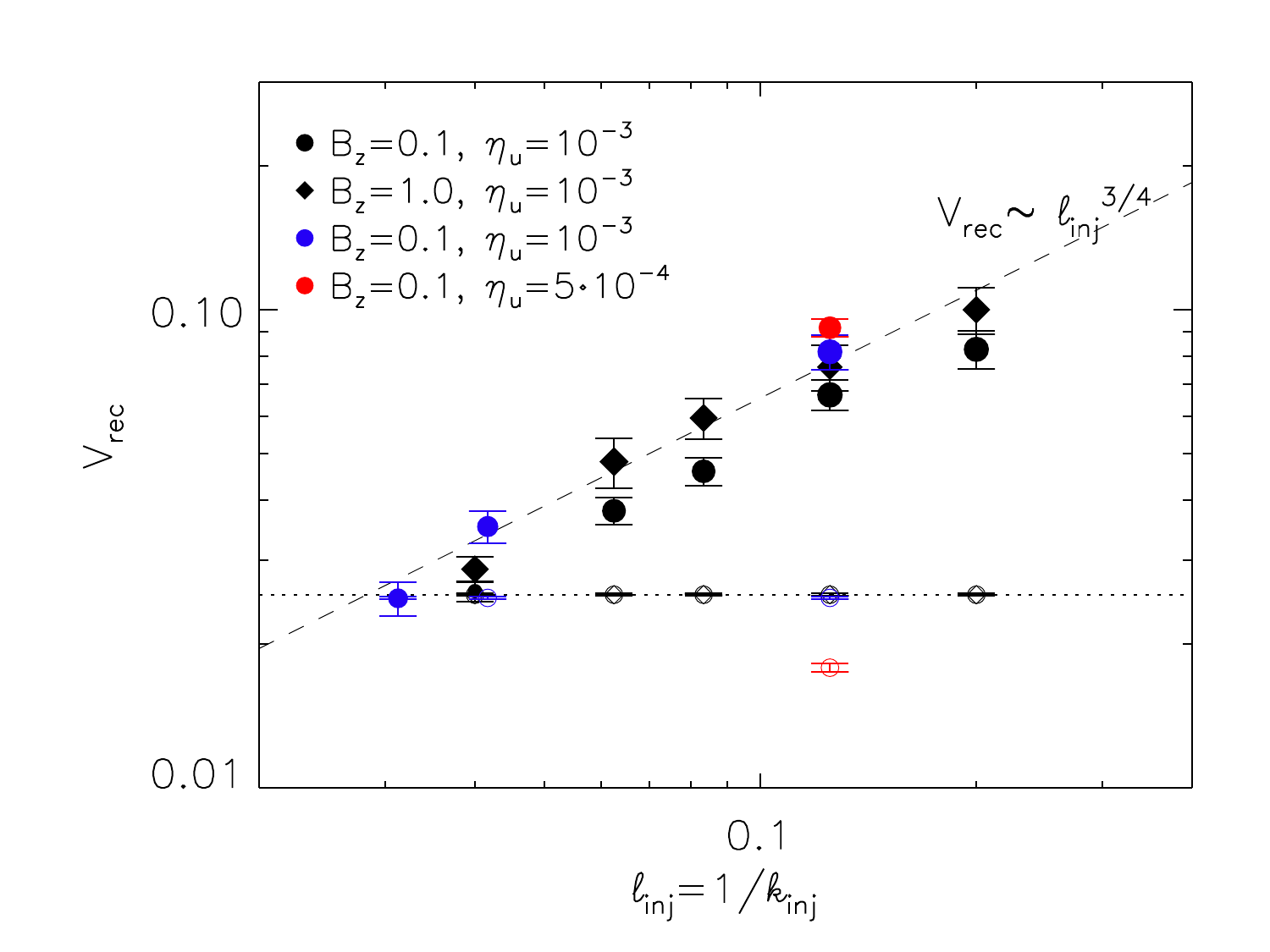}}
\caption{{\it Left Panel} The dependence of the reconnection velocity on the injection power for
different simulations with different drivings. The predicted LV99 dependence is
also shown. $P_{inj}$ and $k_{inj}$ are the injection power and scale,
respectively, $B_z$ is the guide field strength, and $\eta_u$ the value of
unifor resistivity coefficient.
{\it Right Panel} The dependence of the reconnection velocity on the injection scale. From \cite{Kowal_etal:2012}. \label{figure6}}
\end{figure}

As we show below, simulations in \cite{Kowal_etal:2009, Kowal_etal:2012} provided very
good correspondence to the LV99 analytical predictions for the dependence on
resistivity, i.e. no dependence on resistivity for sufficiently strong
turbulence driving, and the injection power, i.e. $V_{rec}\sim P_{inj}^{1/2}$.
The corresponding dependence is shown in Figure~\ref{figure6}, left panel.

The simulations did not reveal any dependence on the strength of the guide field $B_z$
(see Figure~\ref{figure6}). To address this dependence, in the limit
where the parallel wavelength of the strong turbulent eddies is less than the
length of the current sheet, we can rewrite the reconnection speed as
\begin{equation}
V_{rec} \approx \left({P L_x\over  V_{Ax}}\right)^{1/2} {1 \over k_\| V_A}.
\label{eddy}
\end{equation}
Here $P$ is the power in the strong turbulent cascade, $L_x$ and $V_{Ax}$ are
the length scale and Alfv\'{e}n velocity in the direction of the reconnecting
field, and $V_A$ is the total Alfv\'{e}n velocity, including the guide field.

In a physically realistic situation, the dynamics that drive the turbulence,
whatever they are, provide a characteristic frequency and input power.  Since
the guide field enters only in the combination $k_\| V_A$, i.e. through the eddy
turn over rate, this implies that varying the guide field will not change the
reconnection speed. In the simulations the periodicity of the box in the
direction of the guide field complicates the analysis (see more discussion in
\cite{Lazarian_etal:2014}).

The injection of energy in LV99 is assumed to happen at a given scale
and the inverse cascade is not considered in the theory. Therefore
it is not unexpected that the measured dependence on the turbulence scale
differs from the predictions. In fact, it is a bit more shallow compared
to the LV99 predictions (see Figure~\ref{figure6}, right panel).

\begin{figure}[t]
\centering
\raisebox{-0.5\height}{\includegraphics[width=0.48\textwidth]{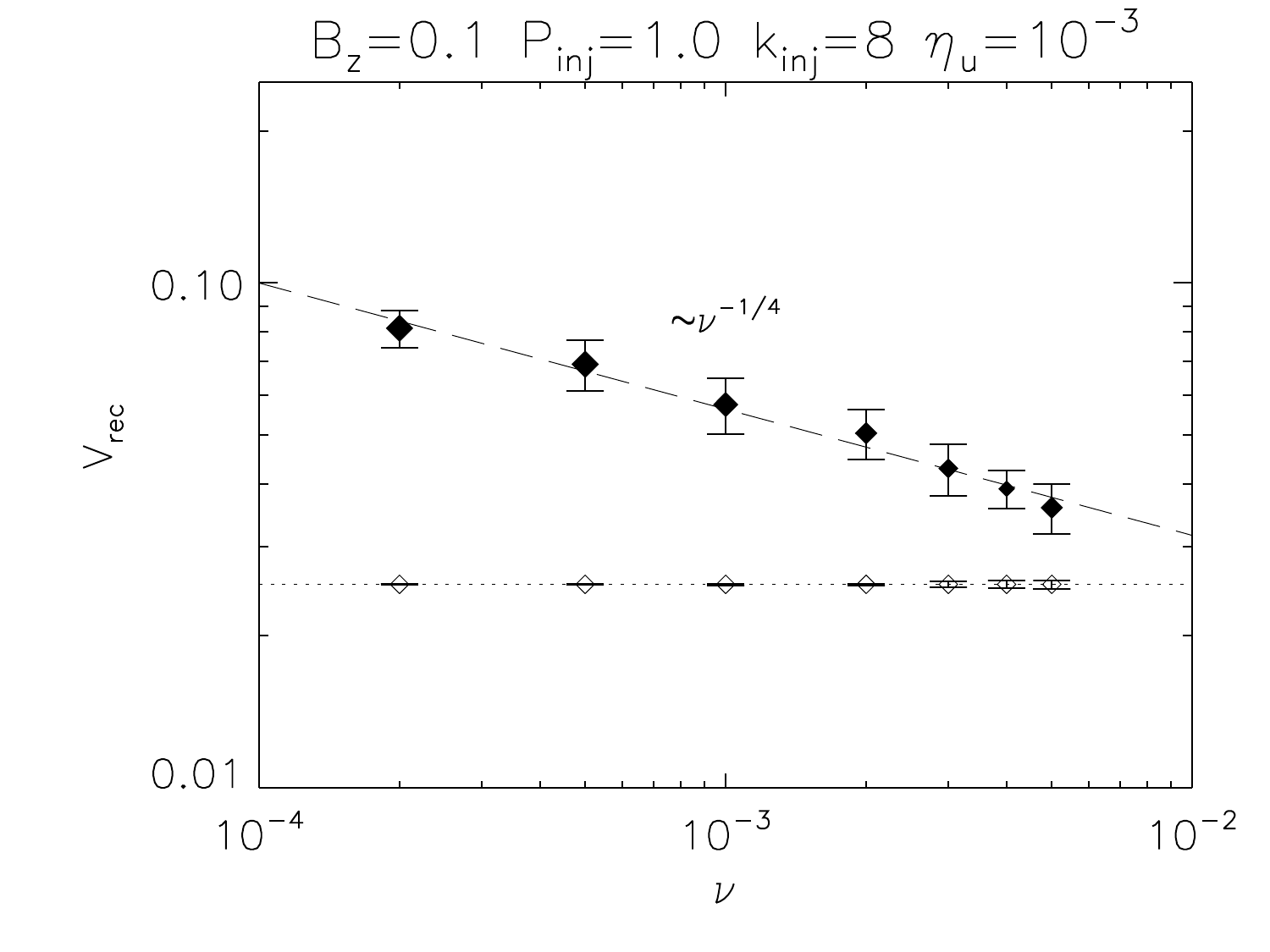}}
\raisebox{-0.5\height}{\includegraphics[width=0.48\textwidth]{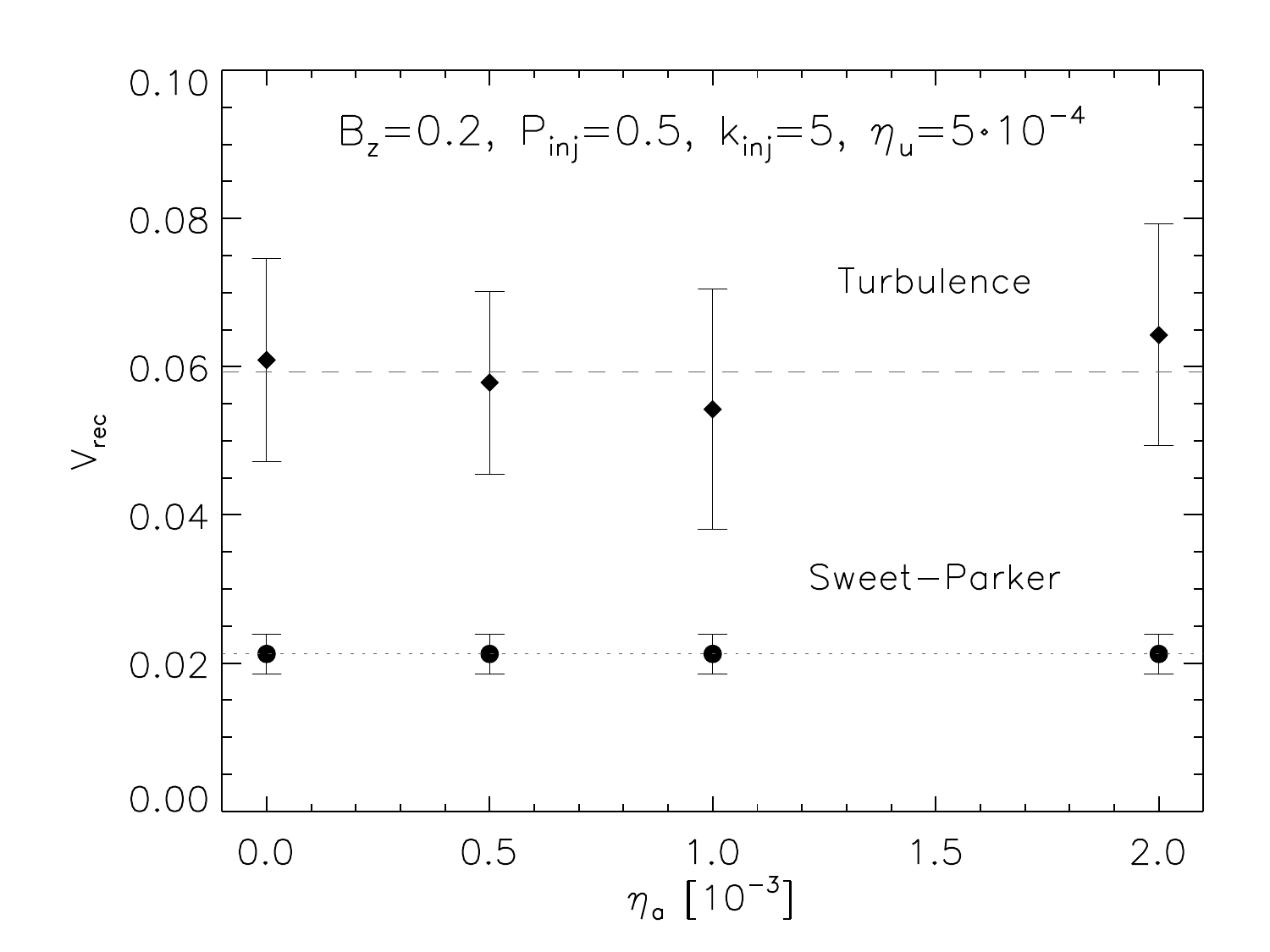}}
\caption{{\it Left panel}. The dependence of the reconnection velocity on
uniform viscosity in the 3D isothermal models of Sweet-Parker reconnection (open
symbols) and reconnection enhanced by the presence of turbulence (closed
symbols) from \cite{Kowal_etal:2012}.  The dependence on viscosity is negligible
in the laminar case, while in the presence of turbulence the reconnection rate
exhibits the scaling $~\nu^{-1/4}$. See text for explanation.}  {\it Right
panel.} The reconnection rate in models with anomalous resistivity for
Sweet-Parker case (filled circles) and in the presence of turbulence (filled
diamonds). We observe no dependence of the reconnection rate on the strength of
anomalous effects. In the Sweet-Parker case the anomalous resistivity is not
turned on, since the maximum current density is below the threashold of the
anomalous resistivity model we used. See \cite{Kowal_etal:2009} for more
detailed description.
\label{fig:viscosity}
\end{figure}

The left panel of Figure~\ref{fig:viscosity} shows the dependence of the reconnection
rate on explicit uniform viscosity obtained from the isothermal simulations of
the magnetic reconnection in the presence of turbulence \cite{Kowal_etal:2012}. The
open symbols show the reconnection rate for the laminar case when there was no
turbulence driving, while closed symbols correspond to the mean values of
reconnection rate in the presence of saturated turbulence.  All parameters in
those models were kept the same, except the uniform viscosity, which varied from
$10^{-4}$ to $10^{-2}$ in the code units.  We notice the lack of any scaling for
the laminar case, which is somewhat in contradiction to the scaling $V_{rec}
\propto \nu^{-1/4}$ derived in \cite{Park_etal:1984}.  We should remind,
that the authors introduced the viscosity dependency using the energy equation
balance, which cannot be applied in the isothermal case.  They also stress that
the proper scaling might be sensitive to the chosen boundaries, which in theirs
numerical tests where closed. In the models presented in Figure~\ref{fig:viscosity} we
use outflow boundaries.  The viscosity scaling for the case when turbulence is
present is shown by closed symbols.  This scaling is also $V_{rec} \propto
\nu^{-1/4}$, but can be explained rather as the effect of the finite inertial
range of turbulence than the effect of energy balance affected by viscosity or
boundary conditions.  For an extended range of motions, LV99 does not predict
any viscosity dependence, if the dissipation scale lays much below the scale of
current sheet.  However, for numerical simulations the range of turbulent
motions is very limited and any additional viscosity decreases the resulting
velocity dispersion and therefore the field wandering thus affecting the reconnection
rate.

LV99 predicted that in the presence of sufficiently strong turbulence, plasma
effects should not play a role.  The accepted way to simulate plasma effects
within MHD code is to use anomalous resistivity.  The results of the
corresponding simulations are shown in the right panel of Figure~\ref{fig:viscosity}
and they confirm that the change of the anomalous resistivity does not change
the reconnection rate.

\begin{figure*}
\centering
\includegraphics[width=0.45\textwidth]{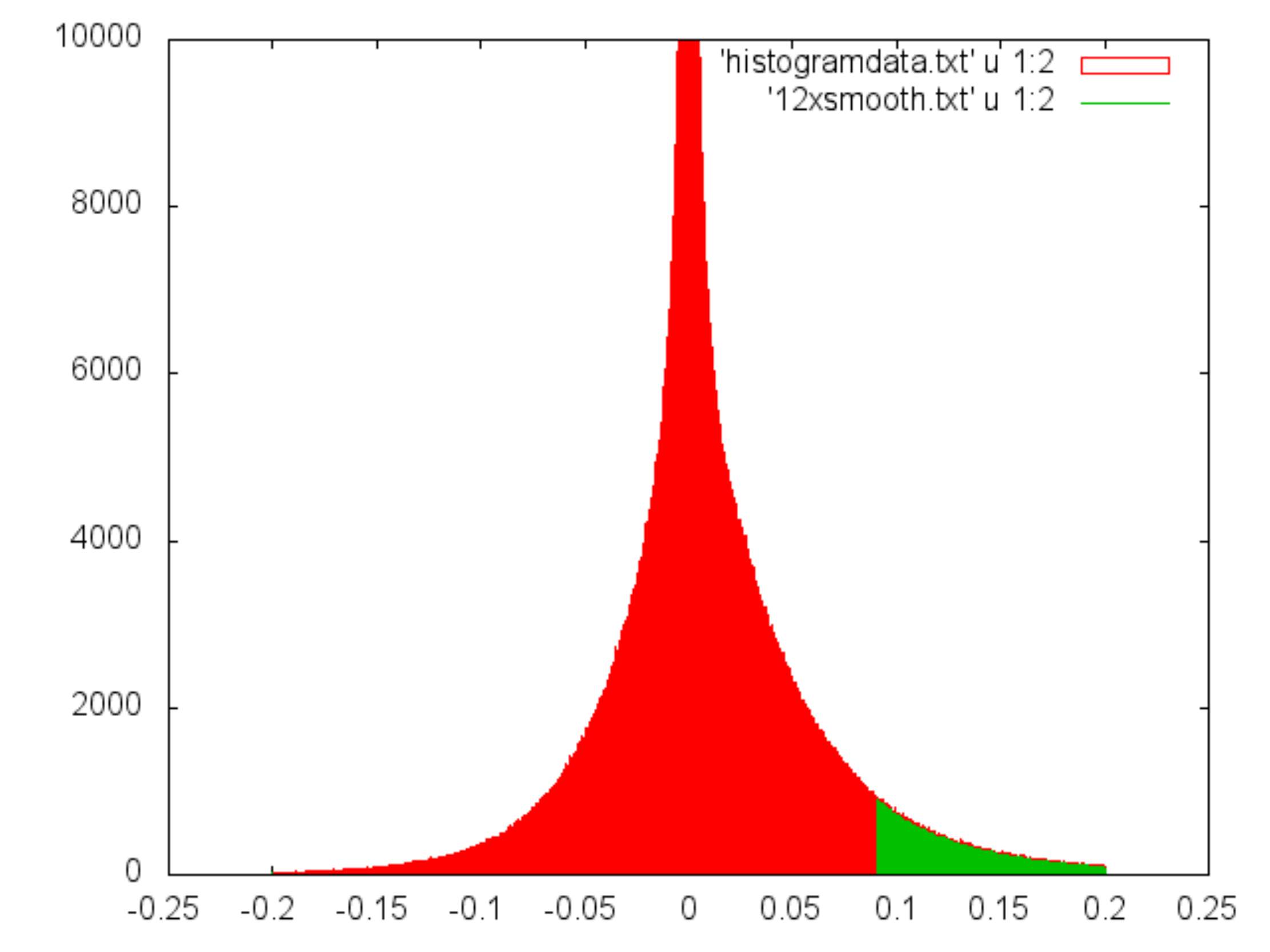}
\includegraphics[width=0.45\textwidth]{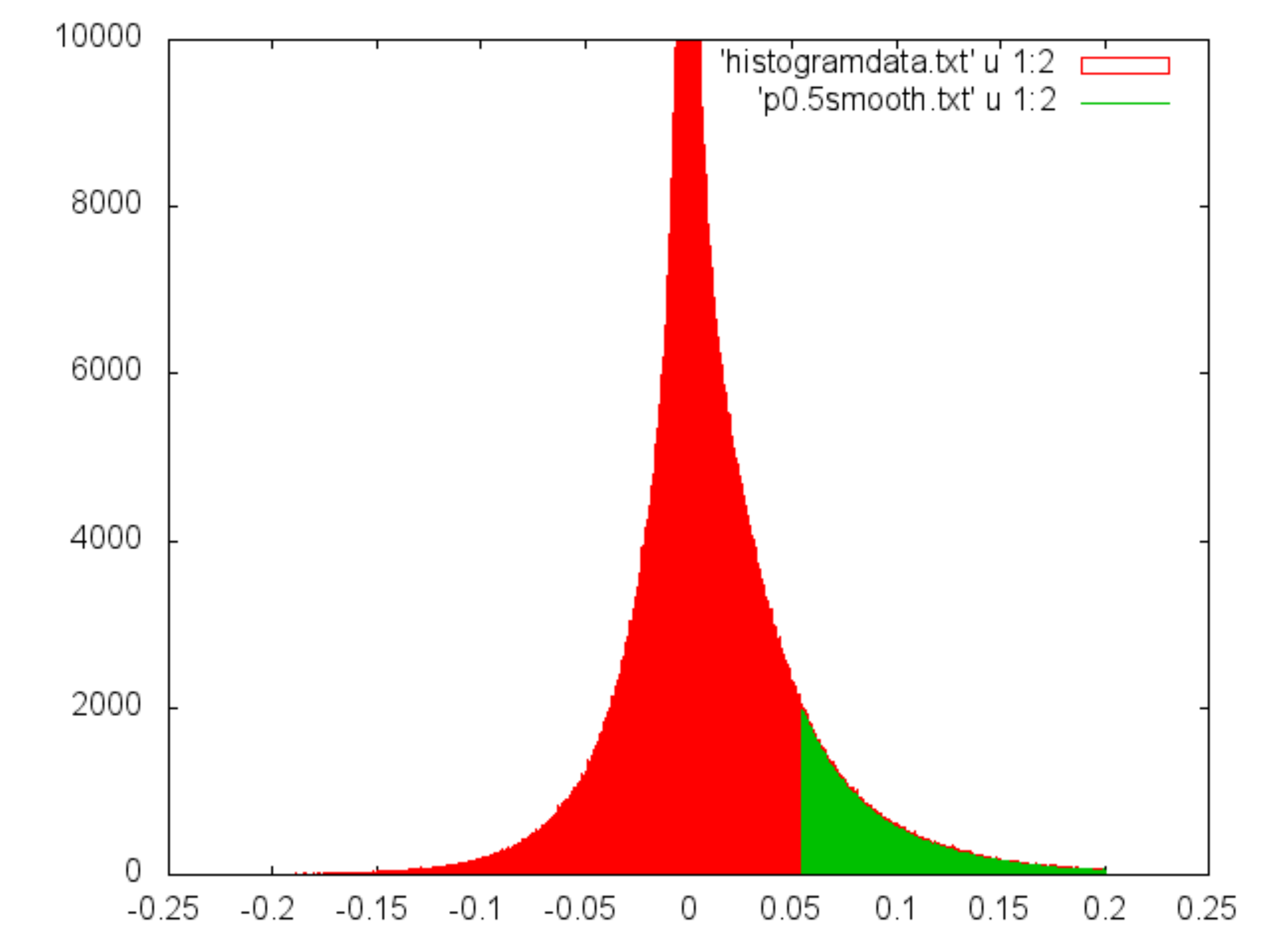}
\includegraphics[width=0.45\textwidth]{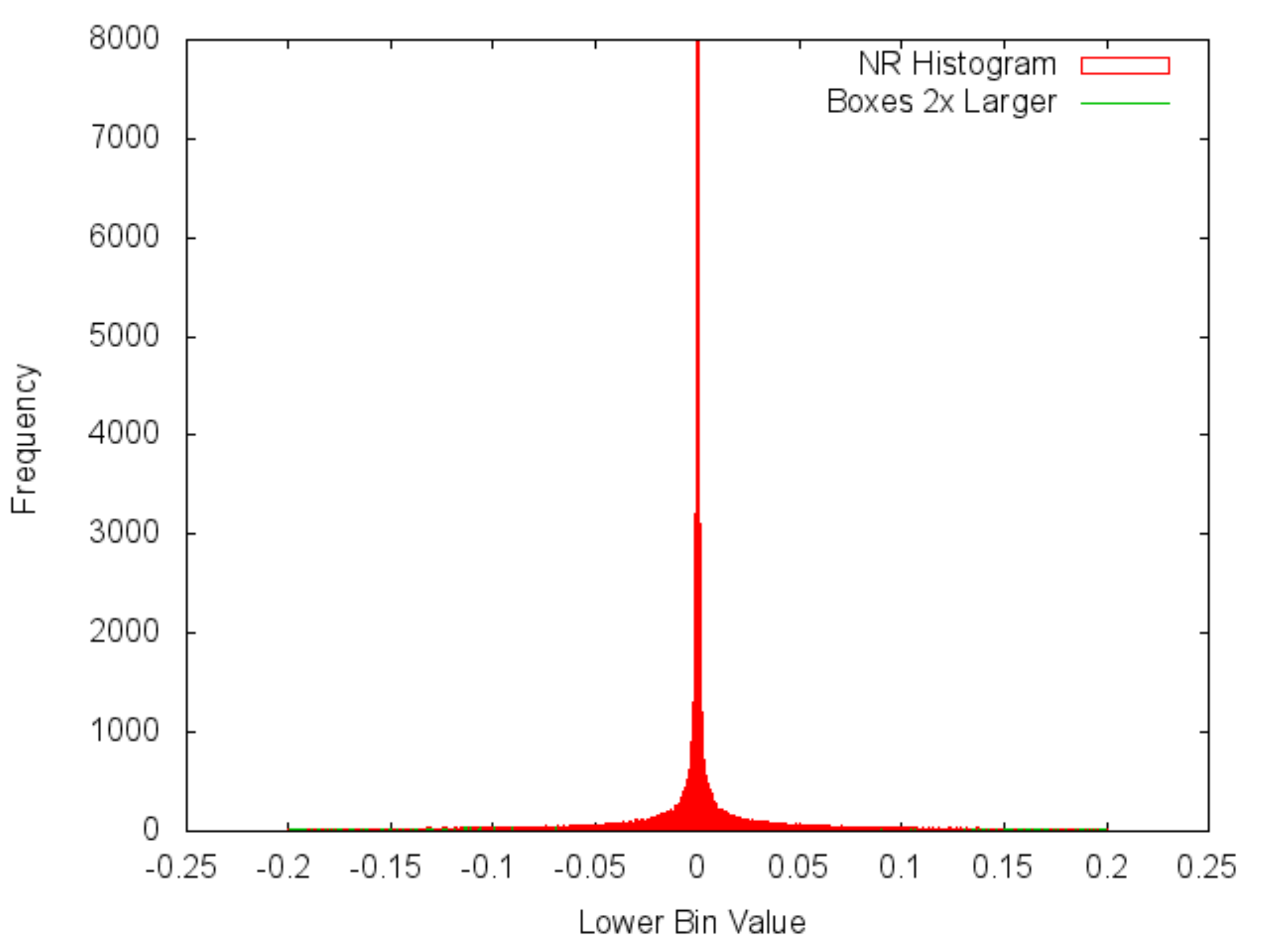}
\includegraphics[width=0.45\textwidth]{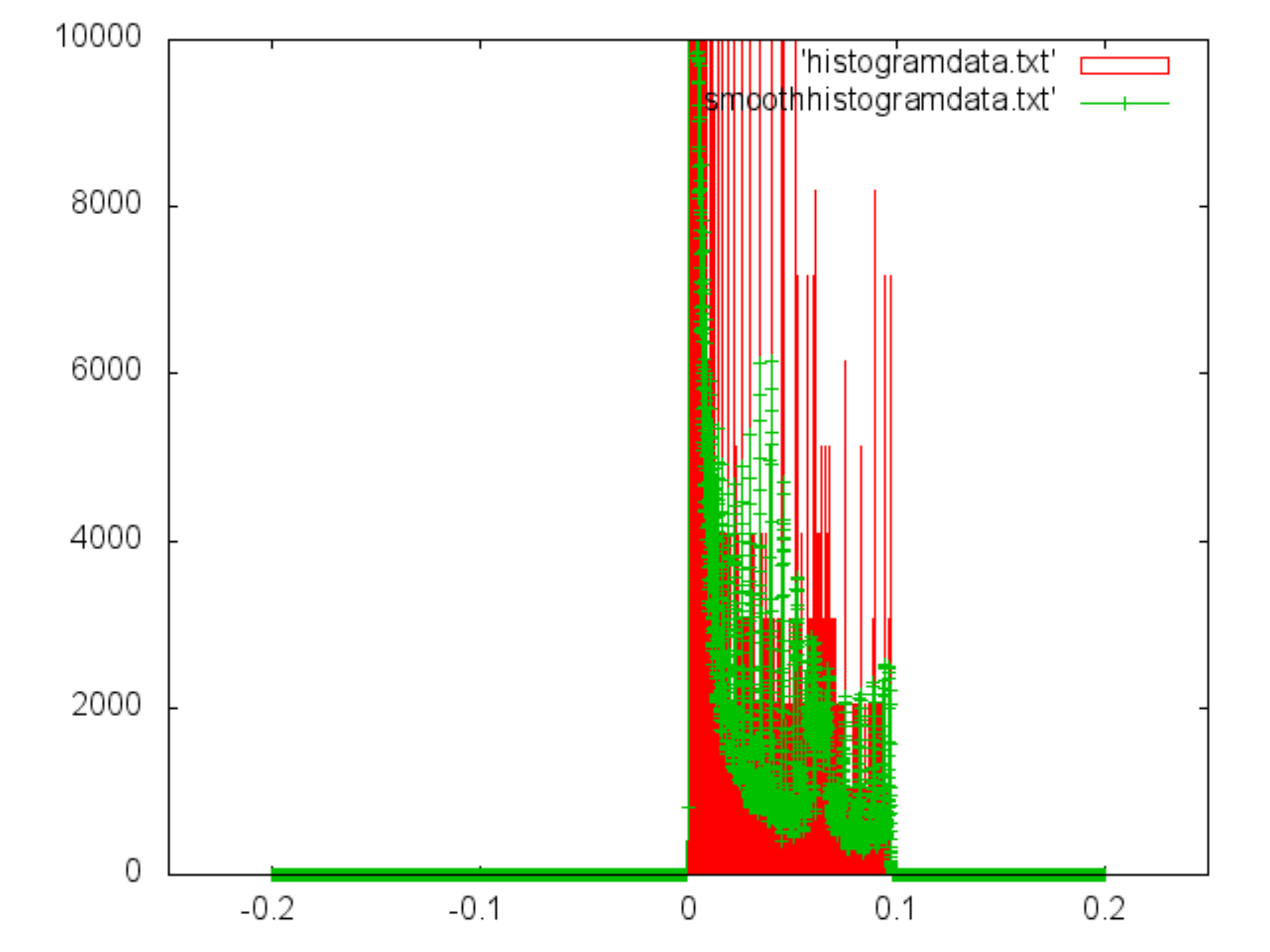}
\caption{These  figures show histograms of the gradient of the reversing
component of the large scale magnetic field in the direction normal to the
unperturbed current sheet, i.e. $\partial_yB_x$.  Upper left panel is for the
highest power simulation, P=1.  Upper right panel is for P=0.5. Lower left is
for P=1 but with no large scale magnetic field reversal, i.e. simply locally
driven strong turbulence.  Bins with twice the number of cells as  the
corresponding bin with the opposite sign of $\partial_yB_x$ are shown in green.
Lower right shows the first simulation in the absence of turbulent forcing.
From \cite{Vishniac_etal:2012}. \label{current}}
\end{figure*}

Within the derivation adopted in LV99 current sheet is broad with individual
currents distributed widely within a three dimensional volum  and the turbulence
within the reconnection region is similar to the turbulence within a
statistically homogeneous volume. Numerically, the structure of the reconnection
region was analyzed by Vishniac et al. \cite{Vishniac_etal:2012} based on the
numerical work by Kowal et al. \cite{Kowal_etal:2009}. The results support LV99
assumptions with reconnection region being broad, the magnetic shear is more or
less coincident with the outflow zone, and the turbulence within it is broadly
similar to turbulence in a homogeneous system.  In particular, this analysis
showed that peaks in the current were distributed throughout the reconnection
zone, and that the width of these peaks were not a strong function of their
strength.  The illustration of the results is shown in Figure~\ref{current}
which shows histograms of magnetic field gradients in the simulations with
strong and moderate driving power, with no magnetic field reversal but with
driven turbulence, and with no driven turbulence at all, but a passive magnetic
field reversal (i.e. Sweet-Parker reconnection).  A few features stand out in
this figure.  First, all the simulations with driven turbulence have a roughly
gaussian distribution of magnetic field gradients.  In the case with no field
reversal (panel c) the peak is narrow and symmetric around zero.  In the
presence of a large scale field reversal the peak is slightly broadened, and
skewed. It is turbulent reconnection does not produce any strong feature
corresponding to a preferred value of the magnetic field gradient.  Instead one
sees a systematic bias towards large positive values.  We conclude from the lack
of coherent features within the outflow zone, and the broad distribution of
values of the gradient of the magnetic field, that the current sheet and the
outflow zone are roughly coincident and this volume is filled with turbulent
structures.

As we discussed, the LV99 model is intrinsically related to the concept of
Richardson dispersion in magnetized fluids. Thus by testing the Richardson
diffusion of magnetic field, one also provides tests for the theory of turbulent
reconnection.

\begin{figure}[!ht]
\centering
\raisebox{-0.5\height}{\includegraphics[width=0.49\textwidth]{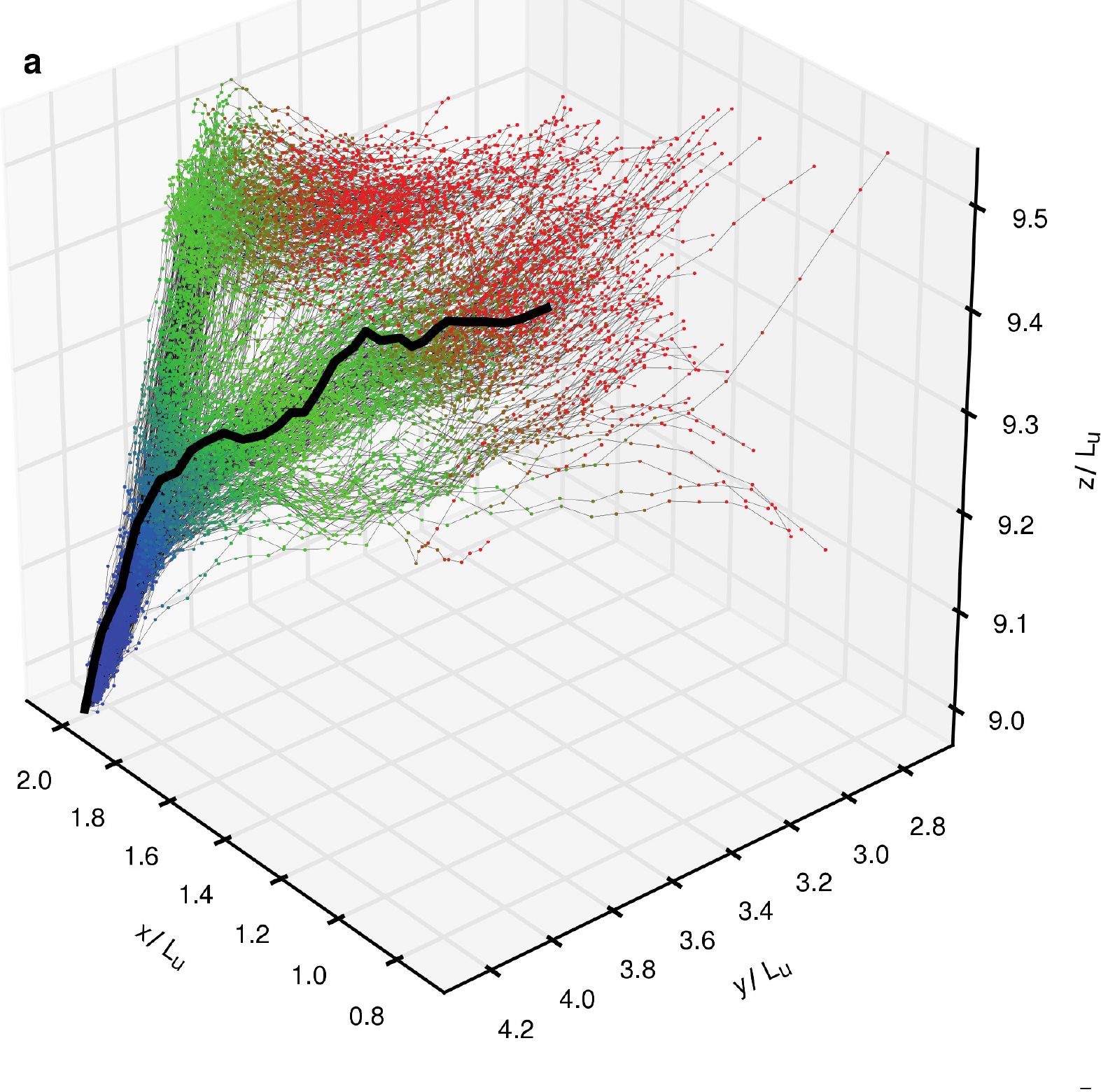}}
\raisebox{-0.5\height}{\includegraphics[width=0.49\textwidth]{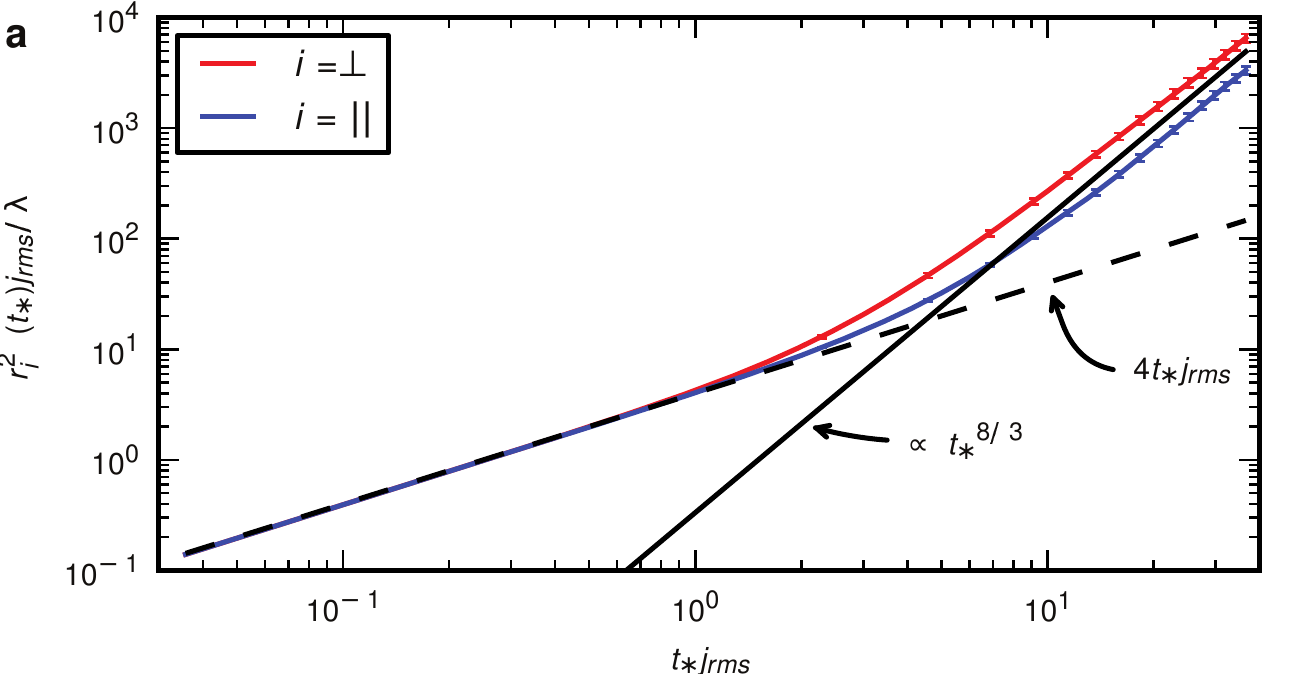}}
\caption{{\it Left panel}. Stochastic trajectories that arrive at a fixed point
in the archived MHD flow, color-coded \textcolor{red}{red},
\textcolor{green}{green}, and \textcolor{blue}{blue} from earlier to later
times.  From \cite{Eyink_etal:2013}.
{\it Right panel.} Mean-square dispersion of field-lines backwards in time, with
\textcolor{red}{red} for direction parallel and \textcolor{blue}{blue} for
direction perpendicular to the local magnetic field.  From \cite{Eyink_etal:2013}.
\label{figure10}}
\end{figure}

The first numerical tests of Richardson dispersion were related to magnetic field
wandering predicted in LV99 \cite{Maron_etal:2004, Lazarian_etal:2004,
Beresnyak:2013b}.  In Figure~\ref{figure10} we show the results obtained in
\cite{Lazarian_etal:2004}.  There we clearly see different regimes of magnetic field
diffusion, including the $y\sim x^{3/2}$ regime.  This is a manifestation of the
spatial Richardson dispersion.

As we discussed in section 3, the LV99 expressions can be obtained by applying  the concept of
Richardson dispersion to a magnetized layer. Thus by testing the Richardson
diffusion of magnetic field, one also provides tests for the theory of turbulent
reconnection.

The numerical tests of Richardson dispersion in space correspond to magnetic
field wandering predicted in LV99.  In Figure~\ref{figure10} we show the results
obtained in \cite{Lazarian_etal:2004}.  There we clearly see the Richardson
regime corresponding to  $y\sim x^{3/2}$ regime (see more discussion in ELV11).

A direct testing of the temporal Richardson dispersion of magnetic field-lines
was performed recently in \cite{Eyink_etal:2013}.  For this experiment,
stochastic fluid trajectories had to be tracked backward in time from a fixed
point in order to determine which field lines at earlier times would arrive to
that point and be resistively ``glued together''.  Hence, many time frames of an
MHD simulation were stored so that equations for the trajectories could be
integrated backward.  The results of this study are illustrated in
Figure~\ref{figure10}.  The left panel shows the trajectories of the arriving
magnetic field-lines, which are clearly widely dispersed backward in time, more
resembling a spreading plume of smoke than a single ``frozen-in'' line.
Quantitative results are presented in the right panel, which plots the
root-mean-square line dispersion in directions both parallel and perpendicular
to the local mean magnetic field.  Times are in units of the resistive time
$1/j_{rms}$ determined by the rms current value and distances in units of the
resistive length $\lambda/j_{rms}$.  The dashed line shows the standard
diffusive estimate $4\lambda t,$  while the solid line shows the Richardson-type
diffusion, the power-law is a bit altered by the numerical effects\footnote{Due
to the bottleneck effect the measured magnetic energy spectrum is $k^{-3/2}$
\cite{BeresnyakLazarian:2010} and this spectrum corresponds to $t^{8/3}$
Richardson dispersion dependence.} We would like to stress that whatever plasma
mechanism of line-slippage holds at scales below the ion gyroradius--- electron
inertia, pressure anisotropy, etc.---will be accelerated and effectively
replaced by the ideal MHD effect of Richardson dispersion.

\begin{figure}[!ht]
\centering
\includegraphics[width=\textwidth]{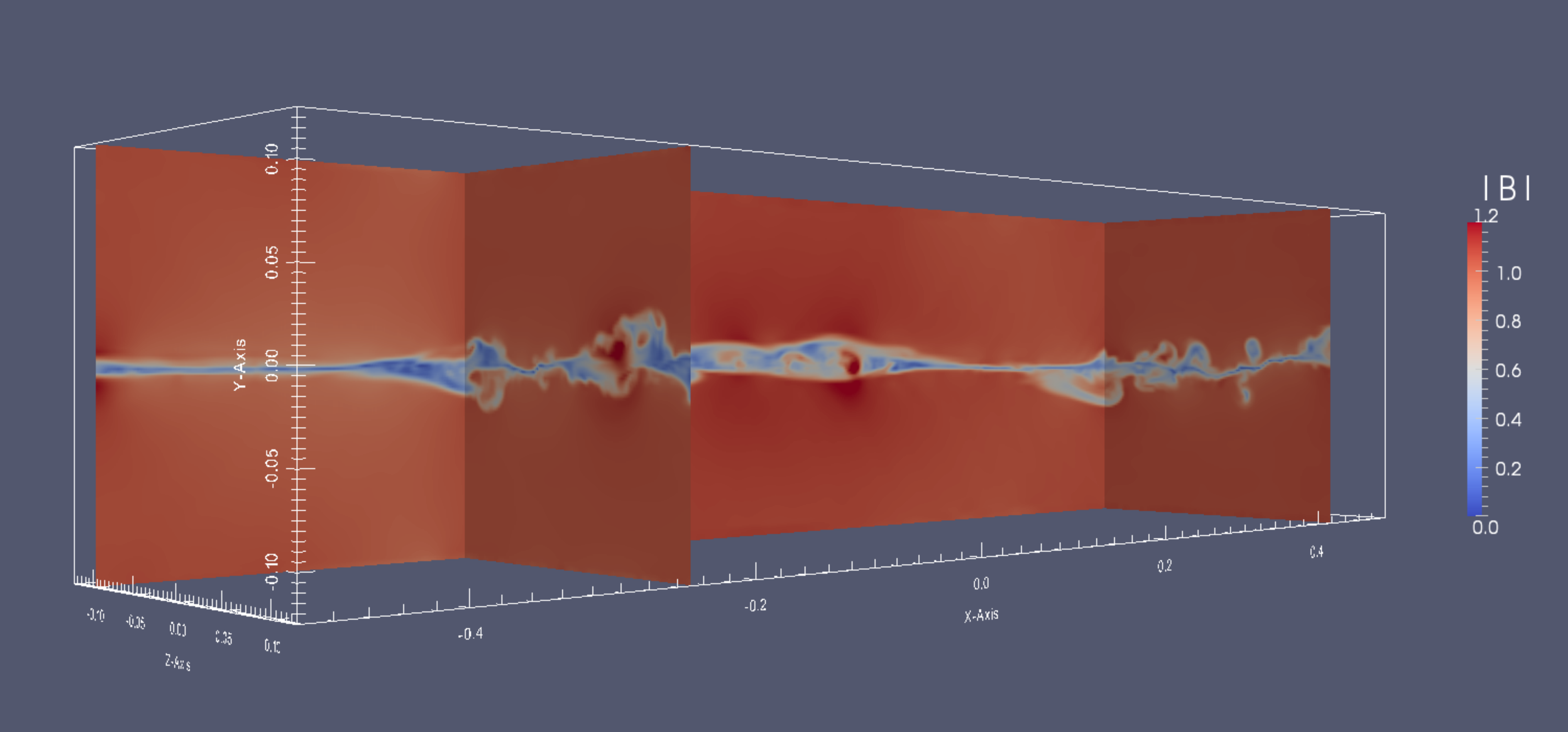}
\caption{Visualization of the model of turbulence generated by the seed
reconnection from \cite{Kowal_etal:2015}. Three different cuts (one XY plane at
Z=-0.1 and two YZ-planes at X=-0.25 and X=0.42) through the computational domain
show the strength of magnetic field $|\vec{B}|$ at the evolution time $t=1.0$.
Kelvin-Helmholtz-type structures are well seen in the planes perpendicular to
the reconnecting magnetic component $B_x$.  In the Z direction, the
Kelvin-Helmholtz instability is slightly suppressed by the guide field of the
strength $B_z=0.1 B_x$ (with $B_x=1.0$ initially). The initial current sheet is
located along the XZ plane at Y=0.0. A weak ($E_{kin} \approx 10^{-4} E_{mag}$)
random velocity field was imposed initialy in order to seed the reconnection.
\label{fig:bmag_cuts}}
\end{figure}

As we discussed in Section 4c the self-sustained turbulent reconnection where
the turbulence is generated by the reconnection itself can be quantified using
the predictions of the LV99 theory. Below we compare the prediction given by
Eq.~(\ref{rec_self}) against the results of recent simulations illustrated by
Figure~\ref{fig:bmag_cuts}.  The figure shows a few slices of the magnetic field
strength $|\vec{B}|$ through the three-dimensional computational domain with
dimensions $L_x=1.0$ and $L_y=L_z=0.25$.  The simulation was done with the
resolution $2048 \times 512 \times 512$.  Open boundary conditions along the X
and Y directions allowed studies of steady state turbulence.  At the presented
time $t=1.0$ the turbulence strength increased by two orders of magnitude from
its initial value of $E_{kin} \approx 10^{-4} E_{mag}$.  Initially, only the
seed velocity field at the smallest scales was imposed (a random velocity vector
was set for each cell).  We expect that most of the injected energy comes from
the Kelvin-Helmholtz instability induced by the local interactions between the
reconnection events, which dominates in the Z-direction, along which a weak
guide field is imposed ($B_z=0.1 B_x$).  As seen in the planes perpendicular to
$B_x$ in Figure~\ref{fig:bmag_cuts}, Kelvin-Helmholtz-like structures are
already well developed at time $t=1.0$. Turbulent structures are also observed
within the XY-plane, which probably are generated by the strong interactions of
the ejected plasma from the neighboring reconnection events.  More detailed
analysis of the spectra of turbulence and efficiency of the Kelvin-Helmholtz
instability as the turbulent injection mechanism are presented in
\cite{Kowal_etal:2015}.

The Kelvin-Helmholtz instability due to the interactions of the outflows
from neighboring reconnection events, which takes place in our simulations, is
somewhat different from that in the current sheet of Sweet-Parker reconnection,
which has been theoretically predicted in \cite{Loureiro_etal:2013}.
In the laminar reconnection, the profile of the outflow velocity has its maximum
in the middle of the current sheet and quickly decays along the direction
parallel to the reconnecting magnetic field component.  This configuration
creates naturally two shear layers in which the Kelvin-Helmholtz instability may
develop if the outflow velocity exceeds the Alfv\'en speed associated with the
upstream magnetic field. In order to confirm the predictions obtained in
\cite{Loureiro_etal:2013} we would need simulations or observations
of the thin current sheets with very large resolutions.

\begin{figure}
 \centering
 \includegraphics[width=0.6\textwidth]{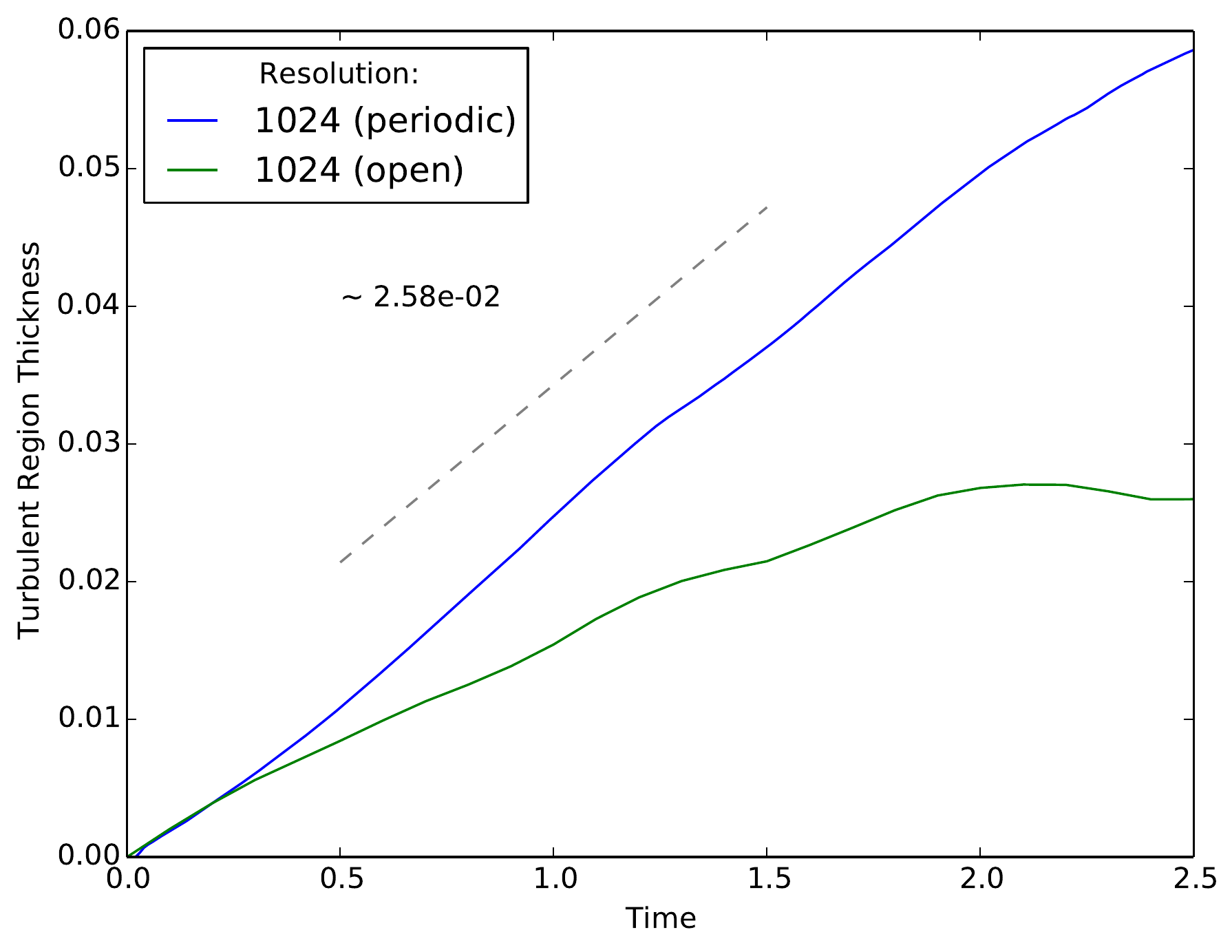}
 \caption{The growth of the turbulent region width for two models similar to one presented in Fig.~\ref{fig:bmag_cuts} but with periodic and open domain. From \cite{Kowal_etal:2015}. \label{fig:thickness}}
\end{figure}

In Figure~\ref{fig:thickness} we show the growth of the turbulent region
thickness for models ran with the same parameters, one in periodic domain, and
another with open boundary conditions.  In the case of periodic box, the
reconnected flux and generated turbulence are accumulated in the region near
current sheet. Since there is no outflow, the thickness of this region grows
linearly with the estimated growth rate of about 0.026 (see blue line in
Fig.~\ref{fig:thickness}). Once we allow the reconnected flux and turbulence to
be ejected along the reconnecting magnetic field, the thickness of turbulent
region saturates after about 2.0 Alfv\'en time units at level of ~0.025 L, where
L is the longitudinal size of the domain.  These values are in agreement with
the estimates from Eqs.~(\ref{growth}) and (\ref{rec_self}).

\section{Observational Testing of Turbulent Reconnection}

\subsection{Solar Reconnection}

To quantify solar reconnection one should accept that the energy is injected
by reconnection and turbulence is driven by magnetic reconnection. In this situation one can
expect substantial changes of the magnetic field direction corresponding to
strong turbulence.  Thus it is natural to identify the velocities measured
during the reconnection events with the strong MHD turbulence regime.  In other
words, one can use:
\begin{equation}
V_{rec}\approx U_{obs, turb} (L_{inj}/L_x)^{1/2},
\label{obs}
\end{equation}
where $U_{obs, turb}$ is the spectroscopically measured turbulent velocity
dispersion. Similarly, the thickness of the reconnection layer should be defined
as
\begin{equation}
\Delta\approx L_x (U_{obs, turb}/V_A) (L_{inj}/L_x)^{1/2}.
\label{delta_obs}
\end{equation}

The expressions given by Eqs.~(\ref{obs}) and (\ref{delta_obs}) can be compared
with observations in \cite{CiaravellaRaymond:2008}.  There, the widths of the
reconnection regions were reported in the range from 0.08$L_x$ up to 0.16$L_x$
while the the observed Doppler velocities in the units of $V_A$ were of the
order of 0.1.  It is easy to see that these values are in a good agreement with
the predictions given by Eq.~(\ref{delta_obs})\footnote{If we associate
the observed velocities with isotropic driving of turbulence, which is
unrealistic for the present situation, then a discrepancy with
Eq.~(\ref{delta_obs}) would appear.  Because of that \cite{CiaravellaRaymond:2008}
did not get quite as good quantitative agreement between observations and theory
as we did, but still within observational uncertainties.}

If we talk about unique predictions that radically differ from LV99 and the {\it
present day} plasma reconnection models then the LV99 prediction of the
triggering of reconnection by wave packets coming from the adjacent
reconnectionsites should be singled out.  Thus a particular series of solar
observations is important. In \cite{Sych_etal:2009}, authors explaining
quasi-periodic pulsations in observed flaring energy releases at an active
region above the sunspot, proposed that the wave packets arising from the
sunspots can trigger such pulsations. This is exactly what is expected within
the LV99 model.

The criterion for the application of LV99 theory is that the outflow region is
much larger than the ion Larmor radius $\Delta \gg \rho_i$.  This is definitely
satisfied for the solar atmosphere where the ratio of $\Delta$ to $\rho_i$ can
be larger than $10^6$. Plasma effects can play a role for small scale
reconnection events within the layer, since the dissipation length based on
Spitzer resistivity is $\sim 1$ cm, whereas $\rho_i\sim 10^3$ cm. However, as we
discussed earlier, this does not change the overall dynamics of turbulent
reconnection.

\subsection{Solar Wind}

Reconnection throughout most of the heliosphere appears similar to that in the
Sun. For example, there are now extensive observations of reconnection jets
(outflows, exhausts) and strong current sheets in the solar wind
\cite{Gosling:2012}. The most intense current sheets observed in the solar wind are
very often not observed to be associated with strong (Alfv\'enic) outflows and
have widths at most a few tens of the proton inertial length $\delta_i$ or
proton gyroradius $\rho_i$ (whichever is larger). Small-scale current sheets of
this sort that do exhibit observable reconnection have exhausts with widths at
most a few hundreds of ion inertial lengths and frequently have small shear
angles (strong guide fields) \cite{Gosling_etal:2007, GoslingSzabo:2008}.  Such
small-scale reconnection in the solar wind requires collisionless physics for
its description, but the observations are exactly what would be expected of
small-scale reconnection in MHD turbulence of a collisionless plasma
\cite{Vasquez_etal:2007}. Indeed, LV99 predicted that the small-scale reconnection
in MHD turbulence should be similar to large-scale reconnection, but with nearly
parallel magnetic field lines and with ``outflows'' of the same order as the
local, shear-Alfv\'enic turbulent eddy motions. It is worth emphasizing that
reconnection in the sense of flux-freezing violation and disconnection of plasma
and magnetic fields is required at every point in a turbulent flow, not only
near the most intense current sheets. Otherwise fluid motions would be halted by
the turbulent tangling of frozen-in magnetic field lines. However, except at
sporadic strong current sheets, this ubiquitous small-scale turbulent
reconnection has none of the observable characteristics usually attributed to
reconnection, e.g. exhausts stronger than background velocities, and would be
overlooked in observational studies which focus on such features alone.

However, there is also a prevalence of very large-scale reconnection events
in the solar wind, often associated with interplanetary coronal mass
ejections and magnetic clouds or occasionally magnetic disconnection events
at the heliospheric current sheet  \cite{Phan_etal:2009, Gosling:2012}.
These events have reconnection outflows with widths up to nearly $10^5$ of the
ion inertial length and appear to be in a prolonged, quasi-stationary regime
with reconnection lasting for several hours. Such large-scale reconnection is as
predicted by the LV99 theory when very large flux-structures with
oppositely-directed components of magnetic field impinge upon each other in the
turbulent environment of the solar wind. The ``current sheet'' producing such
large-scale reconnection in the LV99 theory contains itself many ion-scale,
intense current sheets embedded in a diffuse turbulent background of weaker (but
still substantial) current. Observational efforts addressed to
proving/disproving the LV99 theory should note that it is this broad zone of
more diffuse current, not the sporadic strong sheets, which is responsible for
large-scale turbulent reconnection. Note that the study \cite{Eyink_etal:2013}
showed that standard magnetic flux-freezing is violated at general points in
turbulent  MHD, not just at the most intense, sparsely distributed sheets. Thus,
large-scale reconnection in the solar wind is a very promising area for LV99.

Preliminary comparisons between such events in MHD turbulence and in the
high-speed solar wind have yielded very promising results
\cite{Lalescu_etal:2013}. Criteria can be employed that are designed specifically
to look for large-scale reconnection. For example, the ``partial-variance of
increments'' (PVI) criterion recently proposed by \cite{Osman_etal:2014} can be
adapted for this purpose, by considering magnetic increments over inertial-range
separation distances rather than ion-scale distances and, possibly also, with
coarse-graining of the magnetic field to eliminate smaller-scale features. A
similar modification may be made to the criterion of Gosling \cite{Gosling:2012},
which identifies reconnection events by roughly Alfenic-jetting plasma bounded
on one side by correlated changes in the antiparallel components of ${\bf u}$
and ${\bf B}$ and by anti-correlated changes in those components on the other
side. Here the criterion may be modified by requiring that the two large changes
must be separated spatially by inertial-range lengths, i.e. essentially by
conditioning on a broad outflow jet.

Examples of some events yielded by this latter Gosling-type criterion are shown
in Figure 10. The top panels of the figures show a typical event selected from
the JHU turbulence database, which archives the output of a $1024^3$
pseudo-spectral simulation of the incompressible MHD equations.
The bottom panel shows a similar event obtained from a study of a fast solar
wind stream, 2008 January 14 04:40:00 -- January 21 03:20:00, using three-second
resolution {\it Wind} spacecraft observations from the Magnetic Field
Investigation (MFI) and 3D Plasma Analyzer (3DP) experiments. The left panels
show magnetic field components and the right panels show velocity components,
both rotated into the local minimum-variance-frame \cite{SonnerupCahill:1967}
plotted versus space for 1D cuts through the MHD simulation and versus time for
the spacecraft data. The JHU MHD data are in the arbitrary units of the
simulation, for which the rms magnetic field strength $b'=0.24,$ the magnetic
integral length $L_b=0.35,$ and the resistive dissipation length
$\eta_b=0.0028.$ The units for the {\it Wind} data are nanotesla (nT) for the
magnetic field, kilometer per second (km/s) for velocity, and minutes for time.
Average solar wind conditions were speed u= 660 km/s, magnetic field strength
$B= 4.4$ nT, proton number density ${\rm n}_{p} = 2.4 \,\,{\rm cm}^{-3}$,
Alfv\'en speed $V_A=$62 km/s, and proton beta $\beta_p = 1.2$. The outer scale
of the turbulent inertial-range (boundary with the $1/f$ spectral range) is $33$
mins and the inner scale (a few ion gyroradii) around 10 s.

The event from the MHD database was found by searching for ``Gosling events'' that
show opposing changes in ${\bf u}$ and ${\bf B}$ within a distance of 0.196,
about half an integral length. The event from the high-speed solar wind was
found by applying the same criterion for separation of 400 s. \\  Interestingly,
neither of these events show the ``double-step'' structure, with an intermediate
plateau of reversing magnetic field component, which often characterize the
events identified by Gosling \cite{Gosling:2012}, although other events we have
found do show this structure. Most importantly, both events show the features
expected of large-scale reconnection, with a sizable magnetic reversal over an
inertial-range length and with a corresponding outflow in the same direction and
of the same width. This makes both events likely candidates for turbulent
reconnection. In the case of the MHD database event, this interpretation can be
verified from the simulation data. A detailed study in preparation
(Eyink et al., in prep.) shows that the MHD event presented in the top panels of
Figure 10 accords well with the predictions of the LV99 theory and has the
expected morphological features: a wide (inertial-range scale) outflow jet, a
distribution of small-scale current sheets rather than a single dominant sheet,
turbulent wandering of magnetic field-lines, and Richardson dispersion of
field-lines normal to the reversal direction. It is therefore natural to
identify the similar events in the solar wind as turbulent reconnection as well.
This identification is strengthened by the similar statistical rates of
occurrence of such events at corresponding scales, as observed also in previous
studies of inertial-range magnetic increments in MHD turbulence and the solar
wind \cite{Zhdankin_etal:2012}. The high-speed solar wind is presumably full of such
turbulent reconnection events, across its broad spectrum of inertial-range
length-scales.
\begin{figure}[htp!]
\centering
\begin{tabular}{c}
\includegraphics[width=120mm,height=60mm]{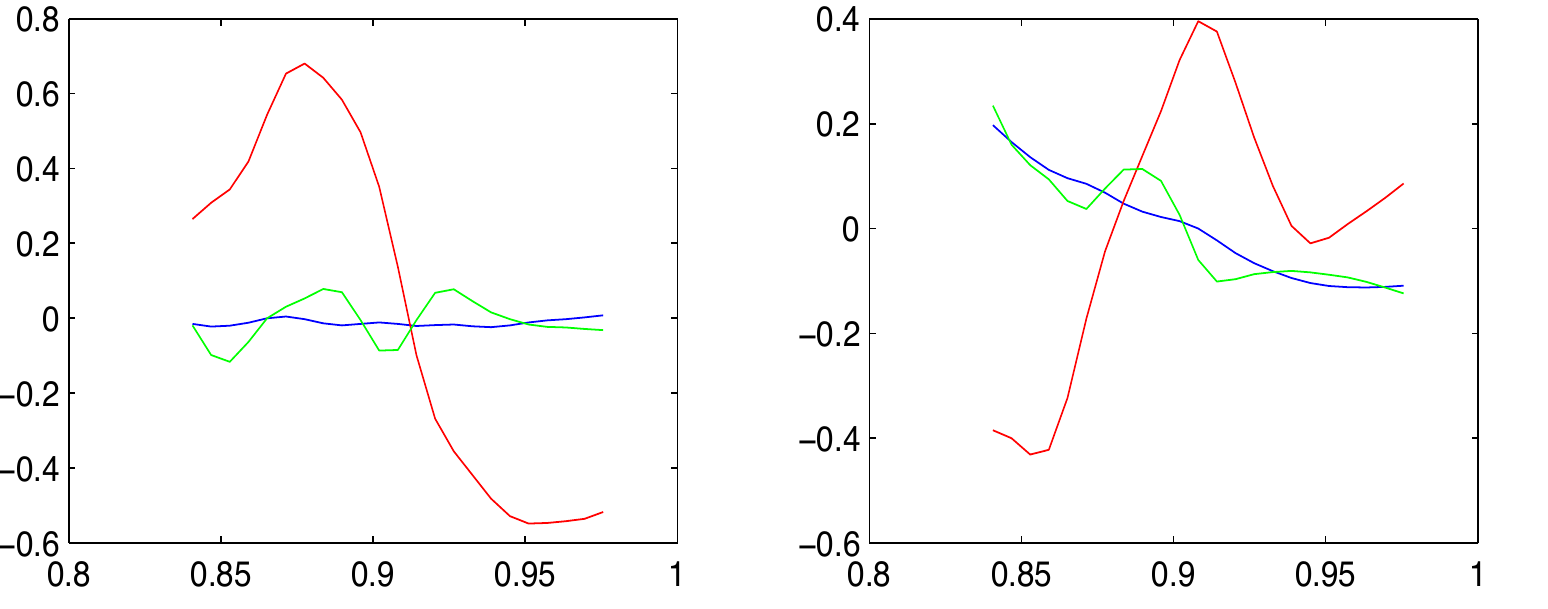}\\
\includegraphics[width=120mm]{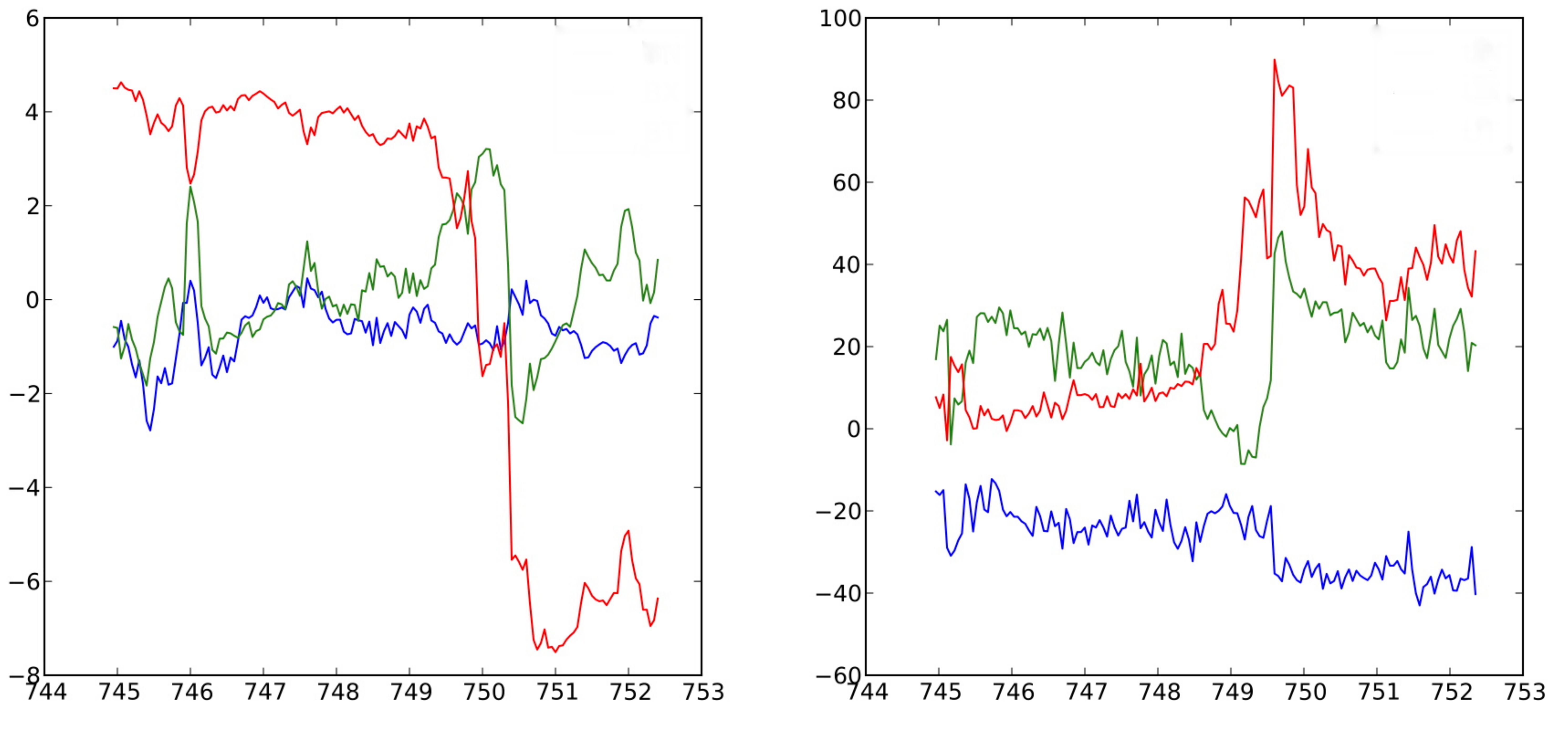}
\end{tabular}
\label{Gosling-criterion}\caption{Candidate Events for Turbulent
Reconnection.} {\it MHD Turbulence Simulation (Top Panels) and High-Speed Solar
Wind (Bottom Panels)}. The left panels show magnetic field components and the
right panels velocity components, both rotated into a local minimum-variance
frame of the magnetic field. The component of maximum variance in
\textcolor{red}{{\bf red}} is the apparent reconnecting component, the component
of medium variance in \textcolor{green}{green} is the nominal guide-field
direction, and the minimum-variance direction in \textcolor{blue}{blue} is
perpendicular to the reconnection layer.
\end{figure}

We note, that the situation for applicability of LV99 generally gets better with
increasing distance from the sun, because of the great increase in scales. For
example, reconnecting flux structures in the inner heliosheath could have sizes
up to $\sim$100 AU, much larger than the ion cyclotron radius $\sim10^3$ km
\cite{LazarianOpher:2009}.

\subsection{Parker spiral and Heliospheric Current Sheet}

More recently, \cite{Eyink:2014} discussed some implications of LV99 for
heliospheric reconnection, in particular for deviations from the Parker spiral
model of interplanetary magnetic field.  Note, that the \cite{Parker:1958}
spiral model of the interplanetary magnetic field, which is one of the most
famous applications in astrophysics and space science of the ``frozen-in''
principle for magnetic field lines. The model has been shown to be approximately
valid when taking into account solar cycle variations in source magnetic field
strength and latitude/time variation in solar wind speeds.  Nevertheless,
\cite{Parker:1958} concluded his paper with a ``warning to the reader against
taking too literally any of the smooth idealized models which we have
constructed in this paper''.

\cite{Burlaga_etal:1982} had studied the magnetic geometry and found ``notable
deviations'' from the spiral model. \cite{Burlaga_etal:1982} studied daily
averages of magnetic field observations of Voyager 1 and 2 in the ecliptic plane
at solar distances R=1-5 AU during a period of increasing solar activity in the
years 1977--1979. In contrast to the Parker predictions for radial magnetic
field component radial dependencies $B_R \sim R^{-2}$ and azimuthal component
$B_T \sim R^{-1}$, \cite{Burlaga_etal:1982} found $B_R \sim R^{-1.56}$
and $B_T \sim R^{-1.20}$. \cite{Burlaga_etal:1982} attributed the observed
deviations to ``temporal variations associated with increasing solar activity,
and to the effects of fluctuations of the field in the radial direction''.
These early observations were recently confirmed by
\cite{KhabarovaObridko:2012}, who presented evidence on the breakdown of the
Parker spiral model for time- and space-averaged values of the magnetic field
from several spacecraft (Helios 2, Pioneer Venus Orbiter, IMP8, Voyager 1) in
the inner heliosphere at solar distances 0.3-5 AU and in the years 1976--1979.
\cite{KhabarovaObridko:2012} interpret their observations as due to ``a
quasi-continuous magnetic reconnection, occurring both at the heliospheric
current sheet and at local current sheets inside the IMF sectors''. They present
extensive evidence that most nulls of BR and BT , where reconnection may occur,
are not associated to the heliospheric current sheet. They as well observe a
rapid disappearance of the regular sector structure at distances past 1 AU,
which they attribute to Òturbulent processes in the inner heliosphere.
\cite{Eyink:2014} estimated the magnetic field slippage velocities and related
the deviation from Parker original predictions to LV99 reconnection.

In addition, \cite{Eyink:2014} analyzed the data relevant to the region
associated with the broadened heliospheric current sheet (HCS), noticed its
turbulent nature and provided arguments on the applicability of LV99 magnetic
reconnection model to HCS. This seems to be a very promising direction of
research to study turbulent reconnection in action using in situ  spacecraft
measurements.

\section{Implications}

\subsection{Magnetic flux freezing in the presence of turbulent reconnection}
\label{ssec:flux_freez}

The concept of flux freezing was first proposed by Hannes Alfv\'{e}n in 1942,
and the principle of frozen-in field lines has provided a powerful heuristic
\cite{Parker:1979, Kulsrud:2005}. However, if strictly valid, flux
would forbid magnetic reconnection, because field-lines frozen into a continuous
plasma flow cannot change their topology. If magnetic reconnection were a
phenomenon isolated to to regions of special magnetic flux topology or other
special conditions, then it would be possible to use flux freezing for generic
magnetic field conditions. However, LV99 model suggests that magnetic
reconnection happens everywhere in magnetized turbulent fluids. This means the
ubiquitous violation of flux freezing in magnetized turbulence.

Standard mathematical proofs of flux-freezing in MHD always assume, implicitly,
that velocity and magnetic fields remain smooth as $\eta \rightarrow 0$.
However, MHD solutions with small resistivities and viscosities (high magnetic
and kinetic Reynolds numbers) are generally turbulent.  These solutions exhibit
long ranges of power-law spectra corresponding to very non-smooth or ``rough''
magnetic and velocity fields.  Fluid particle (Lagrangian) trajectories in such
rough flows are known to be non-unique and stochastic (see
\cite{Bernard_etal:1998, GawedzkiVergassola:2000, EVandenEijnden:2000a,
EVandenEijnden:2000b, EVandenEijnden:2001, Chaves_etal:2003}, and, for reviews,
\cite{Kupiainen:2003} and \cite{Gawedzki:2008}).   It view of the above, it is
immediately clear as a consequence that standard flux-freezing cannot hold in
turbulent plasma flows.  After all, the usual idea is that magnetic field-lines
at high conductivity are tied to the plasma flow and follow the fluid motion.
However, if the latter is non-unique and stochastic, then which fluid element
will the field-line follow?

For a laminar velocity field, this diffusion effect is small.  It is not hard to
see that a pair of field lines will attain a displacement ${\bf r}(t)$ apart
under the combined effect of advection and diffusion obeying $$
\frac{d}{dt}\langle r^2\rangle = 12\lambda + 2\langle {\bf r}\cdot \delta {\bf
u}({\bf r})\rangle $$ where $\delta {\bf u}({\bf r})$ is the relative advection
velocity at separation ${\bf r}$.  Thus, $$ \frac{d}{dt}\langle r^2\rangle \leq
12\lambda + 2\|\nabla {\bf u}\|\langle r^2\rangle, $$ where $\|\nabla {\bf u}\|$
is the maximum value of the velocity-gradient $\nabla {\bf u}$.  It follows that
two lines initially at the same point, by time $t$ can have separated at most
\begin{equation}
\langle r^2(t)\rangle \leq 6\lambda   \frac{e^{2\|\nabla{\bf u}\| t}-1}{\|\nabla{\bf u}\|}. \label{star}
\end{equation}
If we thus consider a smooth laminar flow with a fixed, finite value of
$\|\nabla{\bf u}\|$, then $\langle r^2(t)\rangle\rightarrow 0$ as
$\lambda\rightarrow 0$.  Under such an assumption, magnetic field lines do not
diffuse a far distance away from the solution of the deterministic ordinary differential equation $d{\bf
x}/dt={\bf u}({\bf x},t)$, and the magnetic line-diffusion becomes a negligible
effect.  In that case, magnetic flux is conserved better as $\lambda$ decreases.

However, in a turbulent flow, the above argument fails!  The inequality
(\ref{star}) still holds, of course, but it no longer restricts the dispersion
of field-lines under the joint action of resistivity and advection.  As is
well-known, a longer and longer inertial range of power-law spectrum $E(k)$
occurs as viscosity $\nu$ decreases and the maximum velocity gradient
$\|\nabla{\bf u}\|$ becomes larger and larger.  In fact, energy dissipation
$\varepsilon=\nu\|\nabla{\bf u}\|^2$ is observed to be non-vanishing as
$\nu\rightarrow 0$ in turbulent flow, requiring velocity gradients to grow
unboundedly.  Estimating $\|\nabla {\bf u}\|\sim (\varepsilon/\nu)^{1/2}$, the
upper bound (\ref{star}) becomes
\begin{equation}
\langle r^2(t)\rangle \leq 6\lambda (\nu/\varepsilon)^{1/2} [ \exp(2t (\varepsilon/\nu)^{1/2} ) - 1].
\label{star2}
\end{equation}
This bound allows unlimited diffusion of field-lines. Consider first the case
$\lambda=\nu$ or $Pt=1$, for simplicity, where Richardson's theory implies that
\begin{equation}
\langle r^2(t)\rangle  \sim 12\lambda t +\varepsilon t^3.
\end{equation}
The rigorous upper bound always lies strictly above Richardson's prediction and,
in fact, goes to infinity as $\nu=\lambda\rightarrow 0$!  The case of large
Prandtl number is just slightly more complicated, as previously discussed in
\S\ref{ssec:ionized}.  When $Pt\gg 1,$ the inequality (\ref{star2}) holds as an
equality for times $t\ll t_{trans}$ with
\begin{equation}
 t_{trans}= \frac{\ln(Pt)}{2(\varepsilon/\nu)}.
 \end{equation}
   This is then followed by a Richardson
despersion regime
\begin{equation}
\langle r^2(t)\rangle  \sim 6(\nu^3/\varepsilon)^{1/2} +
\varepsilon (t - t_{trans})^3, \,\,\,\,\,\,\,\,\,\,\,\,\,\,\,\, t\gg
t_{trans},
\end{equation}
assuming that the kinetic Reynolds number is also large and a
Kolmogorov inertial range exists at scales greater than the Kolmogorov length
$(\nu^3/\varepsilon)^{1/4}.$  Once again, the upper bound (\ref{star2}) is much
larger than Richardson's prediction and, at times longer than $t_{trans},$ the
dispersion of field lines is independent of resistivity.

\subsection{Making Goldreich-Sridhar model self-consistent}

Historically, a lot of reconnection research was aimed to obtain the Holy Grail
number of reconnection speed, which on the basis of solar flare observations was
determined to be 0.1$V_A$. This reconnection speed has been recently claimed to
be attained in a number of plasma simulations (see \cite{Yamada_etal:2010}). We
claim, however, that to make any model of strong turbulence self-consistent the
velocity of 0.1$V_A$ is insufficient. Below we show this for the GS95 model by
reproducing the arguments in LV99. Magnetic reconnection is required for free
mixing of magnetic field lines, which is a part of the GS95 picture of
turbulence. In fact, the critical balance that is the corner stone of the GS95
model can be derived from the equality of times for mixing of magnetic field
lines perpendicular to the local direction of magnetic field and the period of
the Alfven wave that this mixing induces. Therefore we consider magnetic
reconnection within magnetic eddies elongated along the local magnetic field
direction.

It is possible to see that within the GS95 picture the reconnection happens with
nearly parallel lines with magnetic pressure gradient $V_A^2/l_{\|}$ being
reduced by a factor $l_{\bot}^2/l_{\|}^2$, since only reversing component is
available for driving the outflow. At the same time the length of the contracted
magnetic field lines is also reduced from $l_{\bot}$ by $l_{\bot}^2/l_{\|}$.
Therefore the acceleration is $\tau_{eject}^{-2} l_{\bot}^2/l_{\|}$. As a
result, the Newtons' law gives $V_A^2 l_{\bot}^2/l_{\|}^3 \approx
\tau_{eject}^{-2} l_{\bot}^2/l_{\|}$. This provides the result for the ejection
rate $\tau_{eject}^{-1}\approx V_A/l_{\|}$. The length over which the magnetic
eddies intersect is $l_{\bot}$ and the rate of reconnection is
$V_{rec}/l_{\bot}$. For the stationary reconnection this gives $V_{rec}\approx
V_{A} l_{\bot}/l_{\|}$, which provides the reconnection rate $V_A/l_{\|}$. The
latter rate  is exactly the rate of the eddy turnovers in GS95 turbulence, which
shows that it is fast magnetic reconnection that makes the GS95 picture
self-consistent. In the case of trans-Alfvenic turbulence this means that the
reconnection velocity should be of the order of $V_A$. This sort of reconnection
rate has never been reported to be attainable within plasma reconnection
simulations (see \cite{Yamada_etal:2010}). However, this is the
reconnection rate that is expected for trans-Alfvenic turbulence within the LV99
model.

\subsection{Reconnection diffusion and star formation}

\begin{figure}[h!]
\centering
\includegraphics[width=0.40\textwidth]{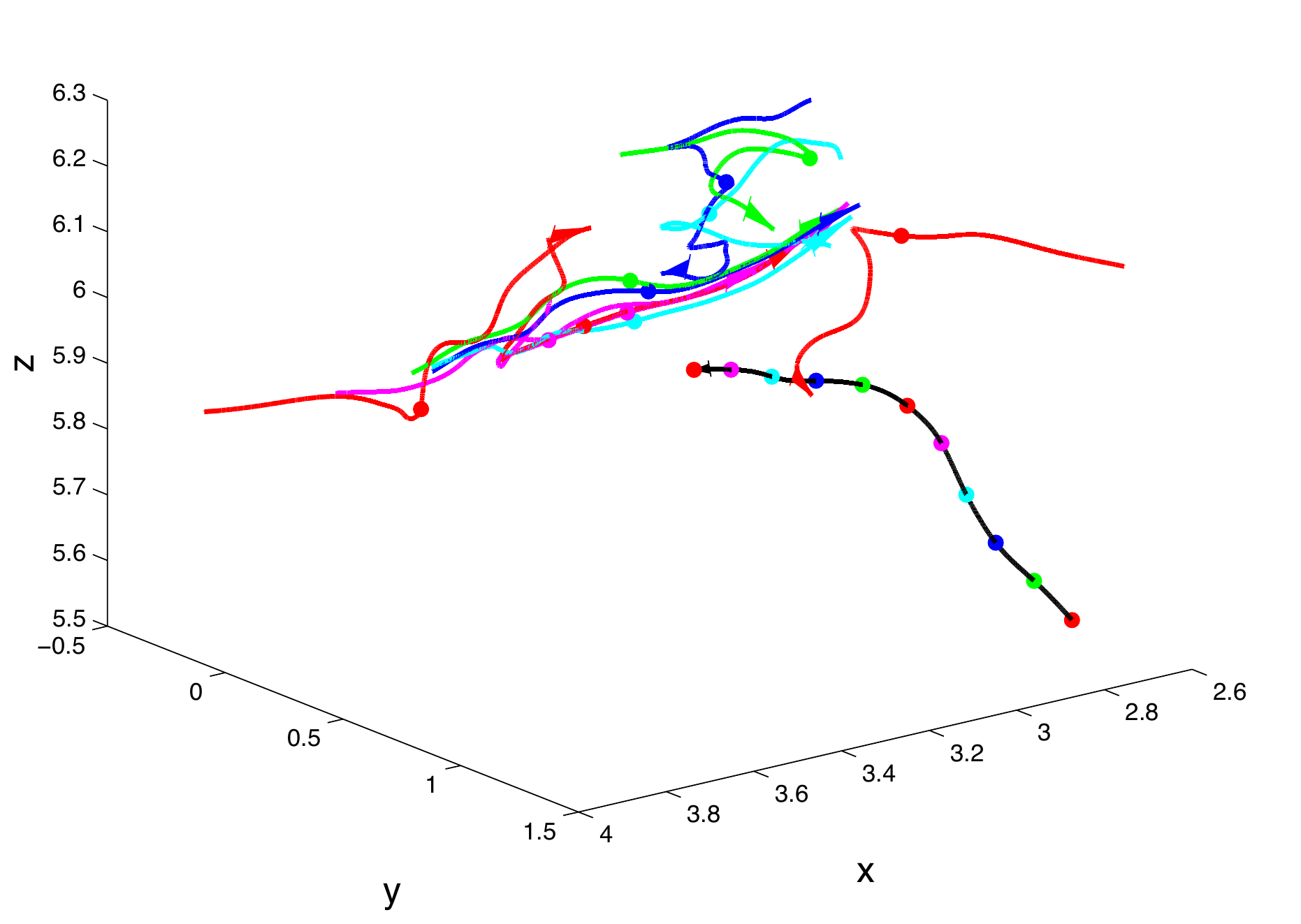}
\includegraphics[width=0.59\textwidth]{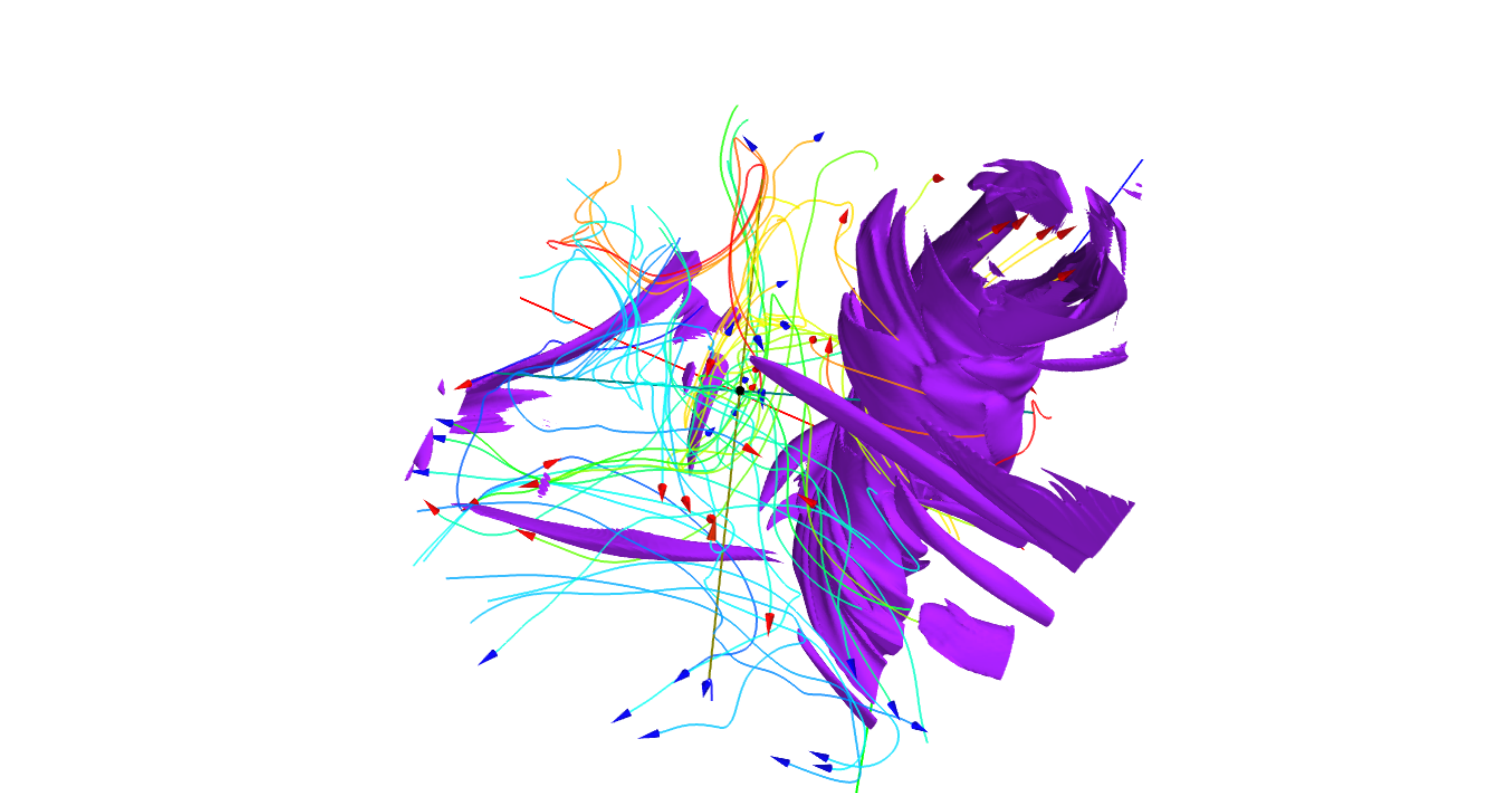}
\caption{{\bf Turbulent Splitting of a Magnetic Field Line.} An
initial magnetic field line in black is decorated with plasma elements, which
are allowed to move with the fluid velocity. The magnetic field-lines that
thread these plasma elements at the later time are drastically different,
effectively ``splitting'' the original line. Left panel. {\bf Reconnection event
within 3D driven turbulence}. The structure of the event corresponds to
the LV99 picture and is consistent with reconnection events studied in solar wind.
 \label{split}}
\end{figure}

As we have argued earlier at length, standard flux-freezing breaks down at every
point and time in a turbulent plasma. In that case, the only objectively
meaningful way to give a magnetic field-line an identity over time is by tagging
it with a certain plasma fluid element. As suggested by Axford
\cite{Axford:1984}, we understand the crucial feature of magnetic reconnection
to be the ``disconnection'' of fluid elements that start on the same field line.
The right panel of Figure \ref{split} below uses data from the JHU MHD
turbulence database archive to illustrate how an initial magnetic field line
changes its connections to plasma fluid elements over time. The figure shows
shows an initial magnetic field line, in black, decorated with eleven plasma
fluid elements, indicated by various colors. The plasma elements are then
evolved with the fluid velocity for about one large-eddy turnover time ($t=2.00$
in units of the simulation). The magnetic field-lines threading these later
plasma elements are drastically different. Indeed, the plasma has ``drifted'' to
distinct lines separated by distances of order the magnetic integral length
(0.35 in the units of the simulation). This drift occurs even though the
conductivity of the simulation is high and the Ohmic electric fields are tiny,
because their small direct effects are greatly magnified by the turbulence.

The violation of flux freezing means that the astrophysical theories based on
the concept of flux freezing must be revised. In particular, the standard star
formation theory assumes that flux freezing is being violated in the partially
ionized gas only due to the relative drift of neutrals and ions. As we discussed
in sections 4(a) and 4(b) in the partially ionized gas the important magnetic
flux violation arises from magnetic diffusion induced by turbulence. This
process that was termed ``reconnection diffusion'' was identified and described
in \cite{Lazarian:2005} (see also \cite{LazarianVishniac:2009}) and successfully
tested in the subsequent publications for the case of molecular clouds and
protostellar disks, e.g. \cite{SantosLima_etal:2010, SantosLima_etal:2012,
SantosLima_etal:2013, deGouveiaDalPino_etal:2012, Leao_etal:2013}. A
comprehensive review dealing with reconnection diffusion is presented in
\cite{Lazarian:2014}.

The left panel illustrates magnetic the evolution of magnetic flux during the
process of reconnection diffusion of magnetic flux out of the circumstellar
accretion disk. The magnetic field lines are smoothed in the picture to
illustrate the evolution of the mean magnetic field.

The theory of transporting matter in turbulent magnetized medium is discussed at
length in \cite{Lazarian:2011} and \cite{Lazarian:2014} and we refer our reader
to these publications. The process was termed ``reconnection diffusion'' to
stress the importance of reconnection in the the diffusive transport.

\begin{figure}
\centering
\includegraphics[width=0.95\textwidth]{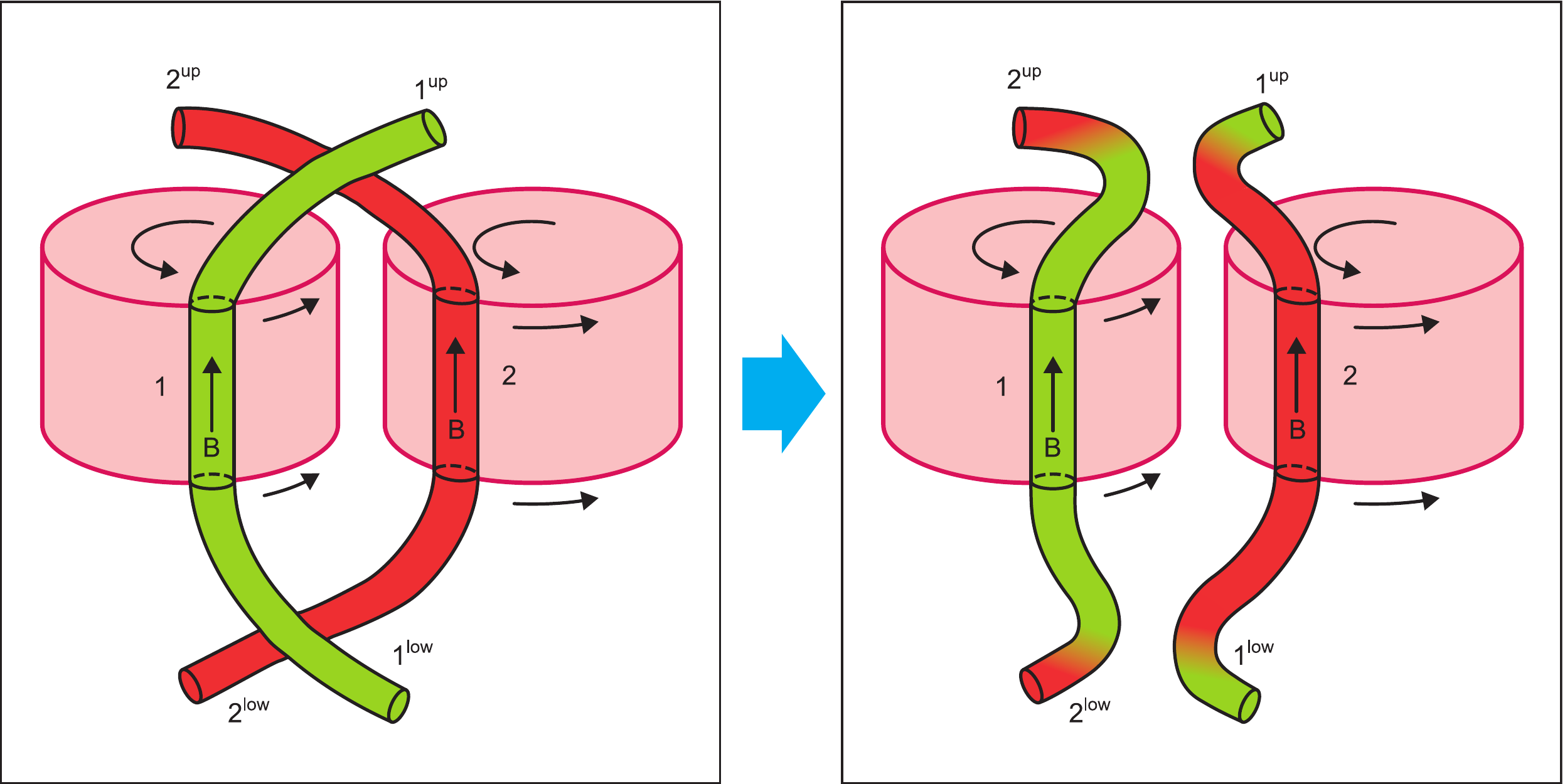}
\includegraphics[width=0.75\textwidth]{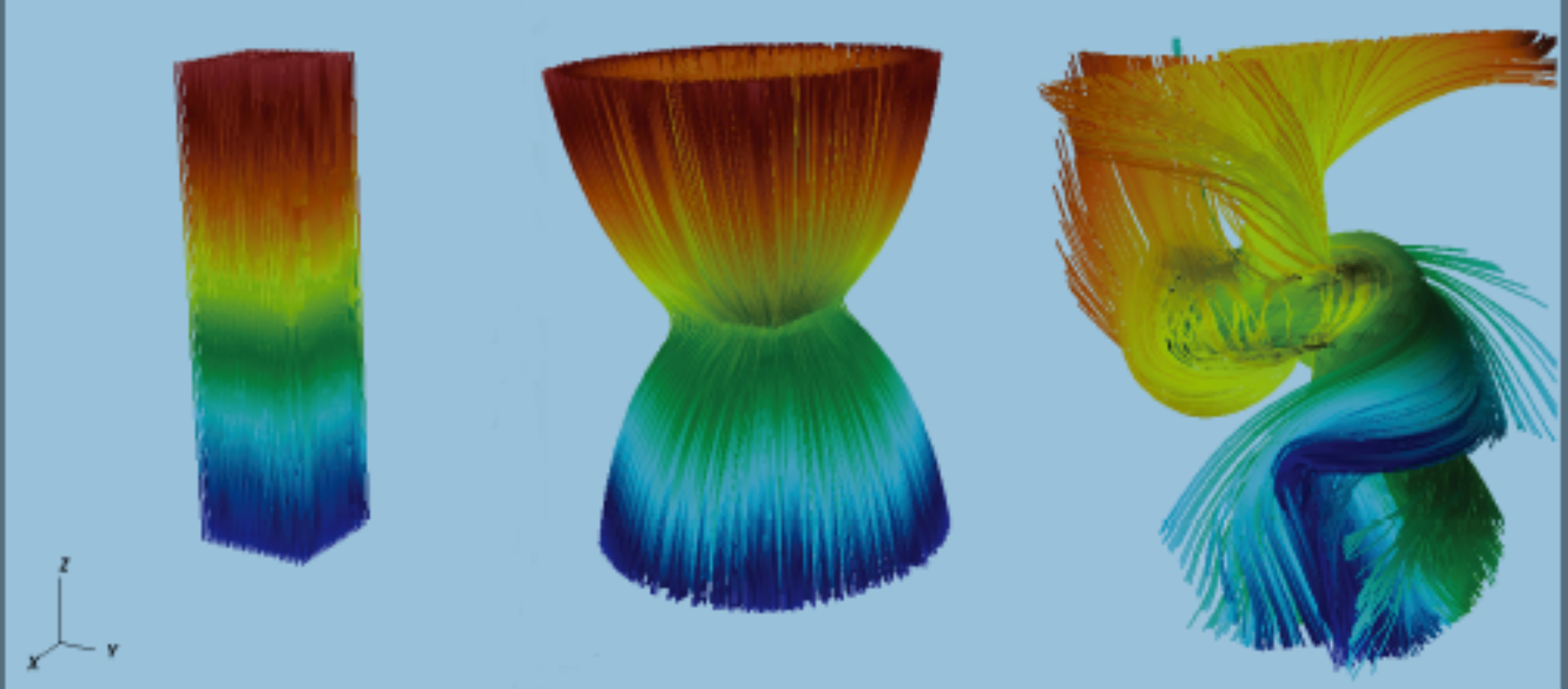}
\caption{Upper panel: {\bf Reconnection diffusion}: exchange of flux with entrained matter.
Illustration of the mixing of matter and magnetic fields due to reconnection as
two flux tubes of different eddies interact. Only one scale of turbulent motions
is shown. In real turbulent cascade such interactions proceed at every scale of
turbulent motions. From \cite{Lazarian:2011}. Lover panel: {\bf Reconnection diffusion in
in an accretion disk.} Illustration of the process using smoothen lines. From Casanova et al. (2015).
\label{mix}}
\end{figure}

The peculiarity of reconnection diffusion is that it requires nearly parallel
magnetic field lines to reconnect, while the textbook description of
reconnection is usually associated with anti-parallel description of magnetic
field lines.  One should understand that the configuration shown in
Figure~\ref{recon} is just a cross section of the magnetic fluxes
depicting the anti-parallel components of magnetic field.  Generically, in 3D
reconnection configurations the sheared component of magnetic field is present.
The process of reconnection diffusion is closely connected with the reconnection
between adjacent Alfv\'{e}nic eddies (see upper panel of Figure~\ref{mix}).  As a result,
adjacent flux tubes exchange their segments with entrained plasmas and flux
tubes of different eddies get connected.  This process involves eddies of all
the sizes along the cascade and ensures fast diffusion which has similarities
with turbulent diffusion in ordinary hydrodynamic flows. The lower panel

The lower panel of Figure \ref{mix} illustrates magnetic the evolution of
magnetic flux during the process of reconnection diffusion of magnetic flux out
of the circumstellar accretion disk. The magnetic field lines are smoothed in
the picture to illustrate the evolution of the mean magnetic field.

Reconnection diffusion should not necessarily be understood as a concept that
makes the earlier theories of star formation invalid. In fact, turbulence in
dark cores giving birth to stars may be reduced and this may make the
traditional ambipolar diffusion, i.e. the drift of neutrals in relation to ions,
important. However, reconnection diffusion must be a part of star formation
paradigm. In fact, it can successfully explain many pieces of observational data
that are completely puzzling within the ambipolar diffusion paradigm. This
includes, for instance, the famous ``magnetic breaking catastrophe'' for accretion
disks which is the inability of removing magnetic flux from accretion disks fast
enough to enable forming of such disks. Similarly, the poor correlation of
density and magnetic field in interstellar media is also impossible to explain
on the basis of ambiplar diffusion. A comprehensive review dealing with
reconnection diffusion is presented in \cite{Lazarian:2014}. Closely
related is the recent development of the ``turbulent general magnetic
reconnection'' (TGMR) theory in \cite{Eyink:2014}. The starting point of
this theory is the understanding that magnetic field-line ``motion'' can be
objectively defined only relative to plasma fluid elements and their magnetic
connectivity. In star formation for example, the magnetic field-lines threading
the protostar will appear to ``slip'' relative to the ambient ISM, whereas the
field-lines embedded in the ISM will likewise appear to ``slip'' through the
collapsing magnetic cloud. Neither picture is more correct than the other and,
indeed, one cannot uniquely define a ``motion'' of the field-lines. However, it
has been shown in \cite{Eyink:2014} that the lines wandering between the
protostar and the surrounding ISM acquire a unique slip-velocity per unit
arc-length of field-line, which is completely independent of which end of the
line is regarded to be the ``foot-point'' tied to the plasma. This so-called
slip-velocity source is given by the expression
\begin{equation}
\Sigma = - (\nabla \times R)_\perp / |B|
\end{equation}
where $R$ is the non-ideal electric field in the generalized Ohm's law $E+u
\times B=R$ for the plasma, and ``$\perp$'' denotes the component perpendicular to B.
It is only by spatially wandering/intersecting a region with non-vanishing
$\Sigma$ that a field-line can evade the ``frozen-in'' condition. Furthermore,
in a turbulent plasma the slip-source $\Sigma$ is enormous, even though the
electric fields R are tiny and the plasma nearly ideal. This approach is
essentially a refinement of the LV99 idea that field-lines must wander into
microscopic ``current sheets'' in order to break the flux-freezing constraint.
\cite{Eyink:2014} applied this theory to explain observed deviations from the
Parker spiral model of the interplanetary magnetic field in our own solar
system, due to ``slippage'' of the field-lines through the turbulent solar wind.

\subsection{Solar flares and gamma ray bursts}

The picture of flares of reconnection described in section 4d
 is broadly supported by current observations and numerical
simulations of solar flares and CME's.  For example, simulations by
\cite{Lynch_etal:2008} of the ``breakout model'' of CME initiation show that an
extremely complex magnetic line structure develops in the ejecta during
and after the initial breakout reconnection phase, even under the severe
numerical resolution constraints of such simulations.  In the very high
Lundquist-number solar environment, this complex field must correspond to a
strongly turbulent state, within which the subsequent ``anti-breakout
reconnection'' and post-CME current sheet occur.  Direct observations of such
current sheets \cite{CiaravellaRaymond:2008, Bemporad:2008} verify the presence of
strong turbulence and greatly thickened reconnection zones, consistent with the
LV99 model.  In the numerical simulations, the ``trigger'' of the initial
breakout reconnection is numerical resistivity and there is no evidence of
turbulence or complex field-structure during the  eruptive flare onset.  This is
very likely to be a result of the limitations on resolution, however, and we
expect that developing turbulence will accelerate reconnection in this phase of
the flare as well.

While the details of the physical processes discussed above can be altered, it
is clear that LV99 reconnection induces bursts in highly magnetized plasmas.
This can be applicable not only to the solar environment but also to more exotic
environments, e.g. to gamma ray bursts.  The model of gamma ray bursts based on
LV99 reconnection was suggested in \cite{Lazarian_etal:2003}.  It was elaborated
and compared with observations in \cite{ZhangYan:2011}.  Currently, the latter
model is considered promising and it attracts a lot of attention of researchers.
 Flares of reconnection that we described above can also be important for
compact sources, like pulsars and black holes in microquasars and AGNs
\cite{deGouveiaDalPinoLazarian:2005}. We would like to note that LV99
reconnection is getting more applications related to emission of astrophysical
objects. For instance, recently it has been discussed to explain the radio and
gamma ray emission  arising  through accretion on black holes
\cite{Singh_etal:2014} as well as for describing the radiation of microquasars
\cite{Khiali_etal:2014}.

\subsection{Turbulent reconnection and particle acceleration}

\begin{figure}[!t]
\centering
\raisebox{-0.5\height}{\includegraphics[width=0.49\textwidth]{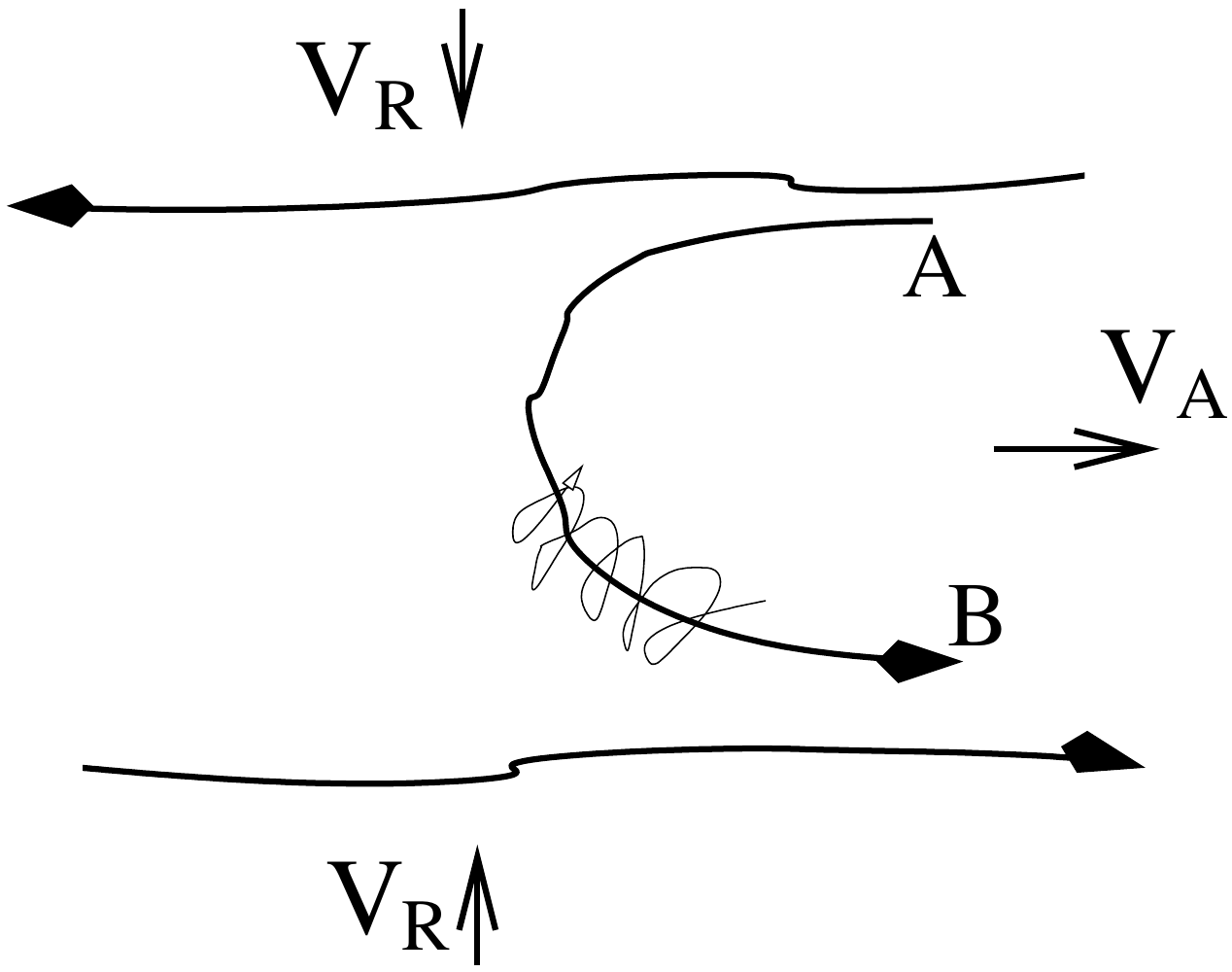}}
\raisebox{-0.5\height}{\includegraphics[width=0.49\textwidth]{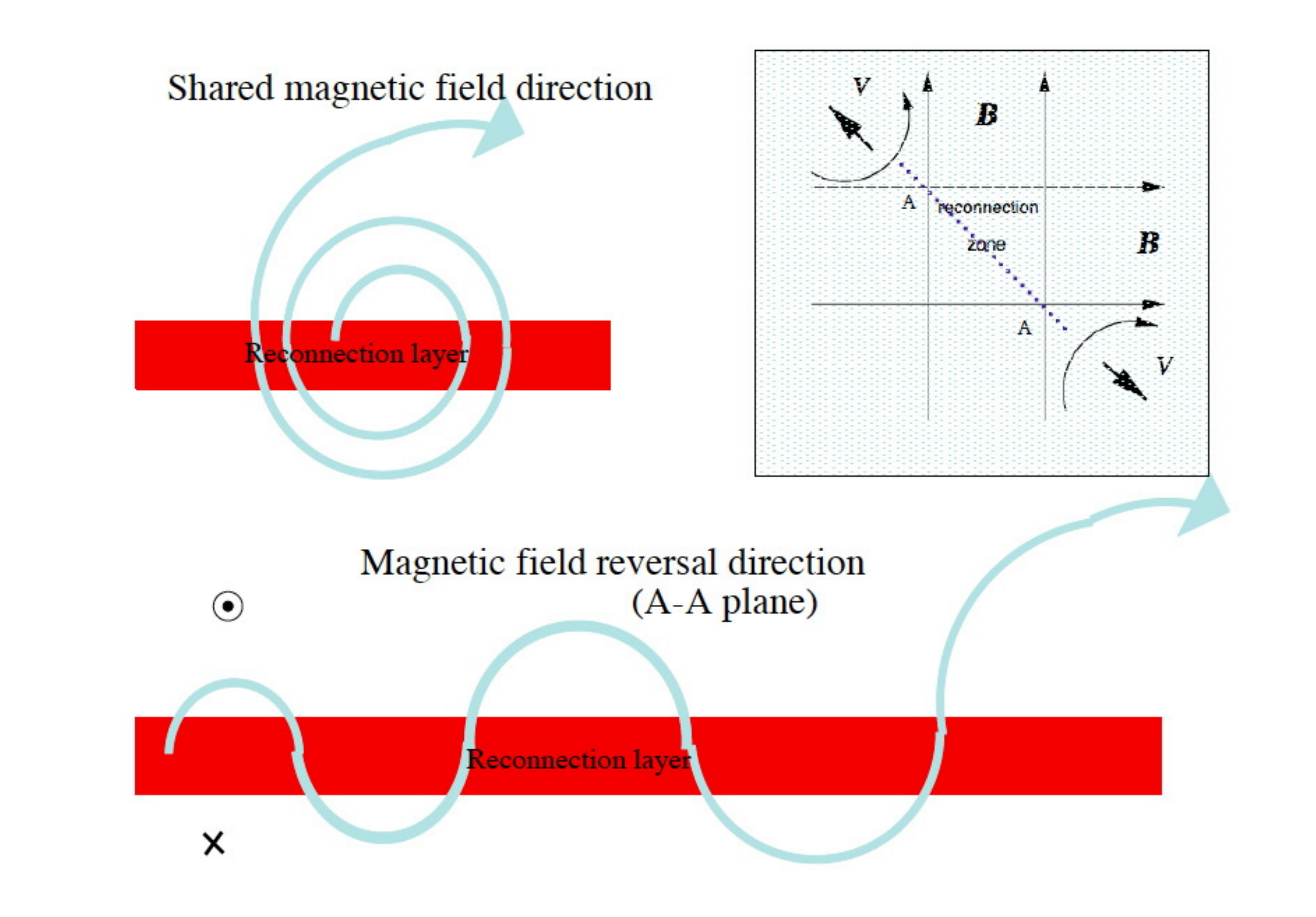}}
\caption{ {\it Left Panel:} Cosmic rays spiral about a reconnected magnetic
field line and bounce back at points A and B. The reconnected regions move
towards each other with the reconnection velocity $V_R$.  From \cite{Lazarian:2005}.
{\it Right Panel:} Particles with a large Larmor radius gyrate about the
magnetic field shared by two reconnecting fluxes (the latter is frequently
referred to as ``guide field''. As the particle interacts with converging
magnetized flow corresponding to the reconnecting components of magnetic field,
the particle gets energy gain during every gyration. From \cite{Lazarian_etal:2012}.
\label{acceler}}
\end{figure}

Turbulent reconnection provides the way of the first order Fermi acceleration as
it is illustrated in Figure \ref{acceler}.  The efficiency of the process is
ensured by LV99 model being the volume-filling reconnection\footnote{We would
like to stress that Figure~\ref{recon} exemplifies only the first moment of
reconnection when the fluxes are just brought together. As the reconnection
develops the volume of thickness $\Delta$ becomes filled with the reconnected 3D
flux ropes moving in the opposite directions.}.

The left panel of Figure \ref{acceler} illustrates a situation when the particle
anisotropy which arises from particles preferentially accelerated in direction
parallel to magnetic field.  Similarly, \cite{Lazarian_etal:2012} showed that the
first order Fermi acceleration can also happen in terms of the perpendicular to
the magnetic field component of particle momentum.  This is illustrated in the
right panel of Figure~\ref{acceler}.  There the particle with a large Larmour
radius is bouncing back and forth between converging mirrors of reconnecting
magnetic field systematically getting an increase of the perpendicular component
of its momentum.  Both processes take place in reconnection layers.

  Disregarding the backreaction one can get the spectrum of accelerated cosmic
rays \cite{deGouveiaDalPinoLazarian:2005, Lazarian:2005}:
\begin{equation}
N(E)dE=const_1 E^{-5/2}dE,
\label{-5/2}
\end{equation}
This result is the result of acceleration in the absence of compression (see
\cite{Drury:2012}). The first order acceleration of particles entrained on the
contracting magnetic loop can be understood from the Liouville theorem.  In the
process of the magnetic tubes contraction a regular increase of the particle's
energies is expected.  The requirement for the process to proceed efficiently is
to keep the accelerated particles within the contracting magnetic loop.  This
introduces limitations on the particle diffusivity perpendicular to the magnetic
field direction.  The process in Figure~\ref{acceler} (left panel) was discussed
in \cite{Drake_etal:2006} in relation to the acceleration of particles in
collisionless reconnection.  There by accounting for the backreaction of
particles the authors obtained a more shallow spectrum.

\begin{figure}
 \centering
 \includegraphics[width=0.9\textwidth]{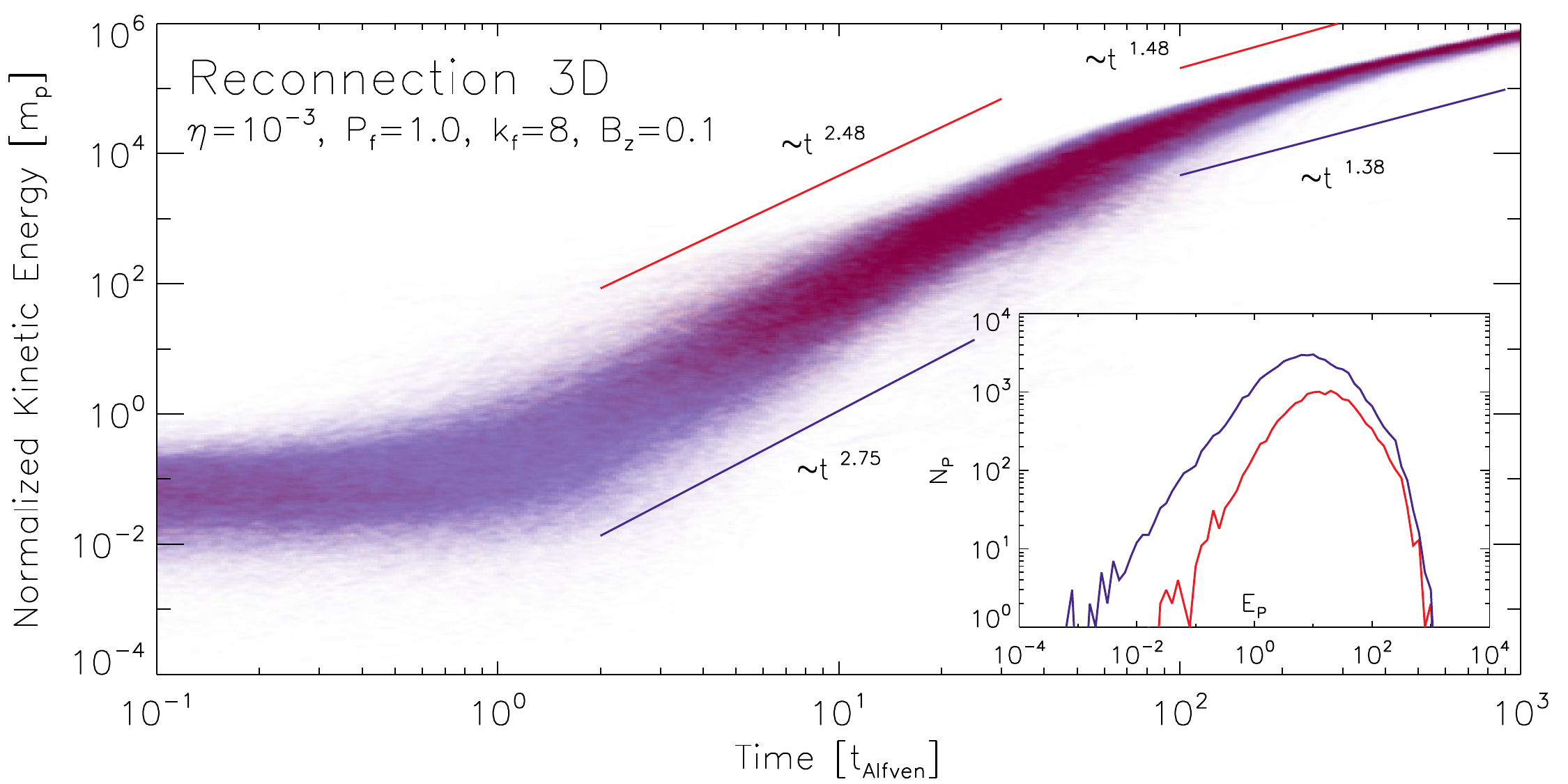}
 \caption{Particle kinetic energy distributions for 10,000 protons injected in
the fast magnetic reconnection domain.  The colors indicate which velocity
component is accelerated (red or blue for parallel or perpendicular,
respectively).  The energy is normalized by the rest proton mass. Subplot shows
the particle energy distributions at $t=5.0$.  Models with $B_{0z} = 0.1$,
$\eta=10^{-3}$, and the resolution 256x512x256 is shown. \label{fig:accel2}}
\end{figure}

  Testing of particle acceleration in turbulent reconnection was performed in
\cite{Kowal_etal:2012} and its results are presented in Figure~\ref{fig:accel2}. The
Figure~\ref{fig:accel2} shows the evolution of the kinetic energy of the
particles.  After injection, a large fraction of test particles accelerates and
the particle energy growth occurs (see also the energy spectrum at $t=5$ in the
detail at the bottom right).  This is explained by a combination of two effects:
the presence of a large number of converging small scale current sheets and the
broadening of the acceleration region due to the turbulence. The acceleration
process is clearly a first order Fermi process, and involves larger number of
particles, since the size of the acceleration zone and the number of scatterers
naturally increases by the presence of turbulence.  Moreover, the reconnection
speed, which in this case is independent of resistivity
\cite{LazarianVishniac:1999, Kowal_etal:2009} and determines the velocity at which the
current sheets scatter particles, has been naturally increased as well (i.e.
$V_{rec}\sim V_A$).  During this stage the acceleration rate is in the range
$2.48-2.75$.

The process of acceleration via turbulent reconnection is expected to be
widespread.  In particular, it has been discussed in \cite{LazarianOpher:2009}
as a cause of the anomalous cosmic rays observed by Voyagers and in
\cite{LazarianDesiati:2010} as a source of the observed cosmic ray anisotropies.
 We expect turbulent reconnection to accelerate energetic particles in
relativistic environments, like those related to accretion disks and
relativistic jets as well as in gamma ray bursts (see
\cite{LyutikovLazarian:2013} for a review).  The latter process discuss first in
\cite{Lazarian_etal:2003}, has been given strong observational support in
\cite{ZhangYan:2011}.

\section{Connection with other ideas of fast reconnection}

As we mentioned in the Introduction the reconnection research is a vast
vigorously developing field and is not limited to reconnection in MHD
approximation in turbulent fluids that we deal in this review. A lot of
experimental research is done for the Earth magnetosphere and laboratory
plasmas, where the MHD description is not valid. Some plasmas may not be
turbulent either. Below we briefly outline the connection of our turbulent model
with other directions of research.

\subsection{Plasmoid/tearing mode reconnection}

Plasmoid/tearing mode reconnection is currently a vibrant direction of
reconnection research (see \cite{Loureiro_etal:2007, Uzdensky_etal:2010}). The
work in this direction shows that SP reconnection is unstable for sufficiently
large Ludquist numbers. What should be kept in mind that in 3D the thicker
outflows induced by plasmoid/tearing reconnection inevitably induce turbulence.
This corresponds to both PIC and MHD simulations that we discussed earlier. This
also inevitable on the theoretical grounds. Indeed, from the mass conservation
constraint requirement in order to have fast reconnection one has to increase
the outflow region thickness in proportion to $L_x$, which means the
proportionality to the Lundquist number $S$. The Reynolds number $Re$ of the
outflow is $\Delta V_A /\nu$, where $\nu$ is viscosity, grows also as $S$. The
outflow gets turbulent for sufficiently large $Re$. It is natural to assume that
once the shearing rate introduced by eddies is larger than the rate of the
tearing instability growth, the instability should get suppressed.

If one assumes that tearing is the necessary requirement for fast reconnection
this entails the conclusion that tearing should proceed at the critically damped
rate, which implies that the $Re$ number and therefore $\Delta$ should not
increase. This entails, however, the decrease of reconnection rate driven by
tearing in proportion $L_x\sim S$ as a result of mass conservation. As a result,
the reconnection should stop being fast. Fortunately, we know that turbulence
itself provides fast reconnection irrespectively whether tearing is involved or
not. Thus one may conclude that the tearing reconnection, similar to the SP
reconnection, should be applicable to a limited range of $S$ for realistic
magnetized plasmas with low viscosity in the perpendicular to magnetic field
direction.

Tearing may be important for initiating turbulence and transiting from the
laminar initial state. To what extend tearing is required is not clear from the
3D simulations that we discussed above. Those suggest the importance of
Kelvin-Helmholdz instability, but whether tearing plays any role must still be
explored. However, if reconnection was excited by tearing/plasmoid instability
generically we expect a transition to the regime of turbulent reconnection.

Another limitation to the applicability of tearing reconnection arises from its
speed. The reported rates do not exceed a small fraction of Alfven speed.
However, as we discuss later magnetic turbulence requires reconnection speeds
which are substantially larger than that value for the theory of MHD turbulence
to be self-consistent (LV99, ELV11). In this situation, the dominance of
turbulent reconnection seems inevitable.

\subsection{Reconnection in 2D in presence of turbulence}

\cite{MatthaeusLamkin:1985, MatthaeusLamkin:1986} explored numerically turbulent
reconnection in 2D.  As a theoretical motivation the authors emphasized
analogies between the magnetic reconnection layer at high Lundquist numbers and
homogeneous MHD turbulence. They also pointed out various turbulence mechanisms
that would enhance reconnection rates, including multiple X-points as
reconnection sites, compressibility effects, motional EMF of magnetic bubbles
advecting out of the reconnection zone. However, the authors did not understand
the importance of ``spontaneous stochasticity'' of field lines and of Lagrangian
trajectories and they did not arrive at an analytical prediction for the
reconnection speed. Although an enhancement of the reconnection rate was
reported in their numerical study, but the setup precluded the calculation of a
long-term average reconnection rate.

The relation of this study with LV99 is not clear, as the nature of turbulence
in 2D is different. In particular, shear-Alfven waves that play the dominant
role in 3D MHD turbulence according to GS95 are entirely lacking in 2D, where
only pseudo-Alfven wave modes exist. We believe that the question whether
turbulence is fast has not been resolved yet if we judge from the available
publications. For instance,  in a more recent study along the lines of the
approach in \cite{MatthaeusLamkin:1985}, i.e. in \cite{Watson_etal:2007}, the
effects of small-scale turbulence on 2D reconnection were studied and no
significant effects of turbulence on reconnection was reported.
\cite{Servidio_etal:2010} have more recently made a study of Ohmic electric
fields at X-points in homogeneous, decaying 2D MHD turbulence. However, they
studied a case of small-scale magnetic reconnection and their results are not
directly relevant to the issue of reconnection of large-scale flux tubes that we
deal in this review.

Instead of studying bulk reconnection in 2D turbulence as the aforementioned
studies did, \cite{Loureiro_etal:2009} and \cite{KulpaDybel_etal:2010} studied
large scale reconnection\footnote{The enhancement of 2D large scale reconnection
was reported starting from 2007 at a few conferences by the authors of
\cite{Kowal_etal:2009}, but for them the 2D study was a testing ground for the
realistic 3D simulations to test LV99. Thus these results were never
published.}, which is advantageous if the determination of the actual
reconnection rates is sought. The two groups reached different conclusions On
the one hand, \cite{Loureiro_etal:2009} had a better resolution but used
periodic boundary conditions, which strongly constrain the ability to do
averaging of the reconnection rate and the attainment of the steady state for
reconnection. They inferred from their data that the 2D turbulent reconnection
rate may be independent of resistivity. On the other hand,
\cite{KulpaDybel_etal:2010} used smaller data cubes but longer averaging, which
is enabled by their outflow boundary conditions. They concluded that the
reconnection does depend on resistivity and therefore is slow.

In view of the difference of MHD turbulence in 2D and 3D we do not view the
reconnection studies in 2D turbulence as directly relevant in any astrophysical
settings. Even if eventually 2D reconnection is proven to be fast, the
reconnection rate is expected to have different dependences on turbulent power.

\subsection{Turbulent reconnection models based on mean field approach}

Guo et al. \cite{Guo_etal:2012} modified and extended ideas originally proposed
in \cite{KimDiamond:2001} and suggested their model of fast turbulent
reconnection. Both papers use mean field approach, but unlike the study in
\cite{KimDiamond:2001} which concluded that turbulence cannot accelerate
reconnection the more recent study obtains expressions for fast reconnection.
These expressions are different from those in LV99 and seem to grossly
contradict to the numerical testing of turbulent reconnection in
\cite{Kowal_etal:2009}. Another model of turbulent reconnection based on the
mean field approach is presented in \cite{HigashimoriHoshino:2012} and it also
suffers with the problems of using the mean field approach for reconnection that
we describe below.

The mean field approach invoked in the aforementioned studies is plagued by poor
foundations and conceptual inconsistencies, however \cite{Eyink_etal:2011}.  In
such an approach effects of turbulence are described using parameters such as
anisotropic turbulent magnetic diffusivity experienced by the fields once
averaged over ensembles. The problem is that it is the lines of the full
magnetic field that must be rapidly reconnected, not just the lines of the mean
field.  ELV11 stress that the former implies the latter, but not conversely. No
mean-field approach can claim to have explained the observed rapid pace of
magnetic reconnection unless it is shown that the reconnection rates obtained in
the theory are strictly independent of the length and timescales of the
averaging. Naturally, it is impossible to get reliable results applying mean
field approach to reconnection (see more discussion in ELV11).

Other attempts to get fast magnetic reconnection from turbulence are related to
the so-called hyper-resistivity concept \cite{Strauss:1986,
BhattacharjeeHameiri:1986, HameiriBhattacharjee:1987, DiamondMalkov:2003},
which is another attempt to derive fast reconnection from turbulence within the
context of mean-field resistive MHD. Apart from the generic problems of using
the mean field approach, we would like to point out that the derivation of the
hyper-resistivity is questionable from a different point of view. The form of
the parallel electric field is derived from magnetic helicity conservation.
Integrating by parts one obtains a term which looks like an effective
resistivity proportional to the magnetic helicity current. There are several
assumptions implicit in this derivation, however. Fundamental to the
hyper-resistive approach is the assumption that the magnetic helicity of mean
fields and of small scale, statistically stationary turbulent fields are
separately conserved, up to tiny resistivity effects. However, this ignores
magnetic helicity fluxes through open boundaries, essential for stationary
reconnection, that vitiate the conservation constraint.

As we discuss further, a common misunderstanding is that ``resistivity arising
from turbulence'' is a real plasma non-ideality ``created'' by the turbulence.
However, such apparent non-ideality is strongly dependent on the length and
timescales of the averaging. It appears only as a consequence of observing the
plasma dynamics at a low resolution, so that the coarse-grained velocity and
magnetic field that are observed will no longer satisfy the microscopic
equations of motion. This coarse-graining or averaging is a purely passive
operation which doesn't change the actual plasma dynamics but only corresponds
to ``taking off one's spectacles''. It is clear that one cannot  create true,
physical non-ideal electric fields by removing one's eyeglasses! Such apparent
non-ideality in a turbulent plasma observed at length-scales in the
inertial-range or larger is a valid representation of the effects of turbulent
eddies at smaller scales. However, such apparent non-ideality is not accurately
represented by an effective ``resistivity'', a representation which in the fluid
turbulence literature has been labelled the ``gradient-transport fallacy''
\cite{TennekesLumley:1972}.  It is also clear that no mean-field or
coarse-graining approach can claim to have explained the observed rapid pace of
magnetic reconnection unless it is shown that the reconnection rates obtained in
the theory are strictly independent of the length and timescales of the
averaging \cite{EyinkAluie:2006, Eyink:2014}.

More detailed discussion of the conceptual problems of the hyper-resistivity
concept and mean field approach to magnetic reconnection is presented in
\cite{Lazarian_etal:2004} and ELV11.

\subsection{Indirect evidence for turbulent reconnection}

A study of tearing instability of current sheets in the presence of background
2D turbulence that observed the formation of large-scale islands was performed
in \cite{Politano_etal:1989}.  While one can argue that observed long-lived
islands are the artifact of adopted 2D geometry, the authors present evidence
for {\it fast energy dissipation} in 2D MHD turbulence and show that this result
does not change as they change the resolution.  More recently
\cite{MininniPouquet:2009} provided evidence for {\it fast dissipation} also in
3D MHD turbulence.  This phenomenon is consistent with the idea of fast
reconnection, but cannot be treated as a direct evidence of the process. Indeed,
fast dissipation and fast magnetic reconnection are rather different physical
processes, dealing with decrease of energy on the one hand and decrease of
magnetic flux on the other.

Works by Galsgaard and Nordlund, e.g. \cite{GalsgaardNordlund:1997}, could also
be interpreted as an indirect support for fast reconnection.  The authors showed
that in their simulations they could not produce highly twisted magnetic fields.
 One possible interpretation of this result could be the fast relaxation of
magnetic field via reconnection\footnote{In this case, these observations could
be related to the numerical finding of \cite{LapentaBettarini:2011} which shows
that reconnecting magnetic configurations spontaneously get chaotic and
dissipate, which, as discussed in \cite{LapentaLazarian:2012}, may be related to
the LV99 model.}  However, in view of many uncertainties of the numerical
studies, this relation is unclear. With low resolution involved in the
simulations the Reynolds numbers could not allow a turbulent inertial-range.

\section{Concluding Remarks}

\subsection{Turbulent reconnection and ``turbulent resistivity''}

As we discussed in the review, the violation of flux freezing and diffusivity of
magnetic field that contradicts to the Alfven theorem follows from the LV99
model of fast reconnection. This, however, is sometimes, misunderstood as our
using some sort of ``turbulent resistivity''. As we mentioned in the review, this
confusion is common for many papers. Therefore we discuss this issue here in
more detail. It is possible to show that ``turbulent resistivity'' description has
fatal problems of inaccuracy and unreliability, due to its poor physical
foundations for turbulent flow. It is true that coarse-graining the MHD
equations by eliminating modes at scales smaller than some length $l$ will
introduce a ``turbulent electric field'', i.e. an effective field acting on the
large scales induced by motions of magnetized eddies at smaller scales. However,
it is well-known in the fluid dynamics community that the resulting turbulent
transport is not ``down-gradient'' and not well-represented by an enhanced
diffusivity. The physical reason is that turbulence lacks the separation in
scales to justify a simple ``eddy-resistivity'' description. As a consequence,
energy is often not absorbed by the smaller eddies, but supplied by them, a
phenomenon called ``backscatter''. In magnetic reconnection, the turbulent
electric field often creates magnetic flux rather than destroys it.

If we know the reconnection rate, e.g. from LV99, then an eddy-resistivity can
always be tuned by hand to achieve that rate. But this is engineering, not
science. While the tuned reconnection rate will be correct by construction,
other predictions will be wrong. The required large eddy-resistivity will smooth
out all turbulence magnetic structure below the coarse-graining scale $l$. In
reality, the turbulence will produce strong small-scale inhomogeneities, such as
current sheets, from the scale $l$ down to the micro-scale. In addition,
field-lines in the flow smoothed by eddy-resistivity will not show the
explosive, super-diffusive Richardson-type separation at scales below $l$. These
are just example of effects that will be lost if the wrong concept of ``eddy
resistivity'' is adopted. Note, that the aforementioned are important for
understanding particle transport/scattering/acceleration in the turbulent
reconnection zone. Continuing with the list, we can point out that in the case
of relativistic reconnection, turbulent resistivities will introduce acausal,
faster than light propagation effects. Nevertheless, the worst feature of the
crude ``eddy-resistivity'' parameterization is its unreliability: because it has
no sound scientific basis whatsoever, it cannot be applied with any confidence
to astrophysical problems. Therefore it is pointless to talk about ``turbulent
resistivity'' for the problems that we discussed in the review, e.g. solar
flares, star formation, gamma ray bursts.

Equivalently, the stochastic flux freezing \cite{Eyink_etal:2011} closely
related to the fast turbulent reconnection concept is definitely not equivalent
to the dissipation of magnetic field by resistivity . While the parametrization
of some particular effects of turbulent fluid may be achieved in models with
different physics, e.g. of fluids with enormously enhanced resistivity, the
difference in physics will inevitably result in other effects being wrongly
represented by this effect. For instance, turbulence with fluid having
resistivity corresponding to the value of ``turbulent resistivity'' must have
magnetic field and fluid decoupled on most of its inertia range turbulent scale,
i.e. the turbulence should not be affected by magnetic field in gross
contradiction with theory, observations and numerical simulations. Magnetic
helicity conservation which is essential for astrophysical dynamo should also be
grossly violated\footnote{This is a serious mistake of a number of numerical
simulations of galactic dynamos where to ``simulate'' effects of turbulent
diffusion the Ohmic resistivity $\nu$ of the order of $V_L L$ is used. Surely
these simulations do not represent the actual fast astrophysical dynamo, but
only a slow one. }.

The approach advocated by us in discussing turbulent reconnection is quite
different. It is not based on coarse-graining. The spontaneous stochasticity of
magnetic field-lines and of Lagrangian trajectories (plasma fluid element
histories) is a real, verified physical phenomenon in turbulent fluids. Whereas
``eddy-resistivity'' ideas predict that magnetic flux is destroyed by turbulence,
our work shows that turbulent spontaneous stochasticity transforms magnetic-flux
conservation into a stochastic conservation law. Because spontaneously
stochastic world-lines in relativistic turbulence must remain within the
light-cone, no acausal effects such as produced by ``eddy-resistivity'' will be
predicted. Our approach is based on fundamental scientific progress in the
understanding of turbulence, not on engineering parameterizations.

\subsection{Goldreich-Sridhar turbulence and turbulent reconnection}

GS95 turbulence is a theory accepted by a substantial part of the astrophysical
community (see \cite{BrandenburgLazarian:2013, BeresnyakLazarian:2015} for
reviews). Born outside the mainstream community of turbulence experts, it was
initially mostly ignored\footnote{The enthusiasm of accepting alternative
theories that, e.g. provide more traditional for the MHD community Kraichnan
index of $-3/2$ \cite{Iroshnikov:1964, Kraichnan:1965}, may also be partially
explained by this fact.} but then was accepted under the pressure of numerical
results. As we mentioned, the debates are not settled about the validity of
possible modifications of the model. In parallel, some part of the community is
still using the so-called 2D plus slab model of turbulence (see
\cite{Matthaeus_etal:1990}) in spite of the fact that it has no support via
numerical simulations with isotropic driving. We consider the latter as some
parametrization of the actual heliospheric turbulence over a limited range. This
parametrization is not physically or numerically motivated and therefore is not
considered within the reconnection model.

The LV99 and our subsequent studies mentioned in the review employed GS95 model.
However, we would like to stress again that none of our principal results on
fast turbulent reconnection and the physics of turbulent reconnection will be
changed if instead of GS95 any other existing model of strong MHD turbulence is
used, provided that this model satisfies the constraints that are given by the
existing numerical simulations. The corresponding expressions for reconnection
rates obtained a wide variety of turbulence indexes are provided in LV99. At the
same time, we discussed in the review that fast LV99 reconnection makes GS95
model self-consistent.

\subsection{2D and 3D reconnection}

Numerical simulations are very demanding in 3D and therefore as therefore the
numerical research attempts initially to attack the problem of reduced
dimensions. In terms of reconnection attempts to attack the problem with 2D
simulations are widely spread. While 2D simulations can get insight to some
processes, the relation between 2D and 3D reconnection is far from trivial. For
instance, from general theoretical positions the importance of 3D for
reconnection was advocated by \cite{Boozer:2012, Boozer:2013}.

In general, a radical change of physics related with the use of the 2D instead
of actual 3D is very common for complex physical systems and the problem of
obtaining misleading results extrapolating those obtained in 2D for the actual
3D systems goes beyond turbulence. Every time when the 2D physics is employed,
it is essential to prove that the results stay the same in 3D. This, for
instance, has not been done in the case of 2D turbulent reconnection
\cite{MatthaeusLamkin:1986} and we believe that this may not be possible due to
fundamental differences of turbulence physics (see ELV11).

Even in the case when 2D reconnection reflects the physics common to 3D, e.g. in
the case of tearing reconnection (e.g. \cite{Loureiro_etal:2007}), we claim that
the development of the instability in 3D and 2D may be very different. The 3D
configurations are more prone to secondary instabilities and to the development
of the fully turbulent state in which the initial instabilities may not be
dominant or even important.

\subsection{Turbulent reconnection and plasma effects}

A substantial part of the reconnection research is based on exploring plasma
physics effects on reconnection (see \cite{Yamada_etal:2010,
UzdenskyRightley:2014} for reviews). LV99 shows that reconnection rates should
not depend on plasma microphysics in the presence of turbulence. This conclusion
was supported by numerical study in \cite{Kowal_etal:2009} where plasma effects
were simulated by introducing anomalous resistivity. The subdominance of
reconnection arising from the Hall effect to that arising due to turbulence was
shown analytically in ELV11. A more rigorous comparison of the turbulence
induced reconnection with that induced by other terms in the Generalized Ohm
Equation was provided in \cite{Eyink:2014} where it was shown that for typical
astrophysical parameters turbulence effects are absolutely dominant.

Nevertheless, the studies that show that magnetic reconnection can be fast in
the absence of turbulence (see \cite{Yamada_etal:2010} for a review and ref.
therein) poses interesting questions on the actual role of turbulence. There
are, for instance, suggestions that tearing of the current sheet may make
collisionless plasma effects applicable to magnetic reconnection on large
astrophysical scales (see \cite{KarimabadiLazarian:2013} for a review). Clearly
in plasma one should consider both small scale plasma turbulence as well as
large scale turbulence that obeys MHD treatment. In addition, the very issue of
plasma collisionality is frequently unclear. Indeed, apart from Coulomb
collisions, ions may be scattered by magnetic inhomogeneities that arise due to
a number of instabilities (e.g. firehose, mirror) in collisionaless plasmas and
this may make plasma effectively collisional (see \cite{Schekochihin_etal:2007,
LazarianBeresnyak:2006}) in agreement with recent simulations
\cite{SantosLima_etal:2013}. In this situation, the MHD description should be
applicable even to plasmas which are formally collisionless.

Our arguments in the review suggest that plasma effects cannot dominate large
scale astrophysical reconnection in the presence of turbulence. However, we
accept that the issue is a subject of interesting debates and more testing is
valuable.

\subsection{Present state of turbulent reconnection theory and outstanding questions}

We would like to emphasize that at present the turbulent reconnection theory
does not amount to the LV99 model only. It is also ELV11 where the LV99
expressions were reproduced using a very different approach that follows from
the recent advances of the Lagrangian description of turbulence. It is also a
very recent paper by \cite{Eyink:2014} where the effects of turbulence were
included within the Generalized Ohm's law and were shown to be in agreement with
results of the two approaches above. The theoretical foundation of turbulent
reconnection got substantial support from the recently developed concept of
``spontaneous stochasticity'' \cite{Eyink_etal:2011}.

The predictions of the turbulent reconnection have been successfully tested both
with direct simulations of turbulent reconnection layer \cite{Kowal_etal:2009,
Kowal_etal:2012} and through the violation of flux conservation that the
turbulent reconnection entailed  \cite{Eyink_etal:2013}. As we discuss in the
review, more promising numerical tests of reconnection, including turbulence
being self-driven are under way.

The turbulent reconnection has shown promise in explaining various astrophysical
problems, as well as in addressing problems facing solar physics and
heliospheric research. Some of these are discussed in this review, while others
are discussed in specialized reviews \cite{BrowningLazarian:2013,
LyutikovLazarian:2013, Lazarian:2014}.

At the same time, a number of questions remains not answered. It is obvious that
LV99 model is a very simplified model. It does not take into account many
effects, e.g. effects of plasma compressibility, turbulence intermittency,
velocity and magnetic field shear. To obtain analytical results it assumes the
turbulence is presented by a single power law and disregards the deviations
arising from multiple scales of energy injection, Ohmic and viscous dissipation
etc. It deals with isothermal MHD description of the process and does not
account for relativistic effects. The role of collisionless plasma effects in
turbulent reconnection is hotly debated.

These limitations of the model are gradually dealt with. For instance, a
modification of our understanding of GS95 cascade was discussed in our review in
relation to describing magnetically dominated perturbations arising from
magnetic reconnection. We also discussed accounting for the partial ionization
of plasmas. Role of plasma effects is also being clarified (see
\cite{Eyink:2014} and ref. therein). Nevertheless, we are at the very beginning
of our studies of turbulent reconnection and its consequences. Therefore we
expect many surprises and discoveries on the way to fully understanding of the
intricate relation of turbulence and reconnection.

\subsection{Turbulence as a converging point for reconnection research}

LV99 model is the one that describes the dynamics of reconnection within
turbulent fluids in MHD regime. By itself, the study of dynamics of magnetic
fields in MHD regime, irrespectively to any plasma physics is a well motivated
direction. However, astrophysical environments are filled with turbulent
plasmas. Thus for astrophysical applications it is important to define the
domain of applicability of turbulence-based versus plasma-based reconnection.
First of all, we should stress that there is no single mode of reconnection.
Astrophysical and laboratory environments present an extensive variety of
conditions for magnetic reconnection.

Numerical experiments show that laminar Sweet-Parker reconnection is feasible in
the regime of low Ludquist numbers, then the laminar picture fails and tearing
gets important. As we further increase the length of the reconnection sheet,
even without external driving, in 3D low viscosity plasma the transition to
turbulence is inevitable. Whether this transfers the reconnection to purely
turbulent reconnection or plasma effects are important for determining the
reconnection rate when the outflow is fully turbulent is a subject of ongoing
debates. In our work we provided arguments in support to the former solution,
i.e. that the transition to the turbulent state when plasma effect do not change
the reconnection rate is most relevant for most astrophysical settings. This
does not exclude that in some particular circumstances, e.g. for the onset of
turbulent reconnection from initially laminar state, for current sheet which
thickness is comparable with ion inertial length, as this is the case of
magnetoshpere\footnote{Plasma turbulence may be still important for such
reconnection, but this type of turbulent reconnection is not described by LV99
model.}, the plasma effects are important. For other opinions and outgoing
debates we refer our reader to \cite{Yamada_etal:2010}. Below we, however, point
out to the tendency that the reconnection research has demonstrated.

Recent years have demonstrated the convergence of turbulent reconnection in LV99
and other directions of reconnection research.  For instance, models of
collisionless reconnection have acquired several features in common with the
LV99 model.  In particular, they have moved to consideration of volume-filling
reconnection (see \cite{Drake_etal:2006}).  While much of the discussion may
still be centered around 2D magnetic islands produced by reconnection, in three
dimensions these islands are expected to evolve into contracting 3D loops or
ropes \cite{Daughton_etal:2008} introducing stochasticity to the reconnection
zone.  Moreover, it is more and more realized that the 3D geometry of
reconnection is essential and that the 2D studies may be misleading.

The departure from the concept of laminar reconnection and the introduction of
magnetic stochasticity is also apparent in a number of recent papers appealing
to the tearing mode instability to drive fast reconnection (see
\cite{Loureiro_etal:2007, Bhattacharjee_etal:2009}). These studies showed that
tearing modes do not require collisionless environments and thus collisionality
is not a necessary ingredient of fast reconnection.  Finally, the development of
turbulence in 3D numerical simulations of reconnection (see section 4c) clearly
testifies that the reconnection induces turbulence even if the initial
reconnection conditions are laminar.

All in all, in the last decade, the models competing with LV99 have undergone a
substantial evolution, from 2D collisionless reconnection based mostly on Hall
effect to 3D reconnection where the collisionless condition is no more required,
Hall effect is not employed, but magnetic stochasticity and turbulence play an
important role in the thick reconnection regions. Nevertheless, we want to
stress that collisionless reconnection may be suitable for the description of
reconnection when the reconnecting flux-structures are comparable with the ion
gyro scale, which is the case of the reconnection studied {\it situ} in the
magnetosphere.  However, this is a special case of magnetic reconnection which
makes its very atypical generic astrophysical settups where reconnection
involves scales many orders of magnetude of the gyroradius involved. Even in
this case we may expect the development of turbulence, but this would not be MHD
turbulence which makes LV99 theory not applicable to it. Conversely, it can be
shown by exact analytical estimates \cite{Eyink:2014} that the direct effect of
the microscopic plasma non-idealities are negligible for reconnection at scales
vastly larger than ion gyroradius.

\section*{Acknowledgment}

A.L. research is supported by the NSF grant AST 1212096, Vilas Associate Award
and also award at the
UFRN (Natal). A.L. thanks Eric Priest for stimulating exchanges. G.K.
acknowledges support from FAPESP (projects no. 2013/04073-2 and 2013/18815-0).


\bibliographystyle{unsrt}
\bibliography{references}

\begin{thebibliography}{100}

\bibitem{Alfven:1943}
H.~{Alfv\'{e}n}.
\newblock {On the Existence of Electromagnetic-Hydrodynamic Waves}.
\newblock {\em Arkiv for Astronomi}, 29:1--7, 1943.

\bibitem{Parker:1979}
E.~N. {Parker}.
\newblock {\em {Cosmical magnetic fields: Their origin and their activity}}.
\newblock 1979.

\bibitem{Parker:1970}
E.~N. {Parker}.
\newblock {The Generation of Magnetic Fields in Astrophysical Bodies. I. The
  Dynamo Equations}.
\newblock {\em \apj}, 162:665, November 1970.

\bibitem{Lovelace:1976}
R.~V.~E. {Lovelace}.
\newblock {Dynamo model of double radio sources}.
\newblock {\em \nat}, 262:649--652, August 1976.

\bibitem{PriestForbes:2002}
E.~R. {Priest} and T.~G. {Forbes}.
\newblock {The magnetic nature of solar flares}.
\newblock {\em \aapr}, 10:313--377, 2002.

\bibitem{Innes_etal:1997}
D.~E. {Innes}, B.~{Inhester}, W.~I. {Axford}, and K.~{Wilhelm}.
\newblock {Bi-directional plasma jets produced by magnetic reconnection on the
  Sun}.
\newblock {\em \nat}, 386:811--813, April 1997.

\bibitem{YokoyamaShibata:1995}
T.~{Yokoyama} and K.~{Shibata}.
\newblock {Magnetic reconnection as the origin of X-ray jets and H{$\alpha$}
  surges on the Sun}.
\newblock {\em \nat}, 375:42--44, May 1995.

\bibitem{Masuda_etal:1994}
S.~{Masuda}, T.~{Kosugi}, H.~{Hara}, S.~{Tsuneta}, and Y.~{Ogawara}.
\newblock {A loop-top hard X-ray source in a compact solar flare as evidence
  for magnetic reconnection}.
\newblock {\em \nat}, 371:495--497, October 1994.

\bibitem{Shay_etal:1998}
M.~A. {Shay}, J.~F. {Drake}, R.~E. {Denton}, and D.~{Biskamp}.
\newblock {Structure of the dissipation region during collisionless magnetic
  reconnection}.
\newblock {\em \jgr}, 103:9165--9176, May 1998.

\bibitem{Drake:2001}
J.~F. {Drake}.
\newblock {Magnetic explosions in space}.
\newblock {\em \nat}, 410:525--526, March 2001.

\bibitem{Drake_etal:2006}
J.~F. {Drake}, M.~{Swisdak}, H.~{Che}, and M.~A. {Shay}.
\newblock {Electron acceleration from contracting magnetic islands during
  reconnection}.
\newblock {\em \nat}, 443:553--556, October 2006.

\bibitem{Daughton_etal:2006}
W.~{Daughton}, J.~{Scudder}, and H.~{Karimabadi}.
\newblock {Fully kinetic simulations of undriven magnetic reconnection with
  open boundary conditions}.
\newblock {\em Physics of Plasmas}, 13(7):072101, July 2006.

\bibitem{UzdenskyKulsrud:2006}
D.~A. {Uzdensky} and R.~M. {Kulsrud}.
\newblock {Physical origin of the quadrupole out-of-plane magnetic field in
  Hall-magnetohydrodynamic reconnection}.
\newblock {\em Physics of Plasmas}, 13(6):062305, June 2006.

\bibitem{Bhattacharjee_etal:2003}
A.~{Bhattacharjee}, Z.~W. {Ma}, and X.~{Wang}.
\newblock {Recent Developments in Collisionless Reconnection Theory:
  Applications to Laboratory and Astrophysical Plasmas}.
\newblock In E.~{Falgarone} and T.~{Passot}, editors, {\em {Turbulence and
  Magnetic Fields in Astrophysics}}, volume 614 of {\em {Lecture Notes in
  Physics, Berlin Springer Verlag}}, pages 351--375, 2003.

\bibitem{ZweibelYamada:2009}
E.~G. {Zweibel} and M.~{Yamada}.
\newblock {Magnetic Reconnection in Astrophysical and Laboratory Plasmas}.
\newblock {\em \araa}, 47:291--332, September 2009.

\bibitem{Yamada_etal:2010}
M.~{Yamada}, R.~{Kulsrud}, and H.~{Ji}.
\newblock {Magnetic reconnection}.
\newblock {\em Reviews of Modern Physics}, 82:603--664, January 2010.

\bibitem{Parker:1993}
E.~N. {Parker}.
\newblock {A solar dynamo surface wave at the interface between convection and
  nonuniform rotation}.
\newblock {\em \apj}, 408:707--719, May 1993.

\bibitem{Ossendrijver:2003}
M.~{Ossendrijver}.
\newblock {The solar dynamo}.
\newblock {\em \aapr}, 11:287--367, 2003.

\bibitem{GoldreichSridhar:1995}
P.~{Goldreich} and S.~{Sridhar}.
\newblock {Toward a theory of interstellar turbulence. 2: Strong alfvenic
  turbulence}.
\newblock {\em \apj}, 438:763--775, January 1995.

\bibitem{Sturrock:1966}
P.~A. {Sturrock}.
\newblock {Model of the High-Energy Phase of Solar Flares}.
\newblock {\em \nat}, 211:695--697, August 1966.

\bibitem{LyutikovLazarian:2013}
M.~{Lyutikov} and A.~{Lazarian}.
\newblock {Topics in Microphysics of Relativistic Plasmas}.
\newblock {\em \ssr}, 178:459--481, October 2013.

\bibitem{ShibataMagara:2011}
K.~{Shibata} and T.~{Magara}.
\newblock {Solar Flares: Magnetohydrodynamic Processes}.
\newblock {\em Living Reviews in Solar Physics}, 8:6, December 2011.

\bibitem{Loureiro_etal:2007}
N.~F. {Loureiro}, A.~A. {Schekochihin}, and S.~C. {Cowley}.
\newblock {Instability of current sheets and formation of plasmoid chains}.
\newblock {\em Physics of Plasmas}, 14(10):100703, October 2007.

\bibitem{Lapenta:2008}
G.~{Lapenta}.
\newblock {Self-Feeding Turbulent Magnetic Reconnection on Macroscopic Scales}.
\newblock {\em Physical Review Letters}, 100(23):235001, June 2008.

\bibitem{Daughton_etal:2009a}
W.~{Daughton}, V.~{Roytershteyn}, B.~J. {Albright}, H.~{Karimabadi}, L.~{Yin},
  and K.~J. {Bowers}.
\newblock {Influence of Coulomb collisions on the structure of reconnection
  layers}.
\newblock {\em Physics of Plasmas}, 16(7):072117, July 2009.

\bibitem{Daughton_etal:2009b}
W.~{Daughton}, V.~{Roytershteyn}, B.~J. {Albright}, H.~{Karimabadi}, L.~{Yin},
  and K.~J. {Bowers}.
\newblock {Transition from collisional to kinetic regimes in large-scale
  reconnection layers}.
\newblock {\em Physical Review Letters}, 103(6):065004, August 2009.

\bibitem{Bhattacharjee_etal:2009}
A.~{Bhattacharjee}, Y.-M. {Huang}, H.~{Yang}, and B.~{Rogers}.
\newblock {Fast reconnection in high-Lundquist-number plasmas due to the
  plasmoid Instability}.
\newblock {\em Physics of Plasmas}, 16(11):112102, November 2009.

\bibitem{Cassak_etal:2009}
P.~A. {Cassak}, M.~A. {Shay}, and J.~F. {Drake}.
\newblock {Scaling of Sweet-Parker reconnection with secondary islands}.
\newblock {\em Physics of Plasmas}, 16(12):120702, December 2009.

\bibitem{KarimabadiLazarian:2013}
H.~{Karimabadi} and A.~{Lazarian}.
\newblock {Magnetic reconnection in the presence of externally driven and
  self-generated turbulence}.
\newblock {\em Physics of Plasmas}, 20(11):112102, November 2013.

\bibitem{LaRosaMoore:1993}
T.~N. {Larosa} and R.~L. {Moore}.
\newblock {A Mechanism for Bulk Energization in the Impulsive Phase of Solar
  Flares: MHD Turbulent Cascade}.
\newblock {\em \apj}, 418:912, December 1993.

\bibitem{LazarianVishniac:1999}
A.~{Lazarian} and E.~T. {Vishniac}.
\newblock {Reconnection in a Weakly Stochastic Field}.
\newblock {\em \apj}, 517:700--718, June 1999.

\bibitem{Eyink_etal:2011}
G.~L. {Eyink}, A.~{Lazarian}, and E.~T. {Vishniac}.
\newblock {Fast Magnetic Reconnection and Spontaneous Stochasticity}.
\newblock {\em \apj}, 743:51, December 2011.

\bibitem{Kowal_etal:2009}
G.~{Kowal}, A.~{Lazarian}, E.~T. {Vishniac}, and K.~{Otmianowska-Mazur}.
\newblock {Numerical Tests of Fast Reconnection in Weakly Stochastic Magnetic
  Fields}.
\newblock {\em \apj}, 700:63--85, July 2009.

\bibitem{Kowal_etal:2012}
G.~{Kowal}, A.~{Lazarian}, E.~T. {Vishniac}, and K.~{Otmianowska-Mazur}.
\newblock {Reconnection studies under different types of turbulence driving}.
\newblock {\em Nonlinear Processes in Geophysics}, 19:297--314, April 2012.

\bibitem{Eyink:2014}
G.~L. {Eyink}.
\newblock {Turbulent General Magnetic Reconnection}.
\newblock {\em ArXiv e-prints}, December 2014.

\bibitem{ZhangYan:2011}
B.~{Zhang} and H.~{Yan}.
\newblock {The Internal-collision-induced Magnetic Reconnection and Turbulence
  (ICMART) Model of Gamma-ray Bursts}.
\newblock {\em \apj}, 726:90, January 2011.

\bibitem{Lazarian:2012}
A.~{Lazarian}.
\newblock {Reconnection of Weakly Stochastic Magnetic Field and Flares of
  Magnetic Reconnection}.
\newblock In {\em {39th COSPAR Scientific Assembly}}, volume~39 of {\em {COSPAR
  Meeting}}, page 1046, July 2012.

\bibitem{PriestForbes:2000}
E.~{Priest} and T.~{Forbes}.
\newblock {\em {Magnetic Reconnection}}.
\newblock June 2000.

\bibitem{Speiser:1970}
T.~W. {Speiser}.
\newblock {Conductivity without collisions or noise}.
\newblock {\em Planetary Space Science}, 18:613--622, April 1970.

\bibitem{JacobsonMoses:1984}
A.~R. {Jacobson} and R.~W. {Moses}.
\newblock {Nonlocal dc electrical conductivity of a Lorentz plasma in a
  stochastic magnetic field}.
\newblock {\em \pra}, 29:3335--3342, June 1984.

\bibitem{BhattacharjeeHameiri:1986}
A.~{Bhattacharjee} and E.~{Hameiri}.
\newblock {Self-consistent dynamolike activity in turbulent plasmas}.
\newblock {\em Physical Review Letters}, 57:206--209, July 1986.

\bibitem{MatthaeusLamkin:1986}
W.~H. {Matthaeus} and S.~L. {Lamkin}.
\newblock {Turbulent magnetic reconnection}.
\newblock {\em Physics of Fluids}, 29:2513--2534, August 1986.

\bibitem{Strauss:1986}
H.~R. {Strauss}.
\newblock {Hyper-resistivity produced by tearing mode turbulence}.
\newblock {\em Physics of Fluids}, 29:3668--3671, November 1986.

\bibitem{Guo_etal:2012}
Z.~B. {Guo}, P.~H. {Diamond}, and X.~G. {Wang}.
\newblock {Magnetic Reconnection, Helicity Dynamics, and Hyper-diffusion}.
\newblock {\em \apj}, 757:173, October 2012.

\bibitem{Armstrong_etal:1995}
J.~W. {Armstrong}, B.~J. {Rickett}, and S.~R. {Spangler}.
\newblock {Electron density power spectrum in the local interstellar medium}.
\newblock {\em \apj}, 443:209--221, April 1995.

\bibitem{ChepurnovLazarian:2010}
A.~{Chepurnov} and A.~{Lazarian}.
\newblock {Extending the Big Power Law in the Sky with Turbulence Spectra from
  Wisconsin H{$\alpha$} Mapper Data}.
\newblock {\em \apj}, 710:853--858, February 2010.

\bibitem{Lazarian:2009}
A.~{Lazarian}.
\newblock {Obtaining Spectra of Turbulent Velocity from Observations}.
\newblock {\em \ssr}, 143:357--385, March 2009.

\bibitem{Leamon_etal:1998}
R.~J. {Leamon}, C.~W. {Smith}, N.~F. {Ness}, W.~H. {Matthaeus}, and H.~K.
  {Wong}.
\newblock {Observational constraints on the dynamics of the interplanetary
  magnetic field dissipation range}.
\newblock {\em \jgr}, 103:4775, March 1998.

\bibitem{Burkhart_etal:2010}
B.~{Burkhart}, S.~{Stanimirovi\'{c}}, A.~{Lazarian}, and G.~{Kowal}.
\newblock {Characterizing Magnetohydrodynamic Turbulence in the Small
  Magellanic Cloud}.
\newblock {\em \apj}, 708:1204--1220, January 2010.

\bibitem{Schekochihin_etal:2009}
A.~A. {Schekochihin}, S.~C. {Cowley}, W.~{Dorland}, G.~W. {Hammett}, G.~G.
  {Howes}, E.~{Quataert}, and T.~{Tatsuno}.
\newblock {Astrophysical Gyrokinetics: Kinetic and Fluid Turbulent Cascades in
  Magnetized Weakly Collisional Plasmas}.
\newblock {\em \apjs}, 182:310--377, May 2009.

\bibitem{LazarianBeresnyak:2006}
A.~{Lazarian} and A.~{Beresnyak}.
\newblock {Cosmic ray scattering in compressible turbulence}.
\newblock {\em \mnras}, 373:1195--1202, December 2006.

\bibitem{BrunettiLazarian:2011}
G.~{Brunetti} and A.~{Lazarian}.
\newblock {Acceleration of primary and secondary particles in galaxy clusters
  by compressible MHD turbulence: from radio haloes to gamma-rays}.
\newblock {\em \mnras}, 410:127--142, January 2011.

\bibitem{Bale_etal:2005}
S.~D. {Bale}, P.~J. {Kellogg}, F.~S. {Mozer}, T.~S. {Horbury}, and H.~{Reme}.
\newblock {Measurement of the Electric Fluctuation Spectrum of
  Magnetohydrodynamic Turbulence}.
\newblock {\em Physical Review Letters}, 94(21):215002, June 2005.

\bibitem{Padoan_etal:2006}
P.~{Padoan}, M.~{Juvela}, A.~{Kritsuk}, and M.~L. {Norman}.
\newblock {The Power Spectrum of Supersonic Turbulence in Perseus}.
\newblock {\em \apjl}, 653:L125--L128, December 2006.

\bibitem{VogtEnsslin:2005}
C.~{Vogt} and T.~A. {En{\ss}lin}.
\newblock {A Bayesian view on Faraday rotation maps Seeing the magnetic power
  spectra in galaxy clusters}.
\newblock {\em \aap}, 434:67--76, April 2005.

\bibitem{NormanFerrara:1996}
C.~A. {Norman} and A.~{Ferrara}.
\newblock {The Turbulent Interstellar Medium: Generalizing to a Scale-dependent
  Phase Continuum}.
\newblock {\em \apj}, 467:280, August 1996.

\bibitem{Ferriere:2001}
K.~M. {Ferri\`{e}re}.
\newblock {The interstellar environment of our galaxy}.
\newblock {\em Reviews of Modern Physics}, 73:1031--1066, October 2001.

\bibitem{Subramanian_etal:2006}
K.~{Subramanian}, A.~{Shukurov}, and N.~E.~L. {Haugen}.
\newblock {Evolving turbulence and magnetic fields in galaxy clusters}.
\newblock {\em \mnras}, 366:1437--1454, March 2006.

\bibitem{EnsslinVogt:2006}
T.~A. {En{\ss}lin} and C.~{Vogt}.
\newblock {Magnetic turbulence in cool cores of galaxy clusters}.
\newblock {\em \aap}, 453:447--458, July 2006.

\bibitem{Chandran:2005}
B.~D.~G. {Chandran}.
\newblock {AGN-driven Convection in Galaxy-Cluster Plasmas}.
\newblock {\em \apj}, 632:809--820, October 2005.

\bibitem{BalbusHawley:1998}
S.~A. {Balbus} and J.~F. {Hawley}.
\newblock {Instability, turbulence, and enhanced transport in accretion disks}.
\newblock {\em Reviews of Modern Physics}, 70:1--53, January 1998.

\bibitem{GalsgaardNordlund:1997}
K.~{Galsgaard} and {\AA}.~{Nordlund}.
\newblock {Heating and activity of the solar corona. 3. Dynamics of a low beta
  plasma with three-dimensional null points}.
\newblock {\em \jgr}, 102:231--248, January 1997.

\bibitem{GerrardHood:2003}
C.~L. {Gerrard} and A.~W. {Hood}.
\newblock {Kink unstable coronal loops: current sheets, current saturation and
  magnetic reconnection}.
\newblock {\em \solphys}, 214:151--169, May 2003.

\bibitem{Kulsrud:1983}
R.~M. {Kulsrud}.
\newblock {MHD description of plasma}.
\newblock In A.~A. {Galeev} and R.~N. {Sudan}, editors, {\em {Basic Plasma
  Physics: Selected Chapters, Handbook of Plasma Physics, Volume 1}}, page~1,
  1983.

\bibitem{Braginskii:1965}
S.~I. {Braginskii}.
\newblock {Transport Processes in a Plasma}.
\newblock {\em Reviews of Plasma Physics}, 1:205, 1965.

\bibitem{Fitzpatrick:2011}
R.~Fitzpatrick.
\newblock {Introduction to Plasma Physics}.
\newblock 2011.
\newblock online lecture notes, URL: {\tt
  http://farside.ph.utexas.edu/teaching/plasma/plasma.html}.

\bibitem{SchekochihinCowley:2006}
A.~A. {Schekochihin} and S.~C. {Cowley}.
\newblock {Turbulence, magnetic fields, and plasma physics in clusters of
  galaxies}.
\newblock {\em Physics of Plasmas}, 13(5):056501, May 2006.

\bibitem{SantosLima_etal:2013}
R.~{Santos-Lima}, E.~M. {de Gouveia Dal Pino}, and A.~{Lazarian}.
\newblock {Disc formation in turbulent cloud cores: is magnetic flux loss
  necessary to stop the magnetic braking catastrophe or not?}
\newblock {\em \mnras}, 429:3371--3378, March 2013.

\bibitem{Schekochihin_etal:2007}
A.~A. {Schekochihin}, S.~C. {Cowley}, and W.~{Dorland}.
\newblock {Interplanetary and interstellar plasma turbulence}.
\newblock {\em Plasma Physics and Controlled Fusion}, 49:195, May 2007.

\bibitem{ChoLazarian:2002}
J.~{Cho} and A.~{Lazarian}.
\newblock {Compressible Sub-Alfv\'{e}nic MHD Turbulence in Low- {$\beta$}
  Plasmas}.
\newblock {\em Physical Review Letters}, 88(24):245001, June 2002.

\bibitem{ChoLazarian:2003}
J.~{Cho} and A.~{Lazarian}.
\newblock {Compressible magnetohydrodynamic turbulence: mode coupling, scaling
  relations, anisotropy, viscosity-damped regime and astrophysical
  implications}.
\newblock {\em \mnras}, 345:325--339, October 2003.

\bibitem{KowalLazarian:2010}
G.~{Kowal} and A.~{Lazarian}.
\newblock {Velocity Field of Compressible Magnetohydrodynamic Turbulence:
  Wavelet Decomposition and Mode Scalings}.
\newblock {\em \apj}, 720:742--756, September 2010.

\bibitem{LithwickGoldreich:2001}
Y.~{Lithwick} and P.~{Goldreich}.
\newblock {Compressible Magnetohydrodynamic Turbulence in Interstellar
  Plasmas}.
\newblock {\em \apj}, 562:279--296, November 2001.

\bibitem{ChoVishniac:2000}
J.~{Cho} and E.~T. {Vishniac}.
\newblock {The Anisotropy of Magnetohydrodynamic Alfv\'{e}nic Turbulence}.
\newblock {\em \apj}, 539:273--282, August 2000.

\bibitem{MaronGoldreich:2001}
J.~{Maron} and P.~{Goldreich}.
\newblock {Simulations of Incompressible Magnetohydrodynamic Turbulence}.
\newblock {\em \apj}, 554:1175--1196, June 2001.

\bibitem{Boldyrev:2002}
S.~{Boldyrev}.
\newblock {Kolmogorov-Burgers Model for Star-forming Turbulence}.
\newblock {\em \apj}, 569:841--845, April 2002.

\bibitem{Kritsuk_etal:2007}
A.~G. {Kritsuk}, M.~L. {Norman}, P.~{Padoan}, and R.~{Wagner}.
\newblock {The Statistics of Supersonic Isothermal Turbulence}.
\newblock {\em \apj}, 665:416--431, August 2007.

\bibitem{Boldyrev:2006}
S.~{Boldyrev}.
\newblock {Spectrum of Magnetohydrodynamic Turbulence}.
\newblock {\em Physical Review Letters}, 96(11):115002, March 2006.

\bibitem{BeresnyakLazarian:2006}
A.~{Beresnyak} and A.~{Lazarian}.
\newblock {Polarization Intermittency and Its Influence on MHD Turbulence}.
\newblock {\em \apjl}, 640:L175--L178, April 2006.

\bibitem{Gogoberidze:2007}
G.~{Gogoberidze}.
\newblock {On the nature of incompressible magnetohydrodynamic turbulence}.
\newblock {\em Physics of Plasmas}, 14(2):022304, February 2007.

\bibitem{Iroshnikov:1964}
P.~S. {Iroshnikov}.
\newblock {Turbulence of a Conducting Fluid in a Strong Magnetic Field}.
\newblock {\em \sovast}, 7:566, February 1964.

\bibitem{Kraichnan:1965}
R.~H. {Kraichnan}.
\newblock {Inertial-Range Spectrum of Hydromagnetic Turbulence}.
\newblock {\em Physics of Fluids}, 8:1385--1387, July 1965.

\bibitem{Beresnyak:2013}
A.~{Beresnyak}.
\newblock {Comment on Perez et al [PRX 2, 041005 (2012), arXiv:1209.2011]}.
\newblock {\em ArXiv e-prints}, January 2013.

\bibitem{Beresnyak:2014}
A.~{Beresnyak}.
\newblock {Reply to Comment on ''Spectra of strong magnetohydrodynamic
  turbulence from high-resolution simulations''}.
\newblock {\em ArXiv e-prints}, October 2014.

\bibitem{BeresnyakLazarian:2010}
A.~{Beresnyak} and A.~{Lazarian}.
\newblock {Scaling Laws and Diffuse Locality of Balanced and Imbalanced
  Magnetohydrodynamic Turbulence}.
\newblock {\em \apjl}, 722:L110--L113, October 2010.

\bibitem{Roberts:2010}
D.~A. {Roberts}.
\newblock {Evolution of the spectrum of solar wind velocity fluctuations from
  0.3 to 5 AU}.
\newblock {\em Journal of Geophysical Research (Space Physics)}, 115:12101,
  December 2010.

\bibitem{Perez_etal:2012}
J.~C. {Perez}, J.~{Mason}, S.~{Boldyrev}, and F.~{Cattaneo}.
\newblock {On the Energy Spectrum of Strong Magnetohydrodynamic Turbulence}.
\newblock {\em Physical Review X}, 2(4):041005, October 2012.

\bibitem{Perez_etal:2014}
J.~C. {Perez}, J.~{Mason}, S.~{Boldyrev}, and F.~{Cattaneo}.
\newblock {Comment on the numerical measurements of the magnetohydrodynamic
  turbulence spectrum by A. Beresnyak (Phys. Rev. Lett. 106 (2011) 075001;
  MNRAS 422 (2012) 3495; ApJ 784 (2014) L20)}.
\newblock {\em ArXiv e-prints}, September 2014.

\bibitem{BrandenburgLazarian:2013}
A.~{Brandenburg} and A.~{Lazarian}.
\newblock {Astrophysical Hydromagnetic Turbulence}.
\newblock {\em \ssr}, 178:163--200, October 2013.

\bibitem{Lazarian:2013}
A.~{Lazarian}.
\newblock {Reconnection Diffusion, Star Formation, and Numerical Simulations}.
\newblock In N.~V. {Pogorelov}, E.~{Audit}, and G.~P. {Zank}, editors, {\em
  {Numerical Modeling of Space Plasma Flows (ASTRONUM2012)}}, volume 474 of
  {\em {Astronomical Society of the Pacific Conference Series}}, page~15, April
  2013.

\bibitem{BeresnyakLazarian:2015}
A.~{Beresnyak} and A.~{Lazarian}.
\newblock {MHD Turbulence, Turbulent Dynamo and Applications}.
\newblock In A.~{Lazarian}, E.~M. {de Gouveia Dal Pino}, and C.~{Melioli},
  editors, {\em {Astrophysics and Space Science Library}}, volume 407 of {\em
  {Astrophysics and Space Science Library}}, page 163, 2015.

\bibitem{Lazarian:2006}
A.~{Lazarian}.
\newblock {Enhancement and Suppression of Heat Transfer by MHD Turbulence}.
\newblock {\em \apjl}, 645:L25--L28, July 2006.

\bibitem{Parker:1957}
E.~N. {Parker}.
\newblock {Sweet's Mechanism for Merging Magnetic Fields in Conducting Fluids}.
\newblock {\em \jgr}, 62:509--520, December 1957.

\bibitem{Sweet:1958}
P.~A. {Sweet}.
\newblock {The topology of force-free magnetic fields}.
\newblock {\em The Observatory}, 78:30--32, February 1958.

\bibitem{Lazarian_etal:2004}
A.~{Lazarian}, E.~T. {Vishniac}, and J.~{Cho}.
\newblock {Magnetic Field Structure and Stochastic Reconnection in a Partially
  Ionized Gas}.
\newblock {\em \apj}, 603:180--197, March 2004.

\bibitem{EyinkBenveniste:2013}
G.~L. {Eyink} and D.~{Benveniste}.
\newblock {Diffusion approximation in turbulent two-particle dispersion}.
\newblock {\em \pre}, 88(4):041001, October 2013.

\bibitem{LazarianYan:2014}
A.~{Lazarian} and H.~{Yan}.
\newblock {Superdiffusion of Cosmic Rays: Implications for Cosmic Ray
  Acceleration}.
\newblock {\em \apj}, 784:38, March 2014.

\bibitem{Kupiainen:2003}
A.~{Kupiainen}.
\newblock {Nondeterministic Dynamics and Turbulent Transport}.
\newblock {\em Annales Henri Poincar\'{e}}, 4:713--726, December 2003.

\bibitem{Schindler_etal:1988}
K.~{Schindler}, M.~{Hesse}, and J.~{Birn}.
\newblock {General magnetic reconnection, parallel electric fields, and
  helicity}.
\newblock {\em \jgr}, 93:5547--5557, June 1988.

\bibitem{HesseSchindler:1988}
M.~{Hesse} and K.~{Schindler}.
\newblock {A theoretical foundation of general magnetic reconnection}.
\newblock {\em \jgr}, 93:5559--5567, June 1988.

\bibitem{Susino_etal:2013}
R.~{Susino}, A.~{Bemporad}, and S.~{Krucker}.
\newblock {Plasma Heating in a Post Eruption Current Sheet: A Case Study Based
  on Ultraviolet, Soft, and Hard X-Ray Data}.
\newblock {\em \apj}, 777:93, November 2013.

\bibitem{KimDiamond:2001}
E.-j. {Kim} and P.~H. {Diamond}.
\newblock {On Turbulent Reconnection}.
\newblock {\em \apj}, 556:1052--1065, August 2001.

\bibitem{Cho_etal:2002}
J.~{Cho}, A.~{Lazarian}, and E.~T. {Vishniac}.
\newblock {Simulations of Magnetohydrodynamic Turbulence in a Strongly
  Magnetized Medium}.
\newblock {\em \apj}, 564:291--301, January 2002.

\bibitem{Cho_etal:2003}
J.~{Cho}, A.~{Lazarian}, and E.~T. {Vishniac}.
\newblock {MHD Turbulence: Scaling Laws and Astrophysical Implications}.
\newblock In E.~{Falgarone} and T.~{Passot}, editors, {\em {Turbulence and
  Magnetic Fields in Astrophysics}}, volume 614 of {\em {Lecture Notes in
  Physics, Berlin Springer Verlag}}, pages 56--98, 2003.

\bibitem{RechesterRosenbluth:1978}
A.~B. {Rechester} and M.~N. {Rosenbluth}.
\newblock {Electron heat transport in a Tokamak with destroyed magnetic
  surfaces}.
\newblock {\em Physical Review Letters}, 40:38--41, January 1978.

\bibitem{LazarianVishniac:2009}
A.~{Lazarian} and E.~T. {Vishniac}.
\newblock {Model of Reconnection of Weakly Stochastic Magnetic Field and its
  Implications}.
\newblock In {\em {Revista Mexicana de Astronomia y Astrofisica Conference
  Series}}, volume~36 of {\em {Revista Mexicana de Astronomia y Astrofisica,
  vol. 27}}, pages 81--88, August 2009.

\bibitem{Karimabadi_etal:2013}
H.~{Karimabadi}, V.~{Roytershteyn}, M.~{Wan}, W.~H. {Matthaeus}, W.~{Daughton},
  P.~{Wu}, M.~{Shay}, B.~{Loring}, J.~{Borovsky}, E.~{Leonardis}, S.~C.
  {Chapman}, and T.~K.~M. {Nakamura}.
\newblock {Coherent structures, intermittent turbulence, and dissipation in
  high-temperature plasmas}.
\newblock {\em Physics of Plasmas}, 20(1):012303, January 2013.

\bibitem{Beresnyak:2013b}
A.~{Beresnyak}.
\newblock {On the Rate of Spontaneous Magnetic Reconnection}.
\newblock {\em ArXiv e-prints}, January 2013.

\bibitem{Kowal_etal:2015}
G.~Kowal, D.~A. Falceta-Gon\, {c}alves, A.~Lazarian, and E.~T. Vishniac.
\newblock {Turbulence generated by reconnection}.
\newblock 2015.

\bibitem{Beresnyak:2012}
A.~{Beresnyak}.
\newblock {Basic properties of magnetohydrodynamic turbulence in the inertial
  range}.
\newblock {\em \mnras}, 422:3495--3502, June 2012.

\bibitem{BeresnyakLazarian:2008}
A.~{Beresnyak} and A.~{Lazarian}.
\newblock {Strong Imbalanced Turbulence}.
\newblock {\em \apj}, 682:1070--1075, August 2008.

\bibitem{Cho:2005}
J.~{Cho}.
\newblock {Simulations of Relativistic Force-free Magnetohydrodynamic
  Turbulence}.
\newblock {\em \apj}, 621:324--327, March 2005.

\bibitem{ChoLazarian:2014}
J.~{Cho} and A.~{Lazarian}.
\newblock {Imbalanced Relativistic Force-free Magnetohydrodynamic Turbulence}.
\newblock {\em \apj}, 780:30, January 2014.

\bibitem{LazarianYan:2012}
A.~{Lazarian} and H.~{Yan}.
\newblock {Magnetic reconnection in turbulent plasmas and gamma ray bursts}.
\newblock In F.~A. {Aharonian}, W.~{Hofmann}, and F.~M. {Rieger}, editors, {\em
  {American Institute of Physics Conference Series}}, volume 1505 of {\em
  {American Institute of Physics Conference Series}}, pages 101--115, December
  2012.

\bibitem{Lazarian_etal:2003}
A.~{Lazarian}, V.~{Petrosian}, H.~{Yan}, and J.~{Cho}.
\newblock {Physics of Gamma-Ray Bursts: Turbulence, Energy Transfer and
  Reconnection}.
\newblock {\em ArXiv Astrophysics e-prints}, January 2003.

\bibitem{deGouveiaDalPinoLazarian:2005}
E.~M. {de Gouveia dal Pino} and A.~{Lazarian}.
\newblock {Production of the large scale superluminal ejections of the
  microquasar GRS 1915+105 by violent magnetic reconnection}.
\newblock {\em \aap}, 441:845--853, October 2005.

\bibitem{Giannios:2013}
D.~{Giannios}.
\newblock {Reconnection-driven plasmoids in blazars: fast flares on a slow
  envelope}.
\newblock {\em \mnras}, 431:355--363, May 2013.

\bibitem{Lazarian_etal:2014}
A.~{Lazarian}, G.~{Eyink}, E.~{Vishniac}, and G.~{Kowal}.
\newblock {Reconnection in Turbulent Astrophysical Fluids}.
\newblock In N.~V. {Pogorelov}, E.~{Audit}, and G.~P. {Zank}, editors, {\em
  {8th International Conference of Numerical Modeling of Space Plasma Flows
  (ASTRONUM 2013)}}, volume 488 of {\em {Astronomical Society of the Pacific
  Conference Series}}, page~23, September 2014.

\bibitem{Park_etal:1984}
W.~{Park}, D.~A. {Monticello}, and R.~B. {White}.
\newblock {Reconnection rates of magnetic fields including the effects of
  viscosity}.
\newblock {\em Physics of Fluids}, 27:137--149, January 1984.

\bibitem{Vishniac_etal:2012}
E.~T. {Vishniac}, S.~{Pillsworth}, G.~{Eyink}, G.~{Kowal}, A.~{Lazarian}, and
  S.~{Murray}.
\newblock {Reconnection current sheet structure in a turbulent medium}.
\newblock {\em Nonlinear Processes in Geophysics}, 19:605--610, November 2012.

\bibitem{Maron_etal:2004}
J.~{Maron}, B.~D. {Chandran}, and E.~{Blackman}.
\newblock {Divergence of Neighboring Magnetic-Field Lines and Fast-Particle
  Diffusion in Strong Magnetohydrodynamic Turbulence, with Application to
  Thermal Conduction in Galaxy Clusters}.
\newblock {\em Physical Review Letters}, 92(4):045001, January 2004.

\bibitem{Eyink_etal:2013}
G.~{Eyink}, E.~{Vishniac}, C.~{Lalescu}, H.~{Aluie}, K.~{Kanov},
  K.~{B\"{u}rger}, R.~{Burns}, C.~{Meneveau}, and A.~{Szalay}.
\newblock {Flux-freezing breakdown in high-conductivity magnetohydrodynamic
  turbulence}.
\newblock {\em \nat}, 497:466--469, May 2013.

\bibitem{Loureiro_etal:2013}
N.~F. {Loureiro}, A.~A. {Schekochihin}, and D.~A. {Uzdensky}.
\newblock {Plasmoid and Kelvin-Helmholtz instabilities in Sweet-Parker current
  sheets}.
\newblock {\em \pre}, 87(1):013102, January 2013.

\bibitem{CiaravellaRaymond:2008}
A.~{Ciaravella} and J.~C. {Raymond}.
\newblock {The Current Sheet Associated with the 2003 November 4 Coronal Mass
  Ejection: Density, Temperature, Thickness, and Line Width}.
\newblock {\em \apj}, 686:1372--1382, October 2008.

\bibitem{Sych_etal:2009}
R.~{Sych}, V.~M. {Nakariakov}, M.~{Karlicky}, and S.~{Anfinogentov}.
\newblock {Relationship between wave processes in sunspots and quasi-periodic
  pulsations in active region flares}.
\newblock {\em \aap}, 505:791--799, October 2009.

\bibitem{Gosling:2012}
J.~T. {Gosling}.
\newblock {Magnetic Reconnection in the Solar Wind}.
\newblock {\em \ssr}, 172:187--200, November 2012.

\bibitem{Gosling_etal:2007}
J.~T. {Gosling}, T.~D. {Phan}, R.~P. {Lin}, and A.~{Szabo}.
\newblock {Prevalence of magnetic reconnection at small field shear angles in
  the solar wind}.
\newblock {\em Geophysical Research Letters}, 34:15110, August 2007.

\bibitem{GoslingSzabo:2008}
J.~T. {Gosling} and A.~{Szabo}.
\newblock {Bifurcated current sheets produced by magnetic reconnection in the
  solar wind}.
\newblock {\em Journal of Geophysical Research (Space Physics)}, 113:10103,
  October 2008.

\bibitem{Vasquez_etal:2007}
B.~J. {Vasquez}, V.~I. {Abramenko}, D.~K. {Haggerty}, and C.~W. {Smith}.
\newblock {Numerous small magnetic field discontinuities of Bartels rotation
  2286 and the potential role of Alfv\'{e}nic turbulence}.
\newblock {\em Journal of Geophysical Research (Space Physics)}, 112:11102,
  November 2007.

\bibitem{Phan_etal:2009}
T.~D. {Phan}, J.~T. {Gosling}, and M.~S. {Davis}.
\newblock {Prevalence of extended reconnection X-lines in the solar wind at 1
  AU}.
\newblock {\em Geophysical Research Letters}, 36:9108, May 2009.

\bibitem{Lalescu_etal:2013}
C.~{Lalescu}, G.~{Eyink}, K.~{Kanov}, R.~{Burns}, C.~{Meneveau}, A.~{Szalay},
  E.~{Vishniac}, H.~{Aluie}, and K.~{B\"{u}rger}.
\newblock {Flux-freezing breakdown observed in high-conductivity
  magnetohydrodynamic turbulence}.
\newblock In {\em {APS April Meeting Abstracts}}, page 2003, April 2013.

\bibitem{Osman_etal:2014}
K.~T. {Osman}, W.~H. {Matthaeus}, J.~T. {Gosling}, A.~{Greco}, S.~{Servidio},
  B.~{Hnat}, S.~C. {Chapman}, and T.~D. {Phan}.
\newblock {Magnetic Reconnection and Intermittent Turbulence in the Solar
  Wind}.
\newblock {\em Physical Review Letters}, 112(21):215002, May 2014.

\bibitem{SonnerupCahill:1967}
B.~U.~O. {Sonnerup} and L.~J. {Cahill}, Jr.
\newblock {Magnetopause Structure and Attitude from Explorer 12 Observations}.
\newblock {\em \jgr}, 72:171, January 1967.

\bibitem{Zhdankin_etal:2012}
V.~{Zhdankin}, S.~{Boldyrev}, J.~{Mason}, and J.~C. {Perez}.
\newblock {Magnetic Discontinuities in Magnetohydrodynamic Turbulence and in
  the Solar Wind}.
\newblock {\em Physical Review Letters}, 108(17):175004, April 2012.

\bibitem{LazarianOpher:2009}
A.~{Lazarian} and M.~{Opher}.
\newblock {A Model of Acceleration of Anomalous Cosmic Rays by Reconnection in
  the Heliosheath}.
\newblock {\em \apj}, 703:8--21, September 2009.

\bibitem{Parker:1958}
E.~N. {Parker}.
\newblock {Dynamics of the Interplanetary Gas and Magnetic Fields.}
\newblock {\em \apj}, 128:664, November 1958.

\bibitem{Burlaga_etal:1982}
L.~F. {Burlaga}, R.~P. {Lepping}, K.~W. {Behannon}, L.~W. {Klein}, and F.~M.
  {Neubauer}.
\newblock {Large-scale variations of the interplanetary magnetic field -
  Voyager 1 and 2 observations between 1-5 AU}.
\newblock {\em \jgr}, 87:4345--4353, June 1982.

\bibitem{KhabarovaObridko:2012}
O.~{Khabarova} and V.~{Obridko}.
\newblock {Puzzles of the Interplanetary Magnetic Field in the Inner
  Heliosphere}.
\newblock {\em \apj}, 761:82, December 2012.

\bibitem{Kulsrud:2005}
R.~M. {Kulsrud}.
\newblock {\em {Plasma physics for astrophysics}}.
\newblock 2005.

\bibitem{Bernard_etal:1998}
D.~{Bernard}, K.~{Gawedzki}, and A.~{Kupiainen}.
\newblock {Slow Modes in Passive Advection}.
\newblock {\em Journal of Statistical Physics}, 90:519--569, February 1998.

\bibitem{GawedzkiVergassola:2000}
K.~{Gaw{\c e}dzki} and M.~{Vergassola}.
\newblock {Phase transition in the passive scalar advection}.
\newblock {\em Physica D Nonlinear Phenomena}, 138:63--90, April 2000.

\bibitem{EVandenEijnden:2000a}
W.~{E} and E.~{Vanden Eijnden}.
\newblock {Another note on forced burgers turbulence}.
\newblock {\em Physics of Fluids}, 12:149--154, January 2000.

\bibitem{EVandenEijnden:2000b}
W.~{E} and E.~{vanden Eijnden}.
\newblock {Generalized flows, intrinsic stochasticity, and turbulent
  transport}.
\newblock {\em Proceedings of the National Academy of Science}, 97:8200--8205,
  July 2000.

\bibitem{EVandenEijnden:2001}
W.~{E} and E.~{Vanden-Eijnden}.
\newblock {Turbulent Prandtl number effect on passive scalar advection}.
\newblock {\em Physica D Nonlinear Phenomena}, 152:636--645, May 2001.

\bibitem{Chaves_etal:2003}
M.~{Chaves}, K.~{Gawedzki}, P.~{Horvai}, A.~{Kupiainen}, and N.~{Vergassola}.
\newblock {Lagrangian dispersion in Gaussian self-similar ensembles}.
\newblock {\em eprint arXiv:nlin/0303031}, March 2003.

\bibitem{Gawedzki:2008}
K.~{Gawedzki}.
\newblock {Stochastic processes in turbulent transport}.
\newblock {\em ArXiv e-prints}, June 2008.

\bibitem{Axford:1984}
W.~I. {Axford}.
\newblock {Magnetic field reconnection}.
\newblock {\em Washington DC American Geophysical Union Geophysical Monograph
  Series}, 30:1--8, 1984.

\bibitem{Lazarian:2005}
A.~{Lazarian}.
\newblock {Astrophysical Implications of Turbulent Reconnection: from cosmic
  rays to star formation}.
\newblock In E.~M. {de Gouveia dal Pino}, G.~{Lugones}, and A.~{Lazarian},
  editors, {\em {Magnetic Fields in the Universe: From Laboratory and Stars to
  Primordial Structures.}}, volume 784 of {\em {American Institute of Physics
  Conference Series}}, pages 42--53, September 2005.

\bibitem{SantosLima_etal:2010}
R.~{Santos-Lima}, A.~{Lazarian}, E.~M. {de Gouveia Dal Pino}, and J.~{Cho}.
\newblock {Diffusion of Magnetic Field and Removal of Magnetic Flux from Clouds
  Via Turbulent Reconnection}.
\newblock {\em \apj}, 714:442--461, May 2010.

\bibitem{SantosLima_etal:2012}
R.~{Santos-Lima}, E.~M. {de Gouveia Dal Pino}, and A.~{Lazarian}.
\newblock {The Role of Turbulent Magnetic Reconnection in the Formation of
  Rotationally Supported Protostellar Disks}.
\newblock {\em \apj}, 747:21, March 2012.

\bibitem{deGouveiaDalPino_etal:2012}
E.~M. {de Gouveia Dal Pino}, M.~R.~M. {Le\~{a}o}, R.~{Santos-Lima},
  G.~{Guerrero}, G.~{Kowal}, and A.~{Lazarian}.
\newblock {Magnetic flux transport by turbulent reconnection in astrophysical
  flows}.
\newblock {\em \physscr}, 86(1):018401, July 2012.

\bibitem{Leao_etal:2013}
M.~R.~M. {Le\~{a}o}, E.~M. {de Gouveia Dal Pino}, R.~{Santos-Lima}, and
  A.~{Lazarian}.
\newblock {The Collapse of Turbulent Cores and Reconnection Diffusion}.
\newblock {\em \apj}, 777:46, November 2013.

\bibitem{Lazarian:2014}
A.~{Lazarian}.
\newblock {Reconnection Diffusion in Turbulent Fluids and Its Implications for
  Star Formation}.
\newblock {\em \ssr}, 181:1--59, May 2014.

\bibitem{Lazarian:2011}
A.~{Lazarian}.
\newblock {Fast Reconnection and Reconnection Diffusion: Implications for Star
  Formation}.
\newblock {\em ArXiv e-prints}, November 2011.

\bibitem{Lynch_etal:2008}
D.~K. {Lynch}, C.~E. {Woodward}, R.~{Gehrz}, L.~A. {Helton}, R.~J. {Rudy},
  R.~W. {Russell}, R.~{Pearson}, C.~C. {Venturini}, S.~{Mazuk}, J.~{Rayner},
  J.-U. {Ness}, S.~{Starrfield}, R.~M. {Wagner}, J.~P. {Osborne}, K.~{Page},
  R.~C. {Puetter}, R.~B. {Perry}, G.~{Schwarz}, K.~{Vanlandingham}, J.~{Black},
  M.~{Bode}, A.~{Evans}, T.~{Geballe}, M.~{Greenhouse}, P.~{Hauschildt},
  J.~{Krautter}, W.~{Liller}, J.~{Lyke}, J.~{Truran}, T.~{Kerr}, S.~P.~S.
  {Eyres}, and S.~N. {Shore}.
\newblock {Nova V2362 Cygni (nova Cygni 2006): Spitzer, Swift, and Ground-Based
  Spectral Evolution}.
\newblock {\em \aj}, 136:1815--1827, November 2008.

\bibitem{Bemporad:2008}
A.~{Bemporad}.
\newblock {Spectroscopic Detection of Turbulence in Post-CME Current Sheets}.
\newblock {\em \apj}, 689:572--584, December 2008.

\bibitem{Singh_etal:2014}
C.~B. {Singh}, E.~M. {de Gouveia Dal Pino}, and L.~H.~S. {Kadowaki}.
\newblock {On the role of fast magnetic reconnection in accreting black hole
  sources}.
\newblock {\em ArXiv e-prints}, November 2014.

\bibitem{Khiali_etal:2014}
B.~{Khiali}, E.~M. {de Gouveia Dal Pino}, and M.~V. {del Valle}.
\newblock {A magnetic reconnection model for explaining the multi-wavelength
  emission of the microquasars Cyg X-1 and Cyg X-3}.
\newblock {\em ArXiv e-prints}, June 2014.

\bibitem{Lazarian_etal:2012}
A.~{Lazarian}, G.~{Kowal}, and B.~{Douveia dal Pino}.
\newblock {Astrophysical Reconnection and Particle Acceleration}.
\newblock In N.~V. {Pogorelov}, J.~A. {Font}, E.~{Audit}, and G.~P. {Zank},
  editors, {\em {Numerical Modeling of Space Plasma Slows (ASTRONUM 2011)}},
  volume 459 of {\em {Astronomical Society of the Pacific Conference Series}},
  page~21, July 2012.

\bibitem{Drury:2012}
L.~O. {Drury}.
\newblock {First-order Fermi acceleration driven by magnetic reconnection}.
\newblock {\em \mnras}, 422:2474--2476, May 2012.

\bibitem{LazarianDesiati:2010}
A.~{Lazarian} and P.~{Desiati}.
\newblock {Magnetic Reconnection as the Cause of Cosmic Ray Excess from the
  Heliospheric Tail}.
\newblock {\em \apj}, 722:188--196, October 2010.

\bibitem{Uzdensky_etal:2010}
D.~A. {Uzdensky}, N.~F. {Loureiro}, and A.~A. {Schekochihin}.
\newblock {Fast Magnetic Reconnection in the Plasmoid-Dominated Regime}.
\newblock {\em Physical Review Letters}, 105(23):235002, December 2010.

\bibitem{MatthaeusLamkin:1985}
W.~H. {Matthaeus} and S.~L. {Lamkin}.
\newblock {Rapid magnetic reconnection caused by finite amplitude
  fluctuations}.
\newblock {\em Physics of Fluids}, 28:303--307, January 1985.

\bibitem{Watson_etal:2007}
P.~G. {Watson}, S.~{Oughton}, and I.~J.~D. {Craig}.
\newblock {The impact of small-scale turbulence on laminar magnetic
  reconnection}.
\newblock {\em Physics of Plasmas}, 14(3):032301, March 2007.

\bibitem{Servidio_etal:2010}
S.~{Servidio}, W.~H. {Matthaeus}, M.~A. {Shay}, P.~{Dmitruk}, P.~A. {Cassak},
  and M.~{Wan}.
\newblock {Statistics of magnetic reconnection in two-dimensional
  magnetohydrodynamic turbulence}.
\newblock {\em Physics of Plasmas}, 17(3):032315, March 2010.

\bibitem{Loureiro_etal:2009}
N.~F. {Loureiro}, D.~A. {Uzdensky}, A.~A. {Schekochihin}, S.~C. {Cowley}, and
  T.~A. {Yousef}.
\newblock {Turbulent magnetic reconnection in two dimensions}.
\newblock {\em \mnras}, 399:L146--L150, October 2009.

\bibitem{KulpaDybel_etal:2010}
K.~{Kulpa-Dybe{\l}}, G.~{Kowal}, K.~{Otmianowska-Mazur}, A.~{Lazarian}, and
  E.~{Vishniac}.
\newblock {Reconnection in weakly stochastic B-fields in 2D}.
\newblock {\em \aap}, 514:A26, May 2010.

\bibitem{HigashimoriHoshino:2012}
K.~{Higashimori} and M.~{Hoshino}.
\newblock {The relation between ion temperature anisotropy and formation of
  slow shocks in collisionless magnetic reconnection}.
\newblock {\em Journal of Geophysical Research (Space Physics)}, 117:1220,
  January 2012.

\bibitem{HameiriBhattacharjee:1987}
E.~{Hameiri} and A.~{Bhattacharjee}.
\newblock {Turbulent magnetic diffusion and magnetic field reversal}.
\newblock {\em Physics of Fluids}, 30:1743--1755, June 1987.

\bibitem{DiamondMalkov:2003}
P.~H. {Diamond} and M.~{Malkov}.
\newblock {Dynamics of helicity transport and Taylor relaxation}.
\newblock {\em Physics of Plasmas}, 10:2322--2329, June 2003.

\bibitem{TennekesLumley:1972}
H.~{Tennekes} and J.~L. {Lumley}.
\newblock {\em {First Course in Turbulence}}.
\newblock 1972.

\bibitem{EyinkAluie:2006}
G.~L. {Eyink} and H.~{Aluie}.
\newblock {The breakdown of Alfv\'{e}n's theorem in ideal plasma flows:
  Necessary conditions and physical conjectures}.
\newblock {\em Physica D Nonlinear Phenomena}, 223:82--92, November 2006.

\bibitem{Politano_etal:1989}
H.~{Politano}, A.~{Pouquet}, and P.~L. {Sulem}.
\newblock {Inertial ranges and resistive instabilities in two-dimensional
  magnetohydrodynamic turbulence}.
\newblock {\em Physics of Fluids B}, 1:2330--2339, December 1989.

\bibitem{MininniPouquet:2009}
P.~D. {Mininni} and A.~{Pouquet}.
\newblock {Finite dissipation and intermittency in magnetohydrodynamics}.
\newblock {\em \pre}, 80(2):025401, August 2009.

\bibitem{LapentaBettarini:2011}
G.~{Lapenta} and L.~{Bettarini}.
\newblock {Spontaneous transition to a fast 3D turbulent reconnection regime}.
\newblock {\em EPL (Europhysics Letters)}, 93:65001, March 2011.

\bibitem{LapentaLazarian:2012}
G.~{Lapenta} and A.~{Lazarian}.
\newblock {Achieving fast reconnection in resistive MHD models via turbulent
  means}.
\newblock {\em Nonlinear Processes in Geophysics}, 19:251--263, April 2012.

\bibitem{Matthaeus_etal:1990}
W.~H. {Matthaeus}, M.~L. {Goldstein}, and D.~A. {Roberts}.
\newblock {Evidence for the presence of quasi-two-dimensional nearly
  incompressible fluctuations in the solar wind}.
\newblock {\em \jgr}, 95:20673--20683, December 1990.

\bibitem{Boozer:2012}
A.~H. {Boozer}.
\newblock {Separation of magnetic field lines}.
\newblock {\em Physics of Plasmas}, 19(11):112901, November 2012.

\bibitem{Boozer:2013}
A.~H. {Boozer}.
\newblock {Tokamak halo currents}.
\newblock {\em Physics of Plasmas}, 20(8):082510, August 2013.

\bibitem{UzdenskyRightley:2014}
D.~A. {Uzdensky} and S.~{Rightley}.
\newblock {Plasma physics of extreme astrophysical environments}.
\newblock {\em Reports on Progress in Physics}, 77(3):036902, March 2014.

\bibitem{BrowningLazarian:2013}
P.~{Browning} and A.~{Lazarian}.
\newblock {Notes on Magnetohydrodynamics of Magnetic Reconnection in Turbulent
  Media}.
\newblock {\em \ssr}, 178:325--355, October 2013.

\bibitem{Daughton_etal:2008}
W.~{Daughton}, V.~{Roytershteyn}, B.~J. {Albright}, K.~{Bowers}, L.~{Yin}, and
  H.~{Karimabadi}.
\newblock {Reconnection Dynamics in Semi-Collisional Plasmas}.
\newblock {\em AGU Fall Meeting Abstracts}, page A1705, December 2008.

\end{thebibliography}

\end{document}